\documentclass[aps,showpacs,twocolumn,dvips]{revtex4}
\usepackage{feynmp}
\usepackage{amssymb}
\usepackage{epsfig}
\usepackage{graphicx}
\usepackage{subfigure}
\usepackage{mathrsfs}
\usepackage{hyperref}
\usepackage{cleveref}
\usepackage{cases}
\usepackage{bbm}
\usepackage{xcolor}
\usepackage{tabularx}
\usepackage{array}
\usepackage{multirow}

\usepackage{lmodern}
\usepackage{textcomp}

\begin{document}



\title{Stability of two-dimensional asymmetric materials with a quadratic band crossing point
under four-fermion interaction and impurity scattering}

\date{\today}

\author{Yao-Ming Dong}
\affiliation{Department of Physics, Tianjin University, Tianjin 300072, People's Republic of China}

\author{Ya-Hui Zhai}
\affiliation{Department of Physics, Tianjin University, Tianjin 300072, People's Republic of China}

\author{Dong-Xing Zheng}
\affiliation{Department of Physics, Tianjin University, Tianjin 300072, People's Republic of China}

\author{Jing Wang}
\altaffiliation{Corresponding author: jing$\textunderscore$wang@tju.edu.cn}
\affiliation{Department of Physics, Tianjin University, Tianjin 300072, People's Republic of China}

\begin{abstract}
We investigate the impacts of combination of fermion-fermion interactions and
impurity scatterings on the low-energy stabilities of two-dimensional asymmetric materials
with a quadratic band crossing point by virtue of the renormalization group that allows us to
treat distinct sorts of physical ingredients on the same footing. The coupled flow evolutions of all
interaction parameters which carry the central physical information are derived by taking
into account one-loop corrections. Several intriguing results are manifestly extracted
from these entangled evolutions. At first, we realize that the quadratic band touching
structure is particularly robust once the fermionic couplings flow toward the Gaussian
fixed point. Otherwise, it can either be stable or broken down against
the impurity scattering in the vicinity of nontrivial fixed points.
In addition, we figure out two parameters $\eta$ and $\lambda$ that
measure rotational and particle-hole asymmetries are closely
energy-dependent and exhibit considerably abundant behaviors depending upon the fates
of fermion-fermion couplings and different types of impurities.
Incidentally, as both $\eta$ and $\lambda$ can be remarkably
increased or heavily reduced in the low-energy regime, an asymmetric
system under certain restricted conditions exhibits an interesting phenomenon
in which transitions either from rotational or particle-hole asymmetry to
symmetric situation would be activated.
\end{abstract}


\maketitle


\section{Introduction}

Semimetals with intermediate properties between metals
and insulators have been extensively studied and become one of the
most important fields in condensed matter physics~\cite{Lee2005Nature,
Neto2009RMP,Fu2007PRL,Roy2009PRB,Moore2010Nature,Hasan2010RMP,Qi2011RMP,
Sheng2012book,Bernevig2013book,Herbut2018Science}. These materials, including Dirac~\cite{Wang2012PRB,Young2012PRL,Steinberg2014PRL,Liu2014NM,Liu2014Science,
Xiong2015Science} and Weyl~\cite{Neto2009RMP,Burkov2011PRL,Yang2011PRB,
Wan2011PRB,Huang2015PRX,Xu2015Science,Xu2015NP,Lv2015NP,Weng2015PRX}
semimetals, conventionally have reduced Fermi surfaces that consist
of several discrete Dirac points with gapless low-energy excitations
irrespective of their microscopic details and exhibit linear energy
dispersions along two or three directions~\cite{Lee2005Nature,
Neto2009RMP,Fu2007PRL,Roy2009PRB,Moore2010Nature,Hasan2010RMP,
Qi2011RMP,Sheng2012book,Bernevig2013book,Korshunov2014PRB,Hung2016PRB,
Nandkishore2013PRB,Potirniche2014PRB,Nandkishore2017PRB,Sarma2016PRB,
Herbut2018Science}. Accompanying these developments, the two-dimensional
(2D) electronic system with a quadratic band crossing point (QBCP),
a ``cousin" of semimetal-like family featuring reduced Fermi surface
as well, has been recently attracting intense interest and becoming
one of the hottest topics in this area~\cite{Chong2008PRB,Fradkin2008PRB,
Fradkin2009PRL,Cvetkovic2012PRB,Murray2014PRB,Herbut2012PRB,Mandal2019CMP}.
Such 2D parabolically touching bands can arise on the Lieb lattice~\cite{Tsai2015NJP}
and certain collinear spin density wave states~\cite{Chern2012PRL} as well as the
checkerboard or kagome lattices~\cite{Fradkin2009PRL} at $\frac{1}{2}$ or $\frac{1}{3}$
filling, respectively. Besides, their three-dimensional counterparts have also
received much attention~\cite{Nandkishore2017PRB,Luttinger1956PR,
Murakami2004PRB,Janssen2015PRB,Boettcher2016PRB,Janssen2017PRB,Boettcher2017PRB,
Mandal2018PRB,Lin2018PRB,Savary2014PRX,Savary2017PRB,Vojta1810.07695,
Lai2014arXiv,Goswami2017PRB,Szabo2018arXiv,Foster2019PRB,Wang1911.09654}.

In sharp contrast to the standard 2D Dirac/Weyl materials, the reduced
Fermi surfaces in the 2D QBCP  materials are no longer the Dirac points but
instead replaced by discrete QBCPs in the Brillouin zone, which are
formed by the crossings of up and down quadratic bands~\cite{Fradkin2009PRL,
Murray2014PRB}. As a result, they possess very outlandish low-energy band
structures, namely both parabolical energy dispersions with $E_\mathbf{k}
\propto\mathbf{k}^2$ and gapless excitations. In addition, the density of
states is finite rather than zero~\cite{Fradkin2009PRL,Cvetkovic2012PRB}.
These unusual band structures of 2D QBCP systems together with their unique
low-energy excitations are pivotal to induce a plethora of fascinating
phenomenologies in the low-energy regime~\cite{Fradkin2009PRL,
Vafek2010PRB,Yang2010PRB,Murray2014PRB,Venderbos2016PRB,Wu2016PRL,
Zhu2016PRL,Wang2017PRB}. For instance, it was advocated that both
the quantum anomalous Hall (QAH) with time-reversal symmetry breaking
and quantum spin Hall (QSH) effect protected by time-reversal symmetry
would be generated by electron-electron repulsions in the checkerboard
lattice~\cite{Fradkin2009PRL,Murray2014PRB} or two-valley bilayer graphene
with QBCPs~\cite{Vafek2010PRB,Yang2010PRB}. Besides, Ref.~\cite{Wang2017PRB}
carefully investigated the low-energy topological instabilities against
distinct sorts of impurity scatterings.

It is imperative to point out that these achievements can be only
obtained once the 2D QBCP systems are invariant under both rotational and
particle-hole symmetries. In other words, the QBCP band structure must
be stable and these two kinds of symmetries are preserved during all the
processes. This means that asymmetric situations are insufficiently taken into
account in previous studies. Accordingly, some intriguing questions are
naturally raised if one begins with a 2D QBCP system that does not possess
rotational and particle-hole symmetries. For instance, whether the QBCP
band structure, namely the parabolic dispersion, is adequately robust and
the rotational and particle-hole asymmetries are energy-dependent
in the low-energy regime?  How can we quantificationally characterize
the rotational and particle-hole asymmetries? Unambiguously answering
these questions would be remarkably instructive to deeply understand
the low-energy behaviors of 2D QBCP systems.

To clearly response these inquiries, we within this work put our
focus on the 2D asymmetric QBCP systems. In principle, different types
of short-range fermion-fermion interactions are distinguished by the
Pauli matrixes of the coupling vertexes. Additionally, impurities
are always present in the real systems and are able to trigger
a number of prominent phenomena in the low-energy regime~\cite{Korshunov2014PRB,
Hung2016PRB,Nandkishore2013PRB,Potirniche2014PRB,Nandkishore2017PRB,Sarma2016PRB,
Lee1985RMP,Nersesyan1995NPB,Evers2008RMP,Efremov2011PRB,Efremov2013NJP,
Alavirad2016,Slager2016,Roy2016SR}. Depending on their different couplings with
fermions~\cite{Nersesyan1995NPB,Stauber2005PRB,
Wang2011PRB,Wang2013PRB}, they are clustered into three sorts
in the fermionic systems: random chemical potential, random mass,
and random gauge potential. In order to capture more physical information,
we endeavor to unbiasedly examine the effects of competition between four
types of fermion-fermion interactions and three kinds of impurity
scatterings by means of the momentum-shell renormalization-group (RG)
approach~\cite{Shankar1994RMP,Wilson1975RMP,Polchinski9210046} on
the 2D asymmetric QBCP materials. After collecting all the one-loop
corrections due to the interplay of fermion-fermion interactions and
impurity scatterings, the energy-dependent coupled flow equations of
all related interaction parameters are derived under the standard RG
analysis. To proceed, several intriguing results are extracted from
these RG evolutions. At the outset, we find that the band structure and dispersion
of 2D QBCP systems are considerably stable while fermion-fermion
interactions flow toward the Gaussian fixed point. In comparison,
both of them are robust under the presence of random mass but sabotaged by sufficiently
strong random chemical potential or random gauge potential
once fermionc couplings are governed by nontrivial fixed points.
Afterward, we carefully examine the low-energy behaviors of
rotational and particle-hole asymmetries which are characterized
by two parameters $\eta$ and $\lambda$. They show manifestly
energy-dependent and exhibit distinct fates such as remarkably increased
or heavily reduced based upon the starting values of fermion-fermion interactions
and impurities. Besides above primary results, we figure out that an interesting
transition from either rotational or particle-hole asymmetry to
symmetric situation would be triggered under certain restricted condition in
the 2D QBCP systems.

The rest of paper is organized as follows. In Sec.~\ref{Sec_model},
we provide our model and construct the effective theory for the 2D QBCP
system in the low-energy regime. The one-loop momentum-shell RG analysis is
followed in Sec.~\ref{Sec_RG_analysis}. We within Sec.~\ref{Sec_dispersion}
carefully examine the stability of QBCP's dispersion against distinct sorts
of impurities. In Sec.~\ref{Sec_eta} and Sec.~\ref{Sec_lambda},
we investigate in detail the low-energy fates of rotational and
particle-hole asymmetries under the influence of competitions
between fermion-fermion interactions and impurities, respectively.
Finally, we briefly summarize our primary results
in Sec.~\ref{Sec_summary}.


\section{Model and Effective theory}\label{Sec_model}

We hereby consider electrons on a checkerboard lattice that
is a typical model for the two-dimensional fermionic systems with a
quadratic band crossing point. As for this model, the low-energy
non-interacting Hamiltonian that respects the $C_{4v}$
point group can be derived via expanding the tight-binding model near
the corner of Brillouin zone, namely~\cite{Fradkin2009PRL}
\begin{eqnarray}
H_0=\sum_{|\mathbf{k}|<\Lambda}\sum_{\sigma=\uparrow\downarrow}
\psi^\dagger_{\mathbf{k}\sigma}
\mathcal{H}_0(\mathbf{k})\psi_{\mathbf{k}\sigma}.\label{Eq-H-0}
\end{eqnarray}
Here, $\Lambda$ is the momentum cutoff and $\psi_{\mathbf{k}\sigma}$ is a spinor
that consists of two components corresponding to sublattices A and B of checkerboard
lattice, respectively . In addition, the Hamiltonian density reads
\begin{eqnarray}
\mathcal{H}_0(\mathbf{k})
=t_0\mathbf{k}^2\tau_0+2t_1k_xk_y\tau_1+t_3(k^2_x-k^2_y)\tau_3.\label{Eq-H-01}
\end{eqnarray}
The index $\sigma$ denotes electron spin and $\tau_0$ specifies
the $2\times2$ identity matrix as well as $\tau_i$ with $i=1,3$
serves as Pauli matrixes. The parameters
$t_0,t_1,t_3$ are related to the hopping amplitudes of continuum
Hamiltonian. With respect to this free Hamiltonian~(\ref{Eq-H-0}),
the energy eigenvalues can be directly obtained and compactly
written as~\cite{Fradkin2009PRL,Murray2014PRB}
\begin{eqnarray}
E^{\pm}_{\mathbf{k}}=\frac{\mathbf{k}^2}{\sqrt{2}m}\left[\lambda\pm
\sqrt{\cos^{2}\eta\cos^{2}2\theta_{\mathbf{k}}
+\sin^2\eta\sin^{2}2\theta_{\mathbf{k}}}\right],~\label{Eq-E-k}
\end{eqnarray}
with bringing out $m\equiv1/\sqrt{2(t^2_1+t^2_3)}$, $\lambda\equiv t_0/\sqrt{t^2_1+t^2_3}$,
$\cos\eta\equiv t_3/\sqrt{t^2_1+t^2_3}$, and
$\sin\eta\equiv t_1/\sqrt{t^2_1+t^2_3}$ as well as designating $\theta_k\equiv\arctan k_y/k_x$~\cite{Murray2014PRB}. It is worth addressing two interesting quantities that
are closely determined by these parameters, namely the dispersion and
symmetry of QBCP system. On one hand, we highlight that the existence of a QBCP with the parabolical crossing
of up ($E^+$) and down ($E^-$) energy bands at $\mathbf{k}=0$
can only be realized under the constraint $E^{-}<0$, which directly leads to
\begin{eqnarray}
|t_0|<\mathrm{min}(|t_1|,|t_3|).\label{Eq-ti-ineq}
\end{eqnarray}
This implies that the quadratic band crossing point in the
Brillouin zone would vanish and then the dispersion of QBCP be changed
once the inequality~(\ref{Eq-ti-ineq}) is violated. On the other, the parameters
$\eta$ and $\lambda$ generally determine whether the system owns the rotational
and particle-hole symmetries. To be concrete, $\eta=\frac{\pi}{4}$ and $\lambda=0$
correspond to rotational and particle-hole symmetries, respectively.
Otherwise, these two symmetries are absent.
Without loss of generality, we within this work consider them unbiasedly.

In addition to the free part, we also consider the marginally short-range
fermion-fermion interactions that are of form~\cite{Fradkin2009PRL,Cvetkovic2012PRB}
\begin{eqnarray}
H_{\mathrm{int}}=\sum_i \frac{2\pi}{m}u_i\int d^2\mathbf{x}
\left(\sum_{\sigma=\uparrow\downarrow}\psi^{\dag}_{\sigma}
(\mathbf{x})\tau_i\psi_{\sigma}(\mathbf{x})\right)^2,\label{Eq-H-int}
\end{eqnarray}
where $u_i$ with $i=0,1,2,3$ characterizes the strength of fermion-fermion interaction.
To proceed, the impurities are present in nearly all realistic systems and play an
important role in determining the low-energy properties~\cite{Nersesyan1995NPB,Stauber2005PRB}.
This implies that fermion-fermion interactions and impurity scatterings must be treated
on equal footing. To this end, we introduce
the fermion-impurity part~\cite{Wang2017PRB,Nersesyan1995NPB,Stauber2005PRB,Wang2011PRB},
\begin{eqnarray}
S_{\mathrm{imp}}&=&
\sum^3_{i=0}v_i\int^{+\infty}_{-\infty}\frac{d\omega}{2\pi}
\int^\Lambda\frac{d^2\mathbf{k}^\prime d^2
\mathbf{k}}{(2\pi)^4}\psi^{\dag}_{\sigma}(\omega,\mathbf{k})\nonumber\\ \nonumber\\
&&\times M_i\psi_{\sigma}(\omega,\mathbf{k}^\prime)D_{i}
(\mathbf{k}-\mathbf{k}^\prime)\label{Eq-H-imp},
\end{eqnarray}
where the parameter $v_i$ with the index $i=0, 1,2,3$ is adopted to
characterize the strength of fermion-impurity coupling.
We here stress that impurity field $D_{i}(\mathbf{k})$ is a white-noise quenched impurity
designated by the correlation functions $\langle D_{i}(\mathbf{k})\rangle=0$ and
$\langle D_{i}(\mathbf{k})D_{i}(-\mathbf{k})\rangle=\Delta_{i}/\mathbf{k}^2$.
Here, the parameter $\Delta_i$ with $i=0,1,2,3$ that is constant serves as the concentrations of distinct sorts of impurities~\cite{Moon1409.0573,Aharony2018PRD,Wang2011PRB,
Nersesyan1995NPB,Coleman2015Book}, which is marginal
at the tree level in our 2D QBCP systems. Depending upon their
couplings with fermions~(\ref{Eq-H-imp}), $M_0=\tau_0$, $M_2=\tau_2$,
$M_1=\tau_1$, and $M_3=\tau_3$ correspond to random chemical potential, random mass,
random gauge potential (component-X), and random gauge potential (component-Z),
respectively. We hereby emphasize the random gauge potential is not the real ``gauge potential"
that is associated with the gauge invariance in the quantum field theory, but instead some
kind of impurity, which is closely bound up with variation of the density of state around the Fermi surface~\cite{Nersesyan1995NPB}.

It is convenient to work in the momentum space. To this end, gathering the free
terms and fermion-fermion interactions as well as fermion-impurity couplings,
we eventually obtain our low-energy effective theory,
\begin{widetext}
\begin{eqnarray}
S_{\mathrm{eff}}&=&\int^{+\infty}_{-\infty}\frac{d\omega}{2\pi}
\int^{\Lambda}\frac{d^2\mathbf{k}}{(2\pi)^2}\sum_{\sigma=\uparrow\downarrow}
\psi^\dagger_{\sigma}(\omega,\mathbf{k})\left[-i\omega+t_0\mathbf{k}^2\tau_0
+2t_1k_xk_y\tau_1+t_3(k^2_x-k^2_y)\tau_3\right]\psi_{\sigma}(\omega,\mathbf{k})\nonumber\\
&&+\frac{2\pi}{m}\sum^{3}_{i=0}u_i\int^{+\infty}_{-\infty}
\frac{d\omega_1d\omega_2d\omega_3}{(2\pi)^3}\int^{\Lambda}\frac{d^2\mathbf{k}_1d^2\mathbf{k}_2
d^2\mathbf{k}_3}{(2\pi)^6}\sum_{\sigma,\sigma'=\uparrow\downarrow}
\psi^\dagger_\sigma(\omega_1,\mathbf{k}_1)\tau_i
\psi_\sigma(\omega_2,\mathbf{k}_2)\psi^\dagger_{\sigma'}(\omega_3,\mathbf{k}_3)
\tau_i\nonumber\\
&&\times\psi_{\sigma'}(\omega_1+\omega_2-\omega_3,\mathbf{k}_1
+\mathbf{k}_2-\mathbf{k}_3)+
\sum^3_{i=0}v_i\int^{+\infty}_{-\infty}\frac{d\omega}{2\pi}\int^\Lambda\frac{d^2\mathbf{k}^\prime d^2
\mathbf{k}}{(2\pi)^4}\psi^{\dag}_{\sigma}(\omega,\mathbf{k})M_i\psi_{\sigma}
(\omega,\mathbf{k}^\prime)D_{i}(\mathbf{k}-\mathbf{k}^\prime).\label{Eq-S-eff}
\end{eqnarray}
\end{widetext}
According to this effective theory, one can easily extract the free fermionic propagator
\begin{eqnarray}
G_0(i\omega,\mathbf{k})\!=\!\frac{1}{-i\omega+t_0\mathbf{k}^2
+t_1k_xk_y\tau_1+t_3(k^2_x-k^2_y)\tau_3},\label{Eq-G-0}
\end{eqnarray}
which will be employed to derive the one-loop corrections for RG analysis.


\section{RG studies}\label{Sec_RG_analysis}

In order to capture low-energy properties that rely heavily upon
the competition between fermion-fermion interactions and impurities, we suggest
establishing the coupled energy-dependent connections among all interaction parameters
by means of momentum-shell RG approach~\cite{Murray2014PRB,Altland2006Book,Cvetkovic2012PRB}.
Along with the spirit of the RG method, we integrate out the fast modes of fermionic fields
within the momentum shell $b\Lambda<k<\Lambda$, where $\Lambda$ denotes
the energy scale and variable parameter $b$ can be specified as $b=e^{-l}<1$
with a running energy scale $l>0$, then collect these contributions to the slow modes,
and finally rescale the slow modes to new ``fast modes"~\cite{Wang2011PRB,Wang2013PRB,
Wang2017PRB,Huh2008PRB,Kim2008PRB,Maiti2010PRB,She2010PRB,She2015PRB,Cvetkovic2012PRB,
Murray2014PRB,Roy2016PRB}. To clinch the effective contributions
from the fast modes, we need to perform the calculations of one-loop corrections to interaction
parameters, namely the Feynman diagrams shown in Figs.~\ref{Fig_fermion_propagator_correction}-\ref{Fig_fermion-impurity_correction}
of Appendix~\ref{Appendix_one-loop-corrections}. After performing long but straightforward
calculations followed by similar steps in
Refs.~\cite{Murray2014PRB,Wang2017PRB,Wang2018JPCM,Wang2019JPCM},
we can obtain all these one-loop contributions that
are provided together in Appendix~\ref{Appendix_one-loop-corrections} and Appendix~\ref{Appendix_coefficients}.
Before going further, we choose the non-interacting parts of
effective action as a fixed point at which they are invariant during the RG
transformations. This yields to the RG rescaling transformations of fields and momenta~\cite{Murray2014PRB,Wang2011PRB,Huh2008PRB},
\begin{eqnarray}
k_x&\longrightarrow&k'_xe^{-l},\label{rescale_k_x}\\
k_y&\longrightarrow&k'_ye^{-l},\\
\omega&\longrightarrow&\omega'e^{-2l},\\
\psi(i\omega,\mathbf{k})&\longrightarrow&
\psi'(i\omega',\mathbf{k}')e^{\frac{1}{2}\int dl(6-\eta_f)},\\
D(\mathbf{k})&\longrightarrow&D'(\mathbf{k})\label{rescale_D},
\end{eqnarray}
where the parameter $\eta_f$ that is so-called anomalous dimension of fermionic spinor~\cite{Stauber2005PRB,Murray2014PRB,Wang2019JPCM}
collects the higher-order corrections caused by the interplay
between fermion-fermion interactions and impurity scatterings.
To simplify our calculations, one can measure the momenta and
energy with the cutoff $\Lambda_0$ that is linked to
the lattice constant, namely $k\rightarrow k/\Lambda_0$
and $\omega\rightarrow\omega/\Lambda_0$~\cite{Stauber2005PRB,
Murray2014PRB,Wang2011PRB,Huh2008PRB,She2010PRB}.
At this stage, we are in a suitable position to
derive the coupled flow RG equations of interaction parameters
via comparing new ``fast modes" with old
``fast modes" in the effective theory as follows [it is necessary to point out that
both fermion-fermion and fermion-impurity couplings are marginal
at the tree level in the 2D QBCP systems due to
the RG rescalings~(\ref{rescale_k_x})-(\ref{rescale_D})],
\begin{widetext}
\begin{eqnarray}
\frac{dt_0}{dl}
\!\!&=&\!\!-t_0(\Delta_0v^2_0+\Delta_1v^2_1+\Delta_2v^2_2
+\Delta_3v^2_3)\mathcal{N}_5,\label{Eq_t_0}\\
\frac{dt_1}{dl}
\!\!&=&\!\!-t_1\left[(\Delta_0v^2_0+\Delta_1v^2_1)\mathcal{N}_6
+(\Delta_2v^2_2+\Delta_3v^2_3)\mathcal{N}_5\right],\label{Eq_t_1}\\
\frac{dt_3}{dl}
\!\!&=&\!\!-t_3\left[(\Delta_0v^2_0+\Delta_3v^2_3)\mathcal{N}_6
-(\Delta_1v^2_1+\Delta_2v^2_2)\mathcal{N}_5\right],\label{Eq_t_3}\\
\frac{du_0}{dl}
\!\!&=&\!\!-\mathcal{C}_1
(u^2_0+u^2_1+u^2_2+u^2_3)
-\mathcal{C}_2(u_0u_1+u_2u_3)-\mathcal{C}_3(u_0u_3+u_1u_2)-
\mathcal{C}_4(u_0u_1-u_2u_3)
\nonumber\\
&&\!\!-\mathcal{C}_5(u_0u_3-u_1u_2)+(v^2_0\Delta_0\mathcal{D}_0
+v^2_1\Delta_1\mathcal{D}_1+v^2_3\Delta_3\mathcal{D}_2)u_0,\\
\frac{du_1}{dl}
\!\!&=&\!\!(2u_0u_1-2u^2_1-2u_2u_1-3u_3u_1-u_0 u_2)
\mathcal{C}_3+(u_0u_1+u_2u_3)
(\mathcal{C}_2+\mathcal{C}_3)-\frac{1}{2}(u^2_0+u^2_1+u^2_2+u^2_3)
(\mathcal{C}_2+\mathcal{C}_4)
\nonumber\\
&&\!\!-(u_0u_1-u_2u_3)
(\mathcal{C}_4+\mathcal{C}_5)-(u_1u_3-u_0u_2)
\mathcal{C}_5+\left(v^2_0\Delta_0\mathcal{D}_3+v^2_1\Delta_1
\mathcal{D}_4+v^2_2\Delta_2\mathcal{D}_5+v^2_3\Delta_3\mathcal{D}_6\right)u_1,\\
\frac{du_2}{dl}
\!\!&=&\!\!(3u_0u_2-2u_1u_2-2u^2_2-2u_3u_2+u_1u_3)
(\mathcal{C}_2+\mathcal{C}_3)-
(u_0u_2-u_1u_3)
(\mathcal{C}_4+\mathcal{C}_5)-(u_1u_2+u_0u_3)\mathcal{C}_2\nonumber\\
&&\!\!-(u_2u_3+u_0u_1)\mathcal{C}_3
-(u_1u_2-u_0u_3)\mathcal{C}_4-(u_2u_3-u_0u_1)\mathcal{C}_5+\left(v^2_0\Delta_0\mathcal{D}_7
+v^2_1\Delta_1\mathcal{D}_8+v^2_2\mathcal{D}_9
+v^2_3\Delta_3\mathcal{D}_{10}\right)u_2,\\
\frac{du_3}{dl}
\!\!&=&\!\!
(2u_0u_3-3u_1u_3-2u_2u_3-2u^2_3-u_0u_2)
\mathcal{C}_2+(u_1u_2+u_0u_3)
(\mathcal{C}_2+\mathcal{C}_3)
-\frac{1}{2}(u^2_0+u^2_1+u^2_2+u^2_3)(\mathcal{C}_5+
\mathcal{C}_3)\nonumber\\
&&\!\!-(u_0u_3-u_1u_2)(\mathcal{C}_4+\mathcal{C}_5)-
(u_1u_3-u_0u_2)\mathcal{C}_4+\left(v^2_0\Delta_0\mathcal{D}_{11}+v^2_1\Delta_1
\mathcal{D}_{12}-v^2_2\Delta_2\mathcal{D}_{13}+v^2_3\Delta_3\mathcal{D}_{14}\right)u_3,\\
\frac{d v_0}{dl}
\!\!&=&\!\!v_0\left[(v^2_0\Delta_0+v^2_1\Delta_1+v^2_2\Delta_2+v^2_3\Delta_3)
-8\pi\sqrt{2(t^2_1+t^2_3)}(u_0+u_1+u_2+u_3)\right]\mathcal{N}_1,\\
\frac{d v_1}{dl}
\!\!&=&\!\!v_1\Bigl[v^2_0\Delta_0\left(2\mathcal{N}_2-\mathcal{N}_1\right)
+v^2_1\Delta_1\left(2\mathcal{N}_2-\mathcal{N}_1\right)
-v^2_2\Delta_2\left(2\mathcal{N}_2+\mathcal{N}_1\right)
-v^2_3\Delta_3\left(2\mathcal{N}_2+\mathcal{N}_1\right)\nonumber\\
&&\!\!-8\pi\sqrt{2(t^2_1+t^2_3)}
(u_0+u_1-u_2-u_3)\mathcal{N}_2\Big],\\
\frac{d v_2}{dl}
\!\!&=&\!\!v_2\Bigl[-v^2_0\Delta_0(2\mathcal{N}_3+\mathcal{N}_1)
+v^2_1\Delta_1(2\mathcal{N}_3-\mathcal{N}_1)
-
v^2_2\Delta_2(2\mathcal{N}_3+\mathcal{N}_1)
+v^2_3\Delta_3(2\mathcal{N}_3-\mathcal{N}_1)\nonumber\\
&&\!\!
-8\pi\sqrt{2(t^2_1+t^2_3)}(-u_0+u_1-u_2+u_3)\mathcal{N}_3\Big],\\
\frac{d v_3}{dl}
\!\!&=&\!\!v_3\Big[-v^2_0\Delta_0(2\mathcal{N}_4+\mathcal{N}_1)
+v^2_1\Delta_1(2\mathcal{N}_4-\mathcal{N}_1)+
v^2_2\Delta_2(2\mathcal{N}_4-\mathcal{N}_1)
-v^2_3\Delta_3(2\mathcal{N}_4+\mathcal{N}_1)\nonumber\\
&&
\!\!-8\pi\sqrt{2(t^2_1+t^2_3)}(-u_0+u_1+u_2-u_3)
\mathcal{N}_4\Big],\label{Eq_v_3}
\end{eqnarray}
\end{widetext}
where the coefficients $\mathcal{C}$, $\mathcal{D}$, and
$\mathcal{N}$ are provided in Eqs.~(\ref{Eq_C})-(\ref{Eq-N-6}) of
Appendix~\ref{Appendix_coefficients} as well as $\Delta_i$ and $v_{i}$
with $i=0,1,2,3$ shown in Eq.~(\ref{Eq-H-imp}) are associated with the
concentrations of impurities and interactions between impurities and
fermions, respectively. In order to treat all types
of impurities unbiasedly, we take them equally at the starting point
within the following numerical calculations. Without loss of generality,
it is convenient to assume $\Delta_{i}=1$ during our RG analysis in that
the $\Delta_i$ are just some constants and basic results are insensitive
to concrete initial values.

\begin{figure}
\centering
\includegraphics[width=2.6in]{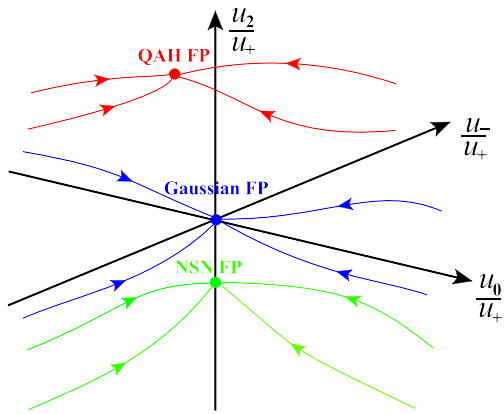}
\vspace{0.2cm}
\caption{(Color online) Schematic flows of
fermion-fermion couplings influenced by the interplay
of fermion-fermion interactions and impurity scatterings
with designating $u_+\equiv\frac{1}{2}(u_3+u_1)$ and
$u_-\equiv\frac{1}{2}(u_3-u_1)$. There are three distinct
types of FPs at the lowest-energy limit, namely Gaussian FP $(u_0,u_-,u_2)/u_+
\rightarrow(0,0,0)$, QAH FP $(u_0,u_-,u_2)/u_+
\rightarrow(0,-3.73,7.46)$, and NSN FP $(u_0,u_-,u_2)/u_+
\rightarrow(0,0,-1.09)$ at the weak impurity~\cite{Wang2017PRB}.}\label{Fig_schem_FPs}
\end{figure}

\section{Stability of QBCP's dispersion}\label{Sec_dispersion}

Reading from Eqs.~(\ref{Eq_t_0})-(\ref{Eq_t_3}), one can directly
realize that the parameters $t_0$, $t_1$, and $t_3$, which are closely
related to the structure of the QBCP system, are
energy-independent constants in the clean limit.
In sharp contrast, the parameters $t_{0,1,3}$ are no longer constants but
intimately hinge upon the evolutions of other interaction parameters
directly or indirectly after taking into account the effects of impurity
scatterings. This implies that the stability of QBCP's dispersion would
be challenged by the effects of impurities with the variation of energy
scale. As the low-energy phenomena are closely associated with its stability,
it is therefore imperative to examine whether QBCP's dispersion is still robust
and how it is changed under the influences of fermion-fermion interactions
and impurities. To this end, we adopt the RG method together with the criterion $|t_0|<\mathrm{min}(|t_1|,|t_3|)$ to judge the stability of band structure in the
whole energy region ranging from the starting point to the lowest-energy limit.
Given our approach is based on the combination of definition of QBCP band structure
and RG analysis, we can not only track the stability of the QBCP band structure
in the whole energy region but also work effectively for both topological
trivial and non-trivial phase transitions~\cite{Fradkin2009PRL,Murray2014PRB,Wang2017PRB}.

\subsection{Distinct sorts of fixed points}

Before going further, we would like to stress that
the fermion-fermion interactions can flow towards strong couplings
after taking into account one-loop corrections. With this respect,
we rescale all the interaction parameters by a combination of two non-sign
changed couplings to overcome the strong couplings and make our
study perturbative~\cite{Murray2014PRB}. Consequently, we are left with
relative evolutions of interaction parameters together with the corresponding
relatively fixed points (FPs) in the parameter space, which are
the Gaussian, quantum anomalous Hall (QAH), and nematic-spin-nematic
(NSN) on sites of bonds, respectively~\cite{Murray2014PRB,Wang2017PRB}.

Considering the low-energy phenomena are conventionally dictated by these fixed points,
it is therefore necessary to put our focus on monitoring the physical behaviors
upon accessing these different fixed points. Given the Gaussian FP is relatively trivial,
we hereby address brief comments on the QAH and NSN FPs.
Generally, the QAH and NSN FPs in the presence of fermion-fermion interactions and impurities
correspond to $(u_0,u_-,u_2)/u_+ = (0, a, b)$ and $(u_0,u_-,u_2)/u_+=(0, 0, c)$, respectively~\cite{Murray2014PRB,Wang2017PRB}. Here, $a$, $b$, and $c$ are
finite constants, which are dependent mildly upon the initial conditions.
However, it is worth highlighting that the instabilities around these two fixed points
are considerably robust although the specific values of $a, b, c$ can be slightly
modified by tuning the starting values of impurity scatterings~\cite{Murray2014PRB,Wang2017PRB}.
In other words, the phase transitions induced nearby these two FPs are still
from QBCP materials to the QAH and NSN states against the variations of these
three constants. For instance, these FPs in the weak impurity respectively
flow to Gaussian FP $(u_0,u_-,u_2)/u_+=(0,0,0)$ and QAH FP $(u_0, u_-, u_2)=(0,-3.73, 7.46)u_+$
as well as NSN FP $(u_0, u_-, u_2)=(0, 0,-1.09)u_+$.
Considering the impurity scatterings only slightly alter the concrete values
but do not change the basic structures and features of these FPs, we provide
a schematic evolutions absorbed by these potential FPs in the weak impurity
as shown in Fig.~\ref{Fig_schem_FPs}.

\begin{figure}
\centering
\hspace{-0.78cm}
\includegraphics[width=2.01in]{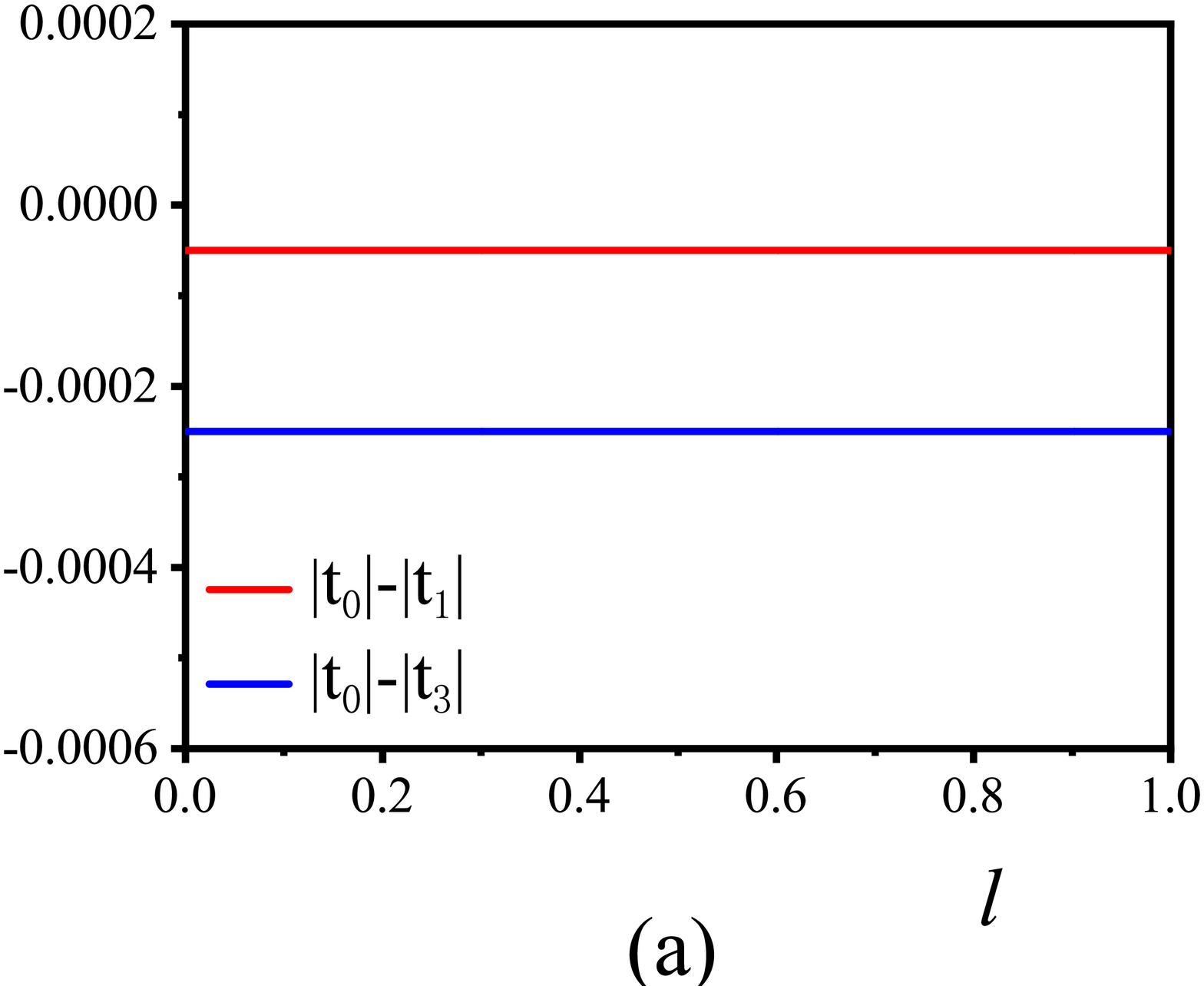}\hspace{-1cm}
\includegraphics[width=2.01in]{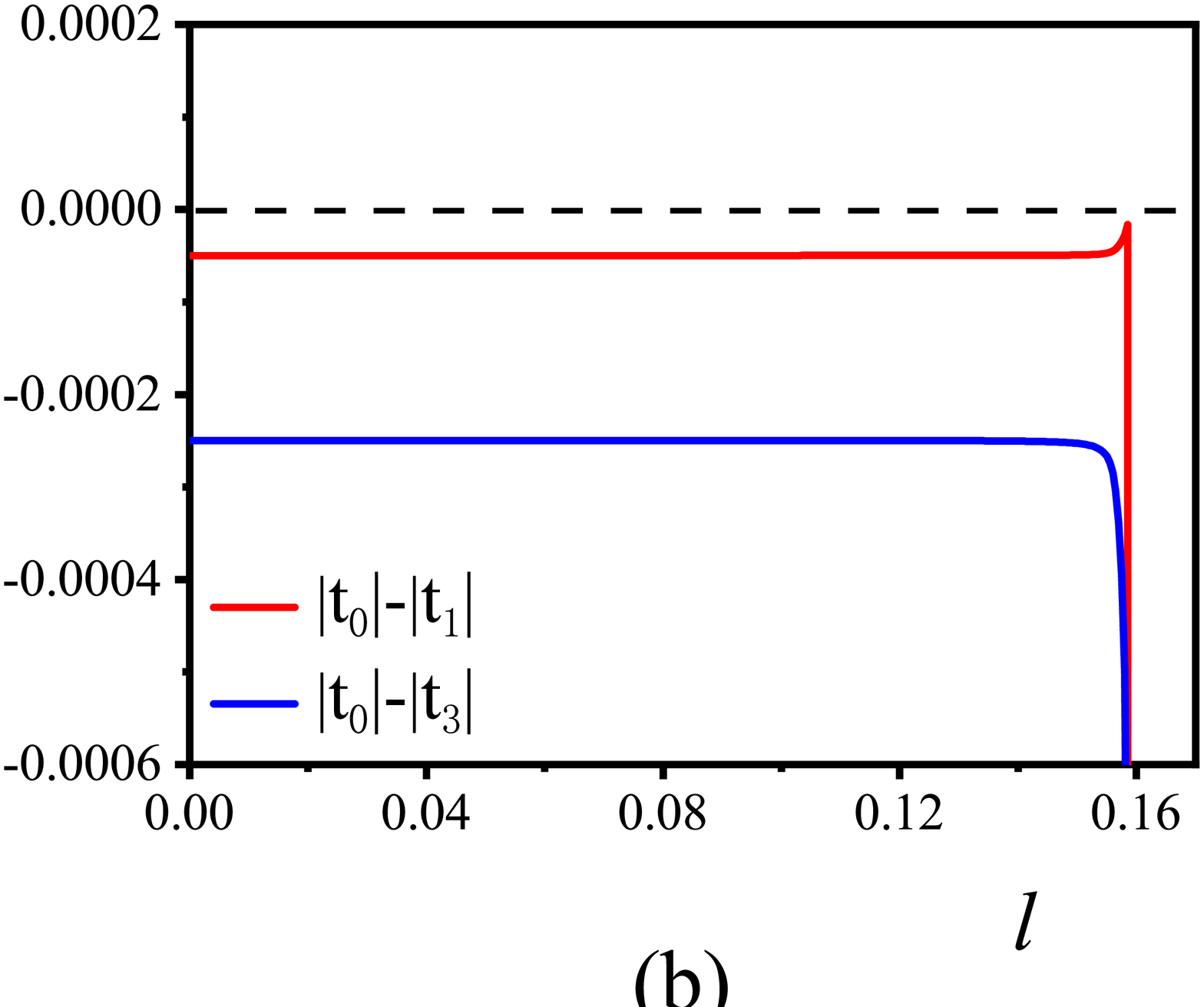}
\vspace{-0.3cm}
\caption{(Color online) Evolutions of $|t_0|-|t_1|$ and
$|t_0|-|t_3|$ once the fermion-fermion interactions are driven to
the Gaussian FP under (a) the sole presence of
$M_0=\sigma_0$ impurity with $v_i(0)=10^{-3}$ (the results for $M_{1,2,3}$ are similar
and hence not shown here) and (b) the presence of all three types of impurities
with $v_i(0)=10^{-4}$ for $t_0(0)>0,t_1(0)>0,t_3(0)>0$.}\label{Fig-2}
\end{figure}

Combining the coupled RG equations with Fig.~\ref{Fig_schem_FPs},
we notice that the fermion-fermion interaction parameters exhibit
distinct energy-dependent behaviors around these FPs, which give rise to distinct
corrections to the evolutions of impurity strengths.
As a result, the parameters $t_0$, $t_1$, and $t_3$ that are directly related to
the flows of impurities~(\ref{Eq_t_0})-(\ref{Eq_t_3}) would receive very distinct
contributions once the systems are approaching different types of FPs.
Without lose of generalities, we will select some typical
starting values of fermion-fermion interactions that can drive $u_i$ into
these FPs and investigate the related physical properties one by one.

\subsection{Gaussian FP}

At the outset, we consider the Gaussian FP. We firstly assume there exists
only one type of impurity in the QBCP system. To simplify our analysis, we from now on
let $\Delta_i=1$ as mentioned at the end of Sec.~\ref{Sec_RG_analysis}
and utilize the parameter $v_i$ with $i=0,1,2,3$ to measure the corresponding
strength of fermion-impurity interaction. After carrying out the numerical
evaluations of Eqs.~(\ref{Eq_t_0})-(\ref{Eq_v_3}), we find that  $|t_0|-|t_1|<0$
and $|t_0|-|t_3|<0$ are always satisfied even the
initial value of the impurity strength is adequately strong. Since the results for
sole presence of $M_i$ impurity with $i=0,1,2,3$ are analogous,
we here only provide the results for presence of $M_0$ as clearly
delineated in Fig.~\ref{Fig-2}(a). Then, we move to the general situation for
the presence of all types of quenched impurities
in the QBCP system. Paralleling similar procedures of $M_0$ impurity
brings out qualitatively analogous corrections to parameters $t_0$, $t_1$, and $t_3$
as manifested in Fig.~\ref{Fig-2}(b). Specifically, $|t_0|<\mathrm{min}(|t_1|,|t_3|)$
cannot be destroyed in the low-energy regime
and hence the QBCP's dispersion is stable against impurities. In addition,
one can check that the relationship among $t_0$, $t_1$ and $t_3$ does not change significantly
even though the impurity strength is increased. In other words,
QBCP' dispersion is stable regardless of strong or
weak impurity around the Gaussian FP. It is hereby necessary to stress
that we within this project employ the impurity scattering rate
$\Gamma^{-1}_i\sim\Delta_i v^2_i/t_0$ (with $i=0,1,2,3$)~\cite{Wang2017PRB} measured
by $\Lambda_0$ to distinguish the weak and strong impurities.
Since the fermion-fermion interactions and fermion-impurity scatterings are considered
on the same footing under the RG analysis, it is of particular
significance to pay attention to the starting point at $l=0$.
To be concrete, we regard it as the ``weak" impurity once the effect of impurity
is less important than that of fermion-fermion interaction, i.e., $\Gamma^{-1}_i(l=0)\ll u_i(l=0)$. On the contrary, it corresponds to
the ``strong" impurity if the influence of the impurity is comparable to
fermion-fermion interaction's with $\Gamma^{-1}_i(l=0)\sim u_i(l=0)$.

\begin{figure}
\centering
\hspace{-0.78cm}
\includegraphics[width=2.01in]{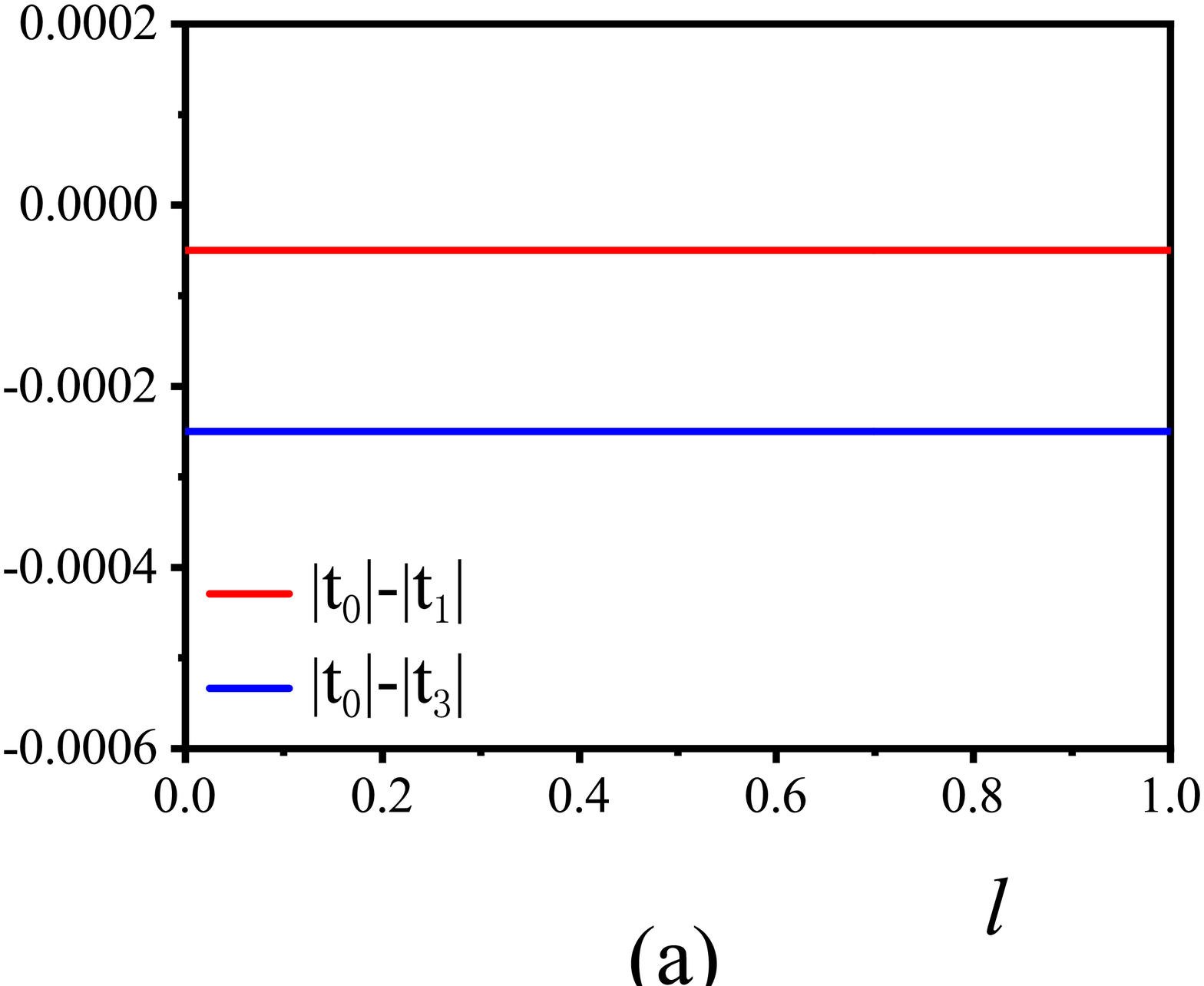}\hspace{-1cm}
\includegraphics[width=2.01in]{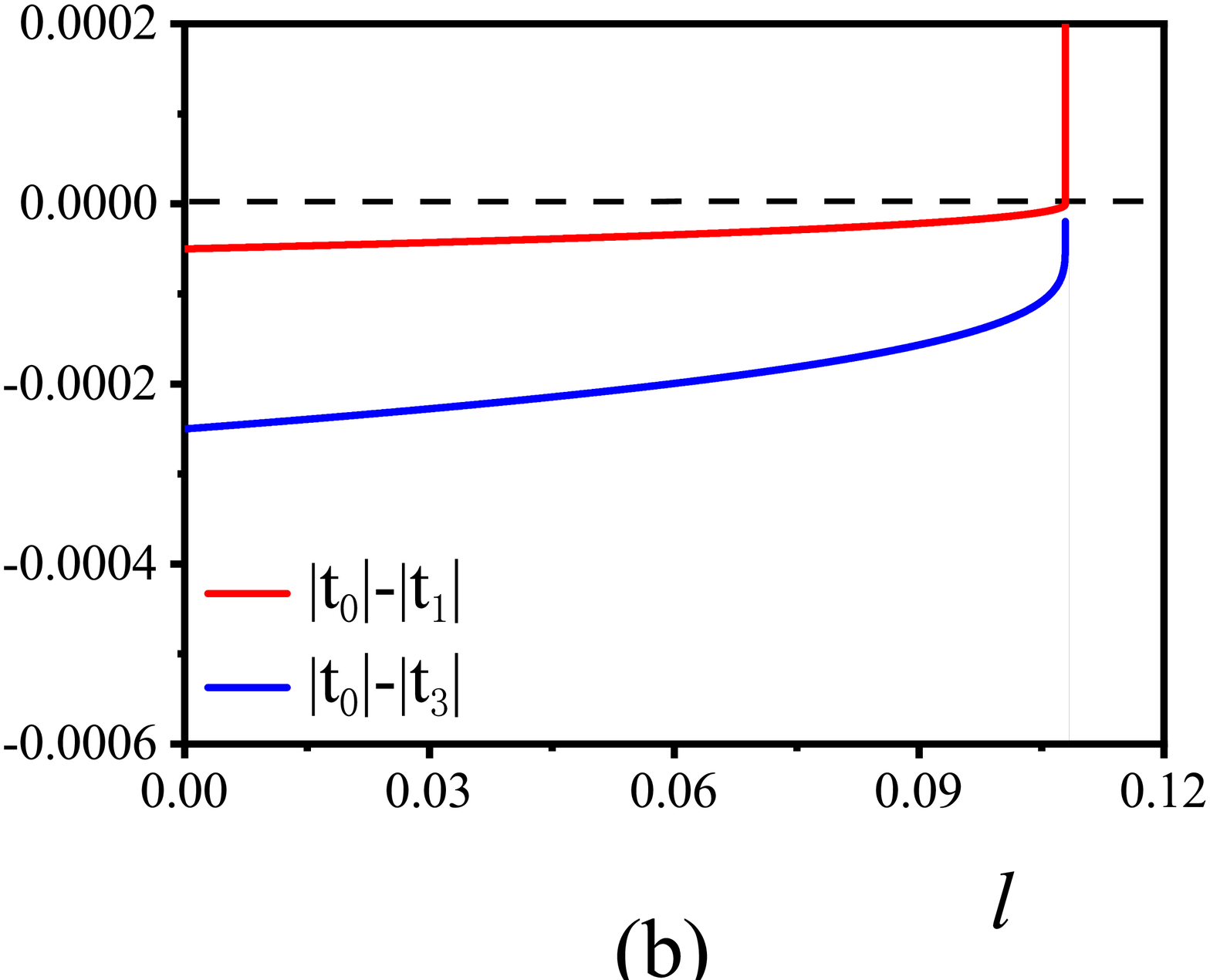}\\
\vspace{-2cm}
\hspace{4.91cm}
\includegraphics[width=0.7in]{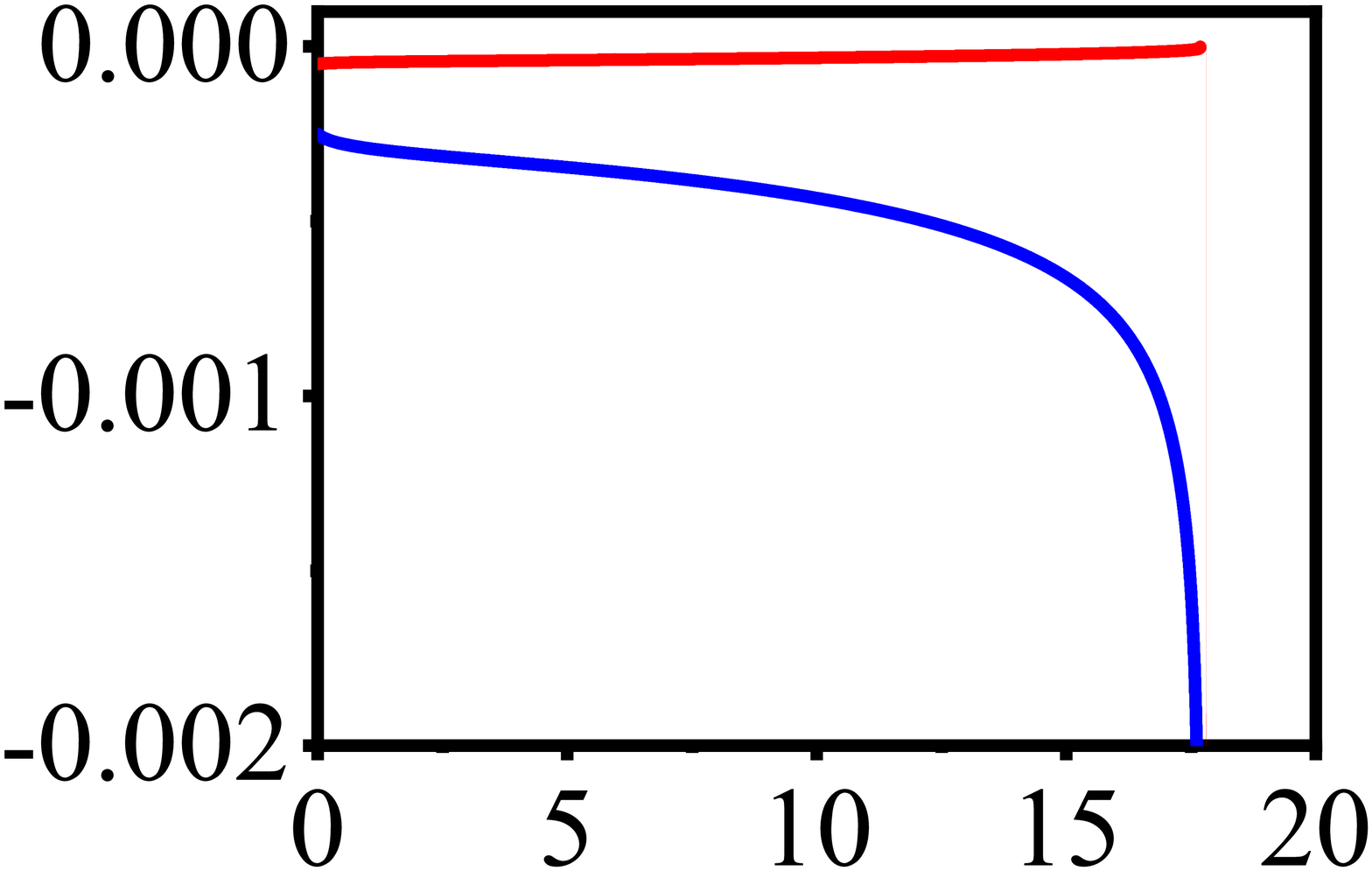}
\vspace{0.9cm}
\caption{(Color online) Evolutions of $|t_0|-|t_1|$ and $|t_0|-|t_3|$
for the presence of $M_0=\sigma_0$ impurity as the fermion-fermion interactions
are driven to the QAH/NSN FP with (a) a weak initial strength of $v_i(0)=10^{-5}$
and (b) a strong initial strength of $v_i(0)=10^{-3}$
(the results for $M_1$ and $M_3$ are similar and hence not shown here).
Inset:  flows for the presence of $M_2=\sigma_2$ impurity
with a strong initial strength of $v_i(0)=10^{-3}$.}\label{Fig-3}
\end{figure}

\begin{figure}
\centering
\hspace{-0.78cm}
\includegraphics[width=2.01in]{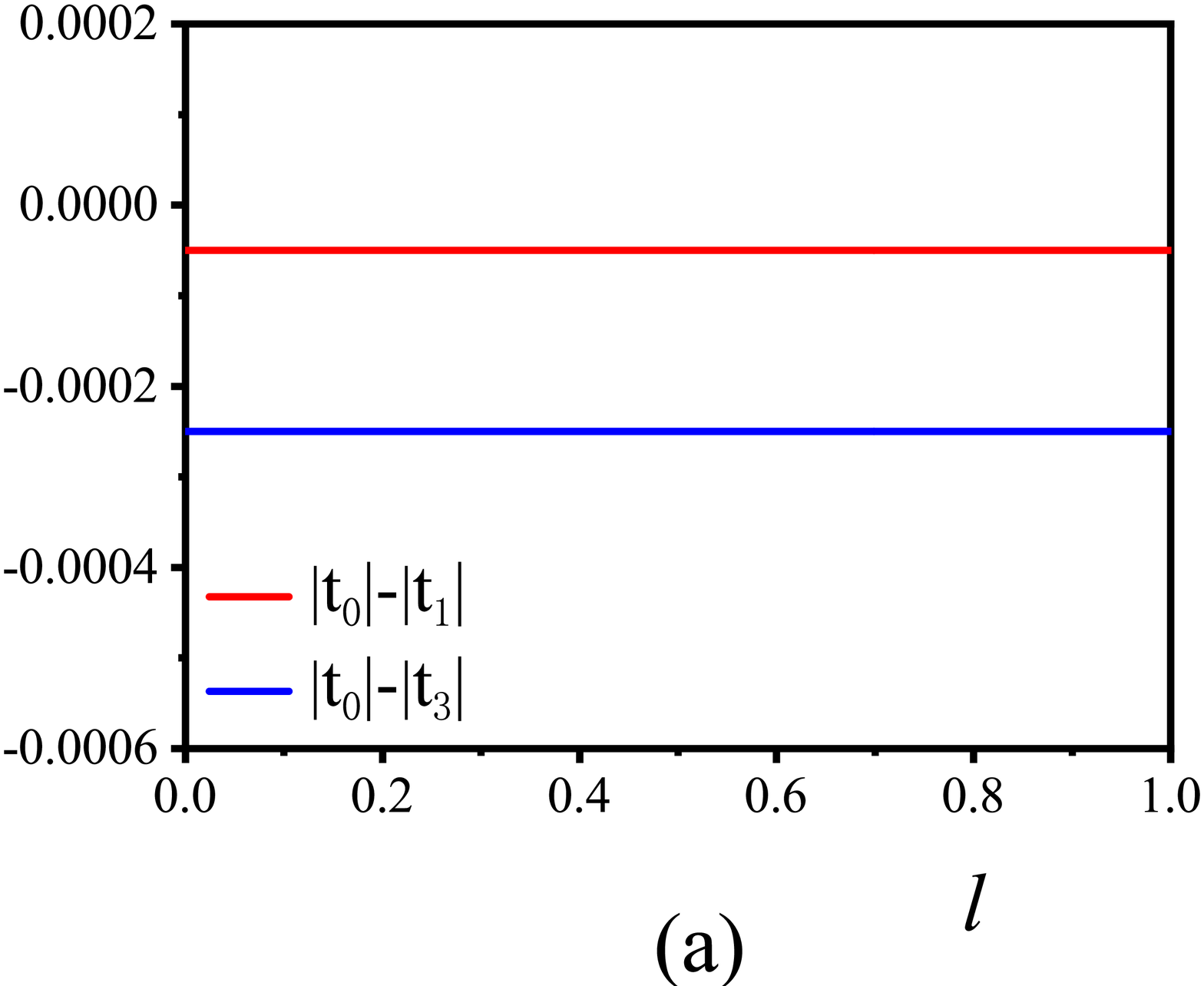}\hspace{-1cm}
\includegraphics[width=2.01in]{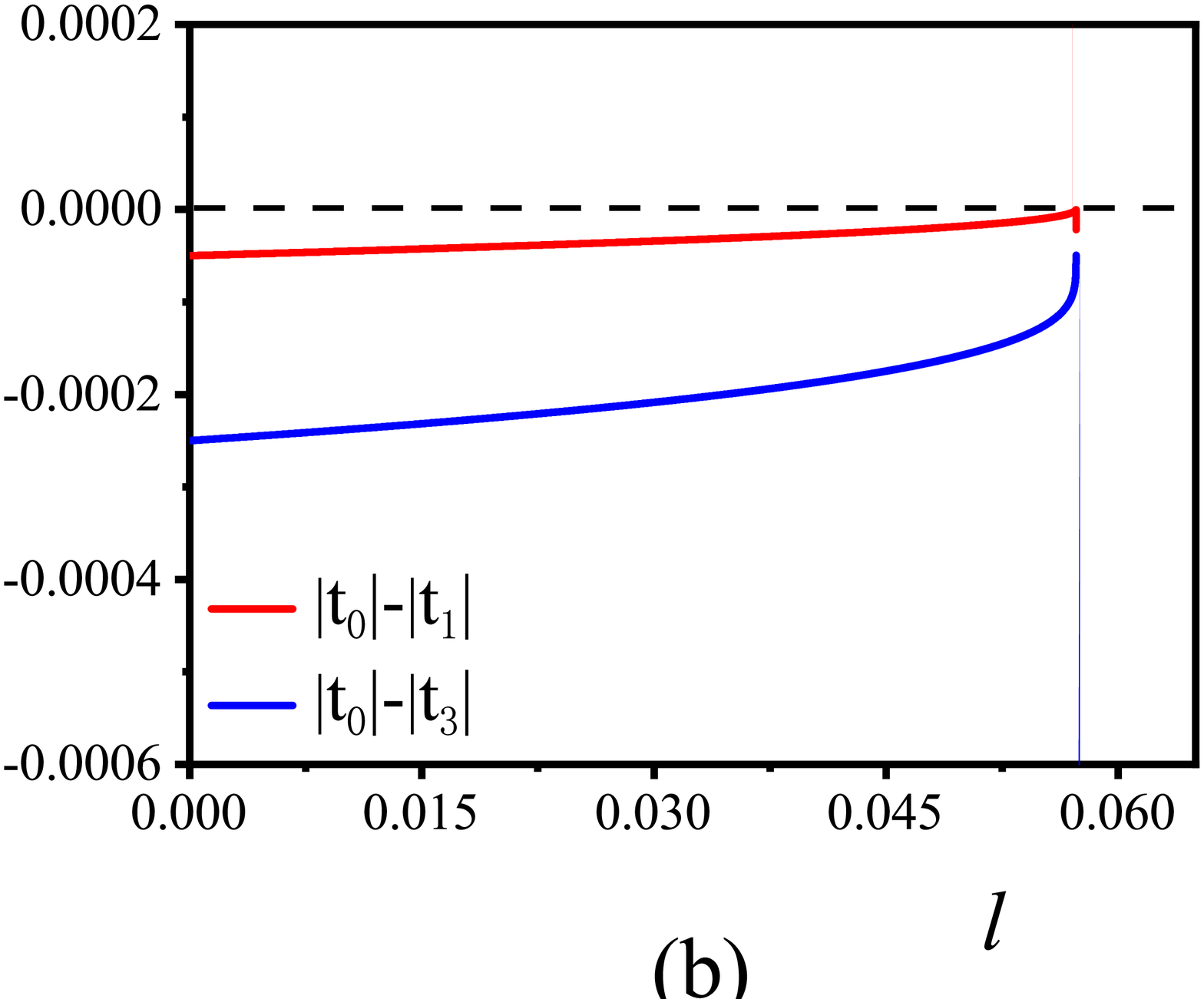}
\vspace{-0.35cm}
\caption{(Color online) Evolutions of $|t_0|-|t_1|$
and $|t_0|-|t_3|$ in the presence of all three types
of impurities while the fermion-fermion interactions are driven to
the QAH/NSN FP with (a) a weak initial strength of $v_i(0)=10^{-5}$ and
(b) a strong initial strength of $v_i(0)=10^{-3}$.}\label{Fig-4}
\end{figure}

\subsection{QAH and NSN FPs}

Subsequently, we move to the case at which fermion-fermion
interaction parameters are attracted and governed by QAH or
NSN FP. To proceed, it is necessary to take into account the
coupled evolutions on the same footing and carry out long but
straightforward RG analysis.

Let us take QAH FP for an example. Again, we begin with visiting the
effects caused by the presence of a single type of impurity and then
consider the presence of all sorts of impurities.
For instance, Fig.~\ref{Fig-3} manifestly exhibits how
$|t_0|-|t_1|$ and $|t_0|-|t_3|$ evolve with lowering
energy scales for the presence of impurity $M_0$
(the conclusions for the presence of $M_1$ or $M_3$ are
qualitatively analogous and hence not shown here).
One can readily find from Fig.~\ref{Fig-3}(a) that the QBCP's dispersion is
robust at weak impurity. In comparison, we find the impurity scattering becomes more
significant while the impurity strength is strong.
To be concrete, $M_0$, $M_1$, or $M_3$ with sufficient impurity strength can trigger
the divergence of $u_i$, indicating emergence of some impurity-induced FP at
certain critical energy scale $l_c$, at which
$|t_0|-|t_1|$ is converted into a positive value
as delineated in Fig.~\ref{Fig-3}(b). This suggests that QBCP's dispersion
is broken at $l_c$ and henceforth the RG evolutions should be stopped before
this critical energy scale. However, as illustrated
in the inset of Fig.~\ref{Fig-3}(b), the restrictions $|t_0|-|t_1|<0$ and
$|t_0|-|t_3|<0$ are always satisfied if only $M_2$ impurity is
turned on even at the strong impurity strength. In other words, QBCP's
dispersion is rather stable for the sole presence of $M_2$ impurity
around the QAH FP. While all three types of quenched impurities are present,
we find that basic results as shown in Fig.~\ref{Fig-4} are consistent with
the sole presence of $M_{0}$, $M_1$, or $M_3$.
Nevertheless, it is necessary to address the following two points
on the basis of Fig.~\ref{Fig-3} and Fig.~\ref{Fig-4}. On one hand, impurity $M_2$
provides the contrary contribution compared to the other types of impurities,
indicating the impurities $M_0,M_1,M_3$ are dominant over the $M_2$ impurity.
On the other, we realize that the critical energy scale denoted by $l_c$ is
slightly lifted, attesting to the competition among all kinds of impurity scatterings.

Additionally, one can check the basic results for NSN FP are similar to
their QAH counterparts. Based on these points, we address that the relationship
among $t_0$, $t_1$, and $t_3$, which is closely associated with
the 2D QBCP's dispersion, can either be robust or qualitatively
changed by the interplay between fermionic interactions and impurities.
Table~\ref{table-stability} summarizes the stability of 2D QBCP's dispersion
against distinct types of impurities around different FPs.
Before closing this section, it is interesting to point out that
the parameters $t_{0,1,3}$ in principle can also be taken some negative
values. As a result, there are in all eight types of starting values. We have
checked that the above results are insusceptible to the signs of $t_{0,1,3}$.
With these respects, we hereafter only consider the $t_{0,1,3}>0$ case.

\begin{table}
\caption{Fates of 2D QBCP's dispersions under the influence
of various impurities around the Gaussian, QAH, and NSN FPs. To be convenient,
 ``$\mathcal{W}$" and ``$\mathcal{S}$" are adopted to characterize weak and strong
strengths of impurities. In addition, ``$\mathbb{S}$" and ``$\mathbb{US}$"
stand for stable and unstable dispersions of 2D
QBCP system, respectively.}\label{table-stability}
\vspace{0.39cm}
\centerline{
\begin{tabular}{p{1.5cm}<{\centering}p{0.55cm}<{\centering}p{0.55cm}<{\centering}
p{0.55cm}<{\centering}
p{0.55cm}<{\centering}p{0.55cm}<{\centering}p{0.55cm}<{\centering}p{0.55cm}<{\centering}
p{0.55cm}<{\centering}p{0.55cm}<{\centering}p{0.55cm}<{\centering}}
\hline
\hline
\rule{0pt}{15pt}&\multicolumn{2}{c}{$M_0$}&\multicolumn{2}{c}{$M_1$}&\multicolumn{2}{c}{$M_2$}&
\multicolumn{2}{c}{$M_3$}&\multicolumn{2}{c}{$M_{0123}$}\\
\hline
\rule{0pt}{15pt}Strength &$\mathcal{W}$&$\mathcal{S}$&$\mathcal{W}$&$\mathcal{S}$&$\mathcal{W}$&$\mathcal{S}$&$\mathcal{W}$&
$\mathcal{S}$&$\mathcal{W}$&$\mathcal{S}$  \\
\rule{0pt}{15pt}Gaussian&$\mathbb{S}$&$\mathbb{S}$&$\mathbb{S}$&$\mathbb{S}$&$\mathbb{S}$&$\mathbb{S}$
&$\mathbb{S}$&$\mathbb{S}$&$\mathbb{S}$&$\mathbb{S}$  \\
\rule{0pt}{15pt}QAH/NSN&$\mathbb{S}$&$\mathbb{US}$&$\mathbb{S}$&$\mathbb{US}$&$\mathbb{S}$&$\mathbb{S}$
&$\mathbb{S}$&$\mathbb{US}$&$\mathbb{S}$&$\mathbb{US}$  \\
\hline
\hline
\end{tabular}
}
\end{table}

\begin{figure}
\centering
\includegraphics[width=3.2in]{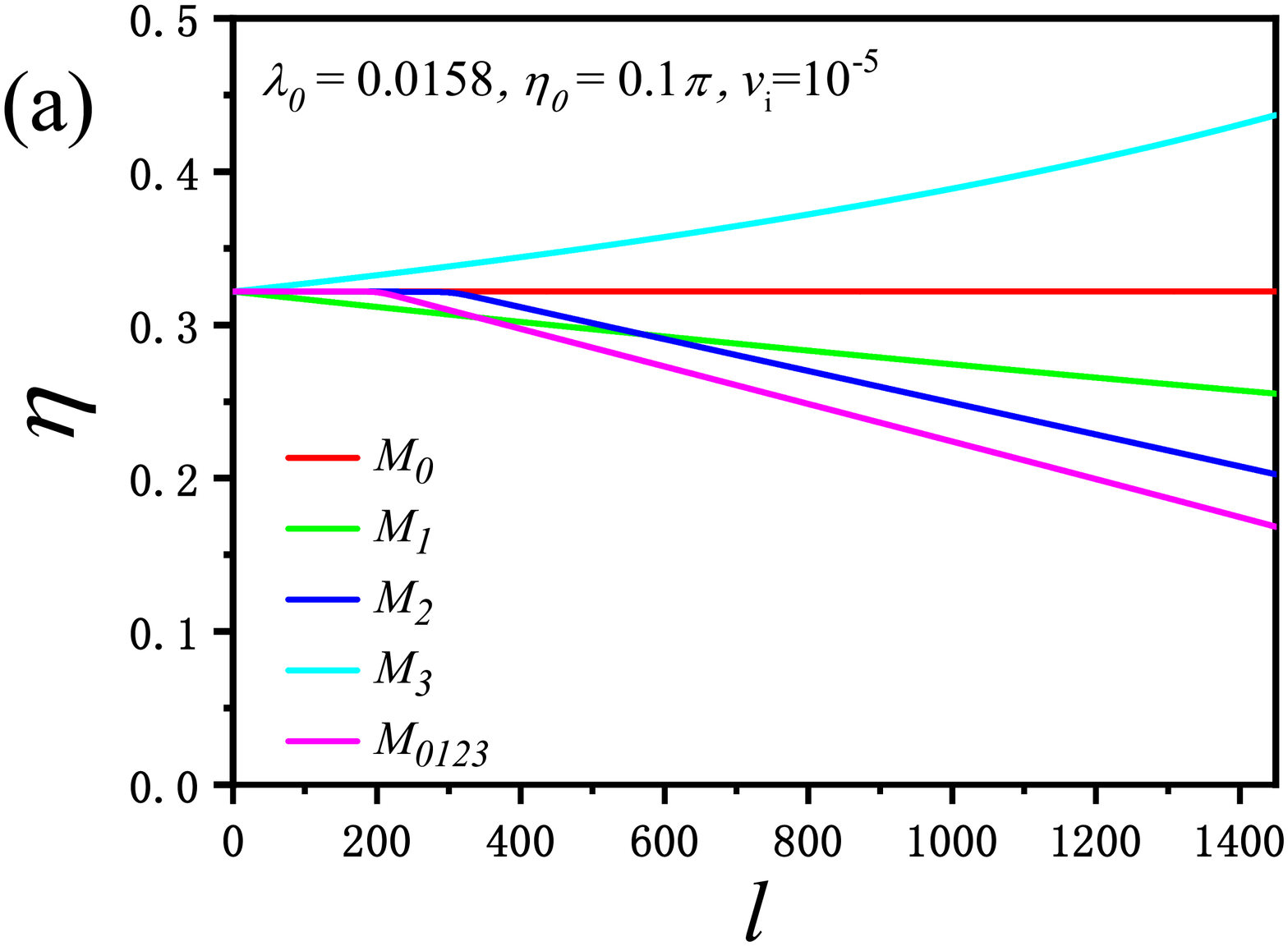}
\vspace{-0.35cm}\\
\includegraphics[width=3.2in]{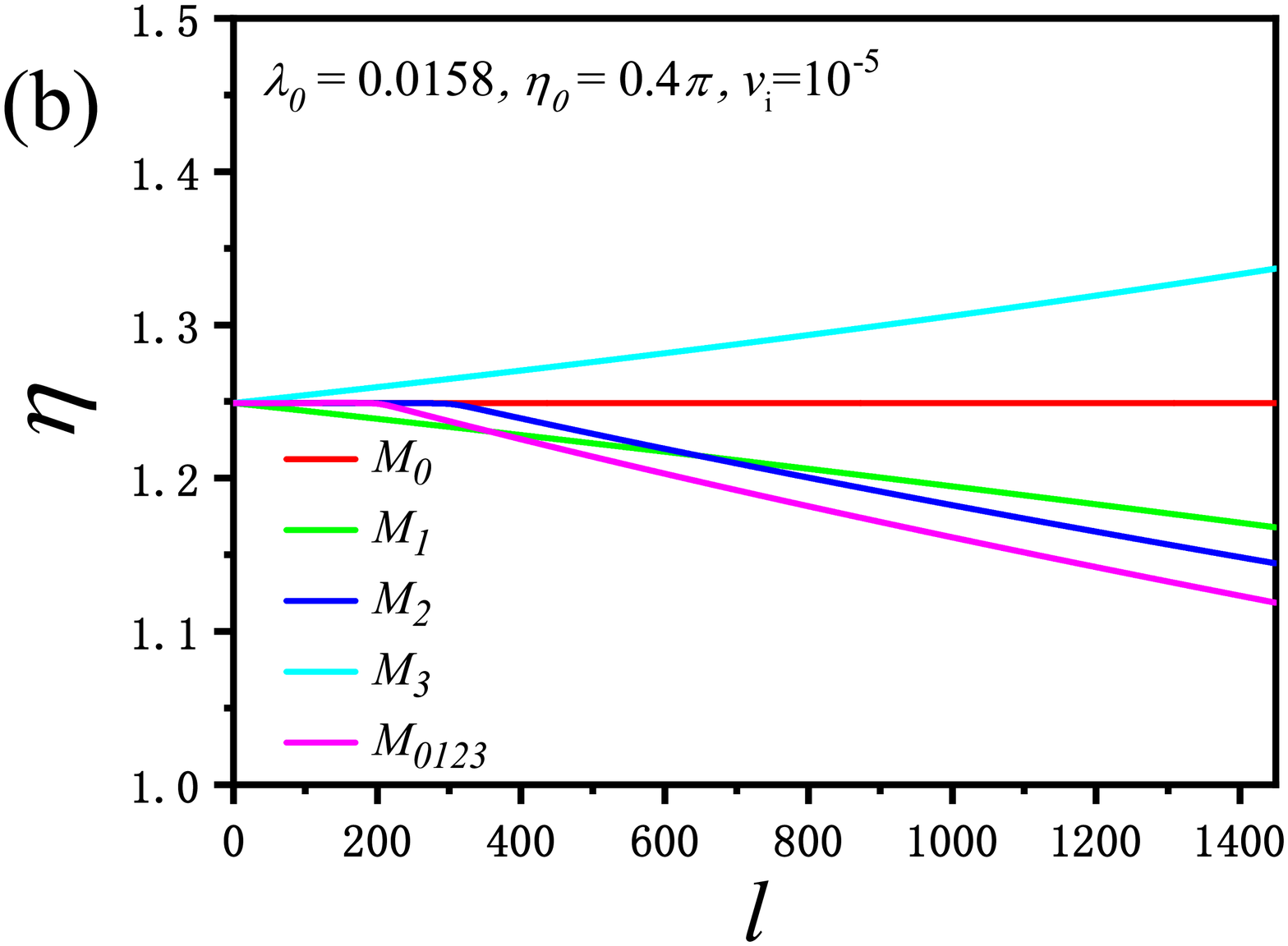}
\vspace{-0.05cm}
\caption{(Color online) Evolutions of $\eta$ against the presence
of impurities at $v_i(0)=10^{-5}$ for (a) $\eta_0=0.1\pi$ and
(b) $\eta_0=0.4\pi$ with fermion-fermion couplings flowing toward
Gaussian FP.}\label{Fig-eta-vi-m5-away-FP}
\end{figure}

\section{Fate of rotational asymmetry}\label{Sec_eta}

As mentioned in Sec.~\ref{Sec_model}, rotational and particle-hole asymmetries are
two quantities of remarkable importance for 2D QBCP systems, which are closely linked to
the low-energy properties. To proceed, we follow Ref.~\cite{Murray2014PRB}
and introduce two parameters $\eta$ and $\lambda$ to account for them,
\begin{eqnarray}
\eta(l)&\equiv&\mathrm{arctan}\frac{t_1(l)}{t_3(l)},\label{Eq_eta-L}\\
\lambda(l)&\equiv&\frac{t_0(l)}{\sqrt{t^2_1(l)+t^2_3(l)}},\label{Eq_lambda-L}
\end{eqnarray}
where $l$ serves as the energy scale and the related coefficients $t_i(l)$ with
$i=0,1,3$ are designated in equations~(\ref{Eq_t_0})-(\ref{Eq_t_3}), which
capture the information of rotational and particle-hole asymmetries (or symmetries) in 2D QBCP
systems. As aforementioned, the 2D QBCP system would be
invariant under rotational symmetry and/or particle-hole symmetry
exactly at $\eta=\frac{\pi}{4}$ and/or $\lambda=0$. However, the parameters
$t_0$, $t_1$, and $t_3$ are intimately intertwined with other
interaction parameters obeying the coupled flow equations. A significant question
is thus naturally raised whether and how $\eta$ and $\lambda$ are affected by the
competition between fermion-fermion interactions and impurity scatterings.

Prior to responding to the above question, it is necessary to present some comments on
the clean-limit situation. Via taking $v_{i}=0$ with $i=0-3$ in Eqs.~(\ref{Eq_t_0})-(\ref{Eq_t_3}),
we can apparently reach that the parameters $t_0$, $t_1$, and $t_3$ are
energy-independent and thus remain some constants at clean limit. As a consequence,
the parameters of asymmetries (symmetries) $\eta$ and $\lambda$ are invariant
with lowering the energy scale in the presence of fermion-fermion interactions.
However, it is well-trodden that the impurities are always present in
realistic systems and play an important role in determining the
low-energy behaviors of fermionic systems~\cite{Lee1985RMP,Evers2008RMP,
Nersesyan1995NPB,Hung2016PRB,Efremov2011PRB,Efremov2013NJP,Korshunov2014PRB}.
In sharp distinction to a clean limit, the parameters $t_0$, $t_1$, and $t_3$
are no longer constants but intimately
evolve and entangle with other interaction parameters due
to the interplay between fermion-fermion interactions and impurities as clearly
delineated in Eqs.~(\ref{Eq_t_0})-(\ref{Eq_v_3}).
Under these respects, it is therefore of remarkable temptation to
explore the energy-dependent hierarchies of parameters $\eta$ and $\lambda$,
whose fates are closely associated with physical behaviors in the low-energy regime.

Learning from the coupled RG equations~(\ref{Eq_t_0})-(\ref{Eq_v_3}),
one can readily realize that $\eta$ and $\lambda$ are manifestly affected
by impurities. In comparison, the fermion-fermion interactions can indirectly
impact $\eta$ and $\lambda$ by modifying the flows of
$v_i$ with $i=0-3$. This signals that asymmetric parameters are
in close conjunction with the evolutions of fermion-fermion interactions
$u_i$ with $i=0-3$. As studied previously~\cite{Murray2014PRB,Wang2017PRB},
it is worth pointing out that the fermion-fermion
strengths $u_i$ in 2D QBCP systems are governed
by the Gaussian, QAH, and NSN FPs induced by the impurity scatterings
as schematically depicted in Fig.~\ref{Fig_schem_FPs}.
Clearly, the energy-dependent $u_i$ would display considerably distinct trajectories
in the vicinity of different types of FPs. This straightforwardly
implies that different evolutions of $u_i(l)$ can bring inequivalent corrections
to $\eta$ and $\lambda$. With these respects, we will separately investigate
the effects of impurities and fermion-fermion interactions on $\eta$ and
$\lambda$ as fermion-fermion strengths are attracted by Gaussian,
QAH, and NSN FPs one by one.

Within this section, we put our focus on the low-energy behaviors of rotational
asymmetry under the influence of both fermion-fermion interactions and impurities.
The evolutions of particle-hole asymmetry will be left for Sec.~\ref{Sec_lambda}.
Since the QBCP's dispersion is stable only in the region of $l<l_c$ as
studied in Sec.~\ref{Sec_dispersion}, we hereafter confine this work within this energy regime.

\begin{figure}
\centering
\includegraphics[width=3.2in]{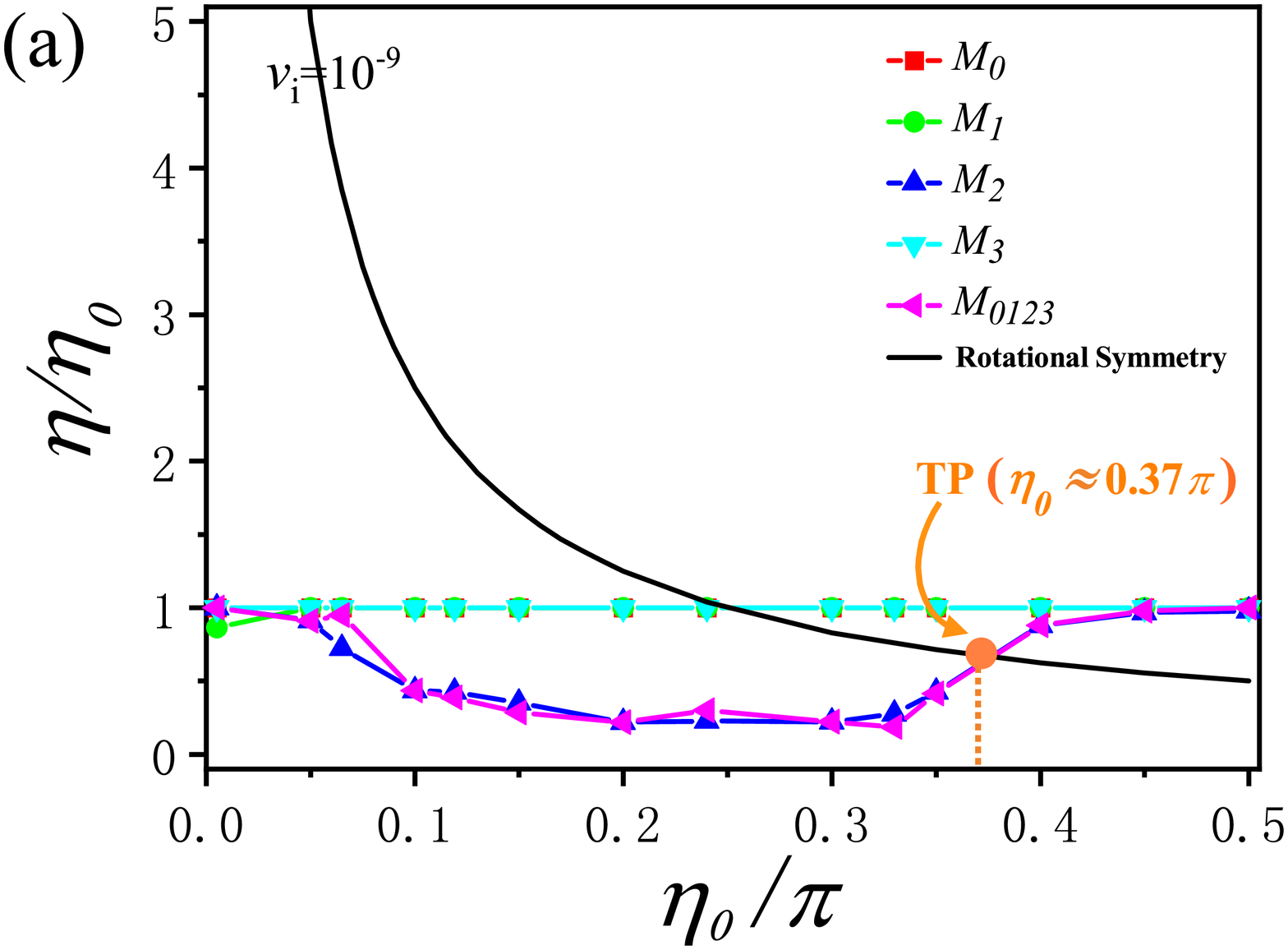}
\vspace{-0.35cm}\\
\includegraphics[width=3.2in]{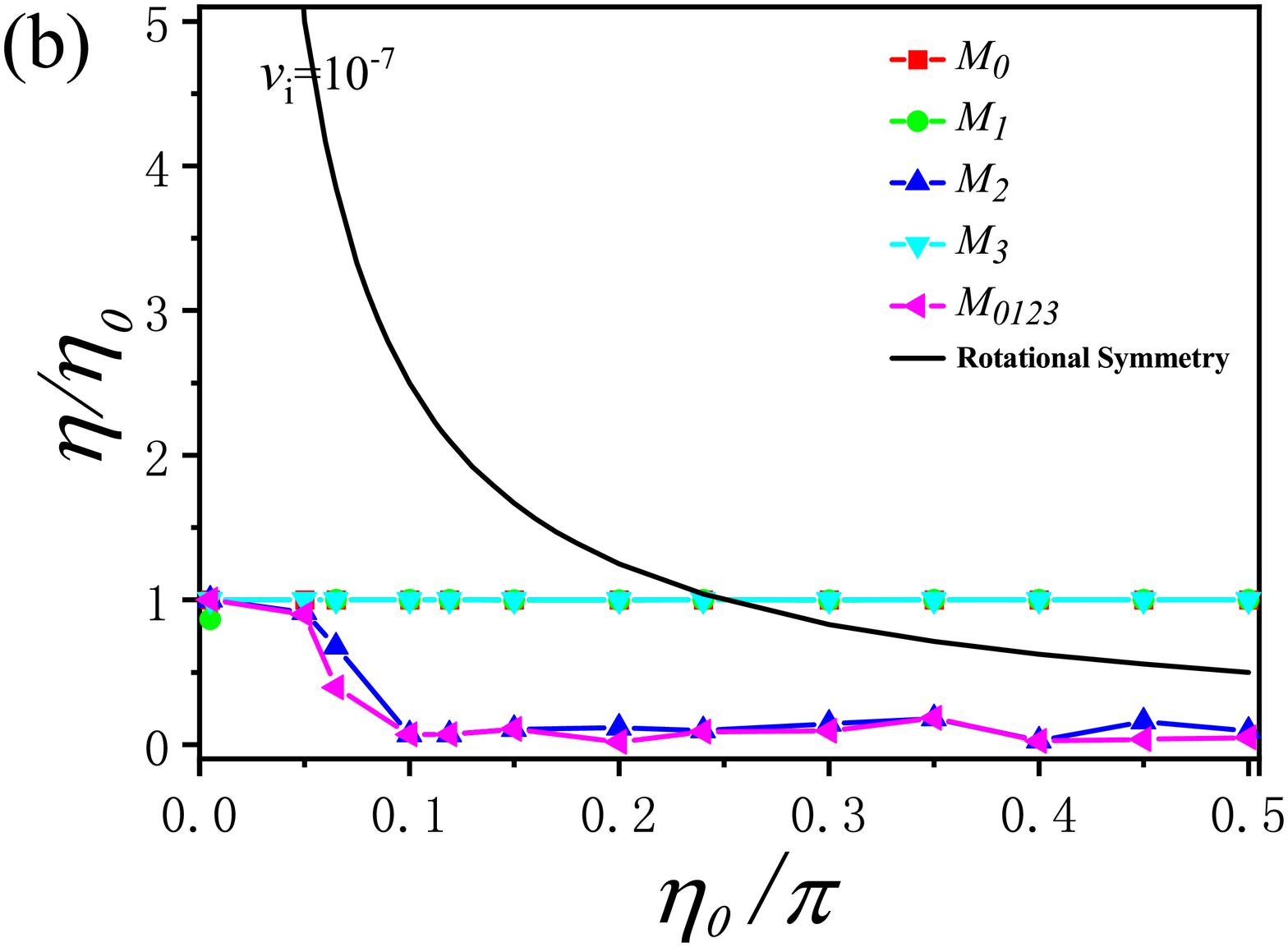}
\vspace{-0.35cm}\\
\includegraphics[width=3.2in]{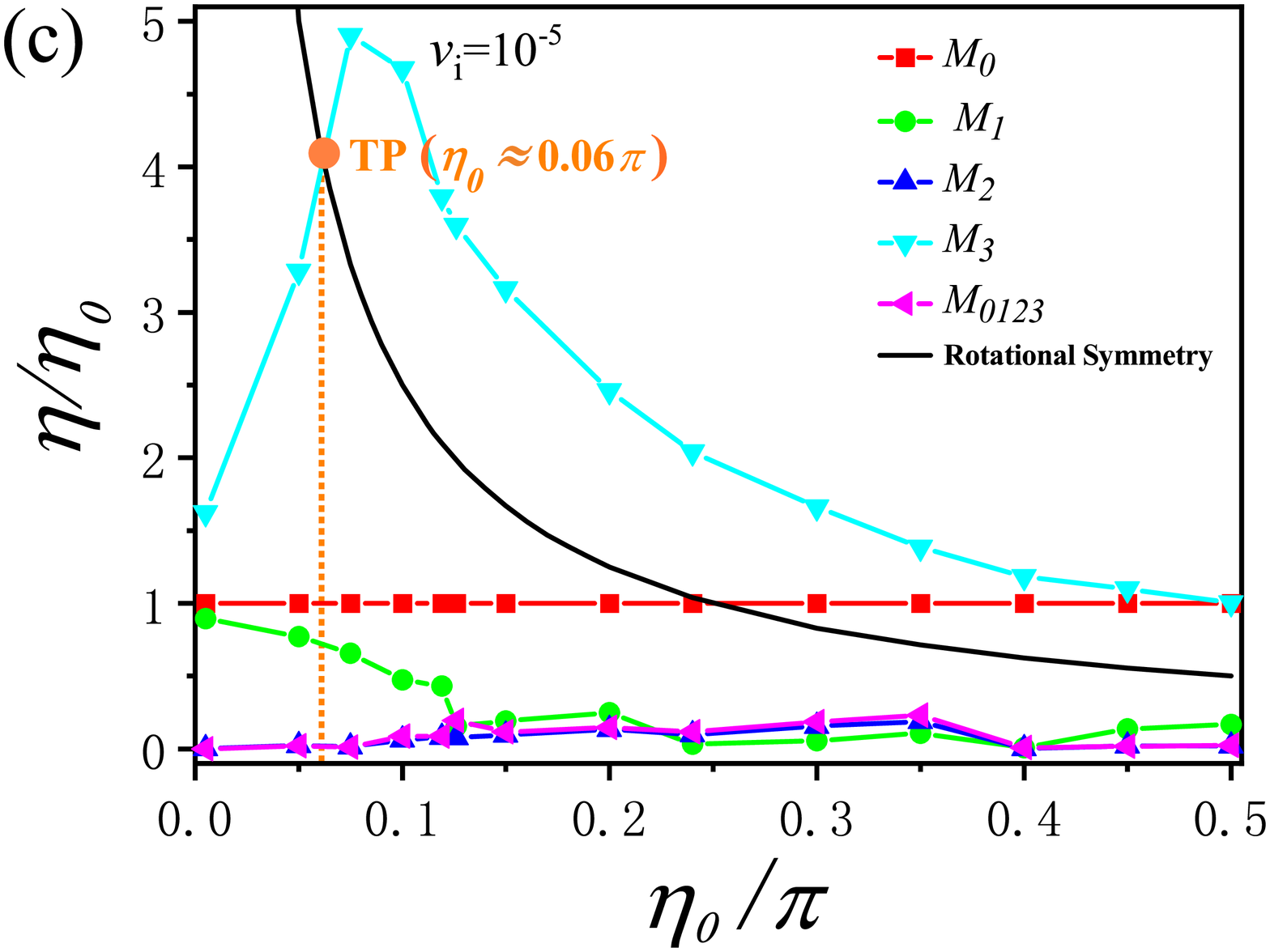}
\vspace{-0.35cm}
\caption{(Color online) Behaviors of $\eta(l\rightarrow l_c)/\eta_0$
with tuning initial values of $\eta_0$ against the impurities for
(a) $v_i(0)=10^{-9}$, (b) $v_i(0)=10^{-7}$, and (c) $v_i(0)=10^{-5}$
as the fermion-fermion couplings go towards Gaussian FP.
The lines for the $M_0$ impurity coincide with their counterparts at
clean limit and ``TP" represents
the very transition point from rotational asymmetry to symmetry.}\label{Fig-vi-M975-eta-eta-0-away-FP}
\end{figure}

\subsection{Warm-up: Tendency of $\eta$ under impurities}\label{Subsection_warm-up}

As a warm-up, we hereby randomly choose several initial values for our
interaction parameters and roughly check the tendency of $\eta$ under the
presence of impurities. With these starting values,
the energy-dependent evolutions of $\eta$ under the influence of distinct types of
impurities are designated in Fig.~\ref{Fig-eta-vi-m5-away-FP} after performing the
numerical analysis of RG equations~(\ref{Eq_t_0})-(\ref{Eq_v_3}). Reading off the information of
Fig.~\ref{Fig-eta-vi-m5-away-FP}, one can straightforwardly figure out that $\eta$
is insensitive to the sole presence of $M_0$ impurity. It is of particular interest
to highlight that this result is well coincident with the analytical analysis.
To be specific, the energy-dependent flow of $\eta$ can be specified as
\begin{eqnarray}
\frac{dt_1/t_3}{dl}&=&\frac{1}{t_3}\frac{dt_1}{dl}-\frac{t_1}{t^2_3}\frac{dt_3}{dl}
=\frac{t_1}{t_3}\left[(\Delta_3v^2_3-\Delta_1v^2_1)\mathcal{N}_6
\right.\nonumber\\
&-&\left.(\Delta_1v^2_1+2\Delta_2v^2_2+\Delta_3v^2_3)\mathcal{N}_5\right],\label{Eq-t-1-t-3}
\end{eqnarray}
where the coefficients $\mathcal{N}_5$ and $\mathcal{N}_6$ are designated in Eq~(\ref{Eq-N-6}).
Combining Eq~(\ref{Eq_eta-L}) and Eq~(\ref{Eq-t-1-t-3}), one can readily draw a conclusion that
the strength of $M_0$ impurity does not enter into the evolution and henceforth
$\eta$ is independent of $M_0$. However, the value of $t_1/t_3$ appearing in
Eq.~(\ref{Eq-t-1-t-3}) is closely related to other types of impurities.
Accordingly, we only need to study the effects of $M_{1,2,3}$ impurities on the
parameter $\eta$. To be concrete, the sole presence of $M_1$ or $M_2$ impurity causes
$\eta$ to decrease with lowering the energy scale. In comparison, $\eta$
would be gradually increased with the reduction of energy scale once there exists
the $M_3$ impurity in a 2D QBCP system. Moreover, as clearly depicted in Fig.~\ref{Fig-eta-vi-m5-away-FP},
$\eta$ is apparently diminished under the influence of all types of impurities.
As a result, it indicates the combination of $M_1$ and $M_2$ dominates over $M_3$
in the low-energy region.

\subsection{Gaussian FP}\label{Subsection_eta-Gaussian}

Before going further, it is necessary to emphasize that
the results of Fig.~\ref{Fig-eta-vi-m5-away-FP} (a) and (b)
are based upon two concrete values, i.e., $\eta_0=0.1\pi$ and $\eta_0=0.4\pi$,
respectively. However, we would like to point out the rotational parameter $\eta$ that
is associated with the parameters $t_1$ and $t_3$ via $\eta=\mathrm{arctan}\frac{t_1}{t_3}$,
in principle, can run through the following range $\eta\in\{[-\frac{\pi}{2},0)
\cup(0,\frac{\pi}{2}]\}$ while the parameters $t_0$, $t_1$ and $t_3$ are
restricted by $|t_0|<\mathrm{min}(|t_1|,|t_3|)$.
At this stage, it therefore naturally raises a question
of where the $\eta$ eventually goes toward at the lowest-energy limit
once its initial value is randomly taken from this restricted range.

\begin{figure}
\centering
\includegraphics[width=3.2in]{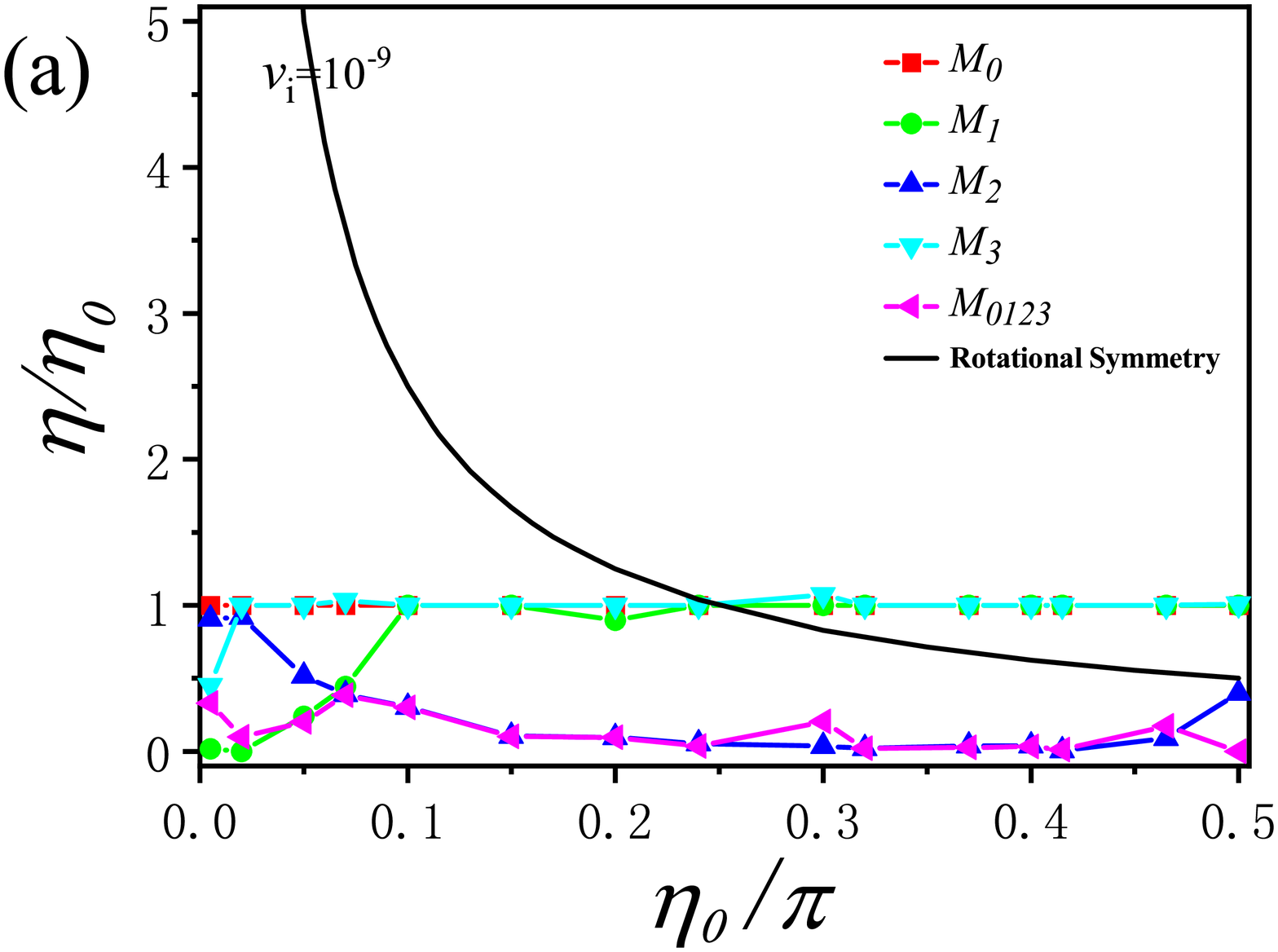}
\vspace{-0.35cm}\\
\includegraphics[width=3.2in]{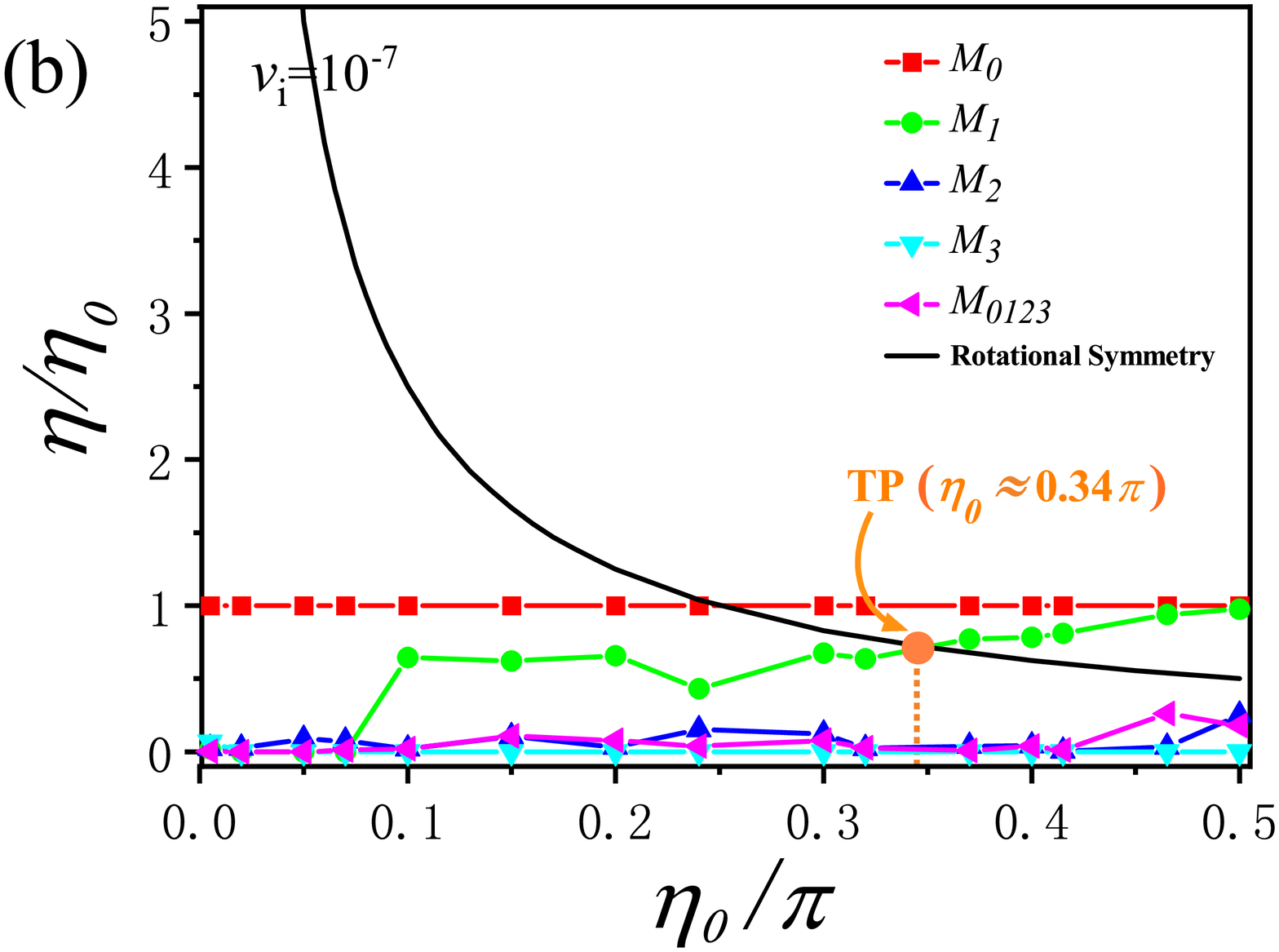}
\vspace{-0.35cm}\\
\includegraphics[width=3.2in]{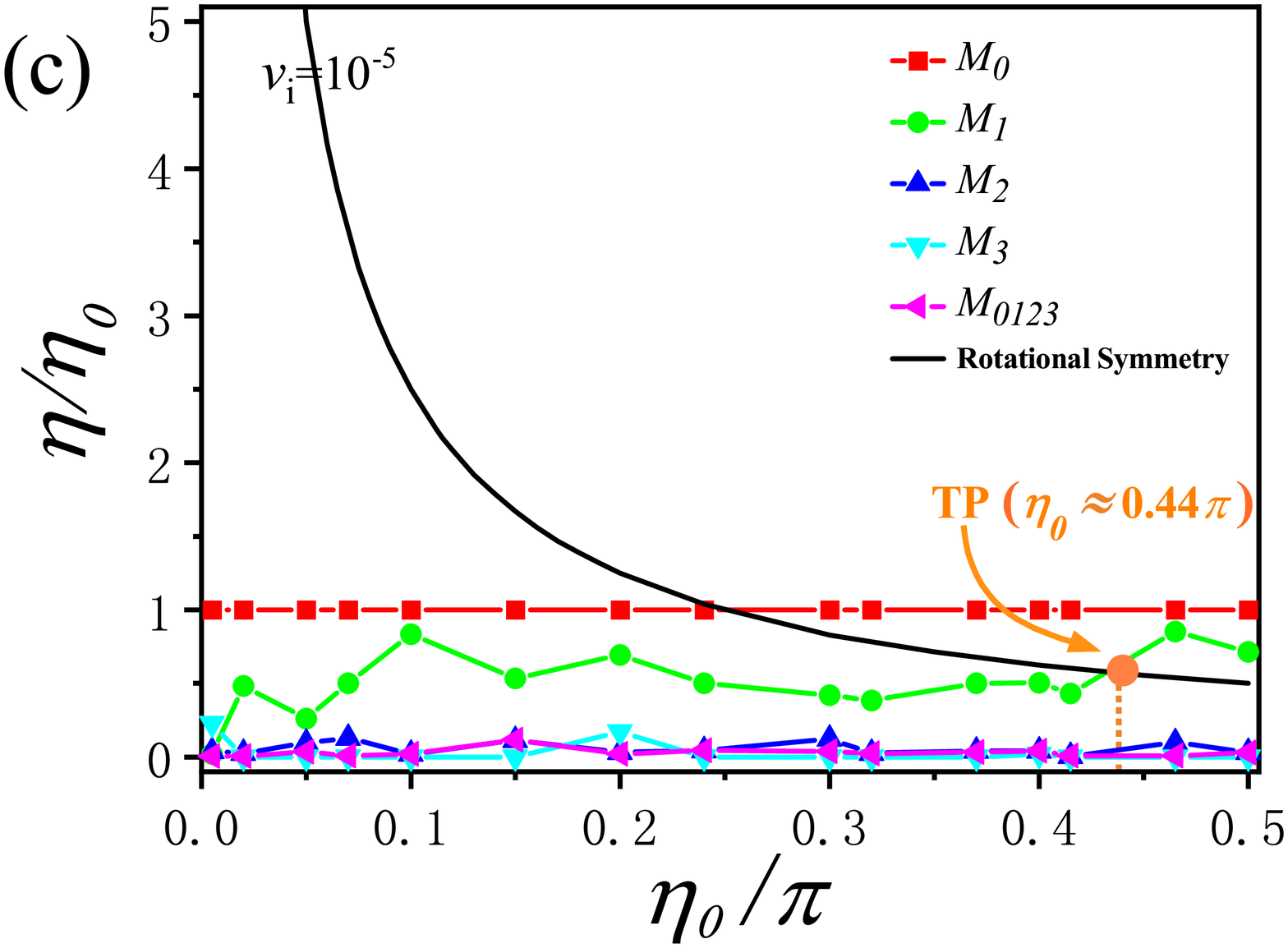}
\vspace{-0.35cm}
\caption{(Color online) Behaviors of $\eta(l\rightarrow l_c)/\eta_0$ with varying initial
values of $\eta_0$ against the impurities
for (a) $v_i(0)=10^{-9}$, (b) $v_i(0)=10^{-7}$, and (c) $v_i(0)=10^{-5}$
as the fermion-fermion couplings flow towards the
QAH FP. The lines for the $M_0$ impurity coincide with their counterparts at
clean limit and ``TP" represents
the very transition point from rotational asymmetry to symmetry.}\label{Fig-vi-M975-eta-eta-0-nearby-FP}
\end{figure}

Without loss of generality, we only consider the right range with
$\eta_0>0$ (the left range can be studied similarly) and then choose
several representative starting values of $\eta$, impurity strengths,
and fermion-fermion couplings that can drive the system into Gaussian FP.
To proceed, the values of $\eta$ around the Gaussian FP can be extracted
from the coupled RG equations that involve the competitions between
fermion-fermion interactions and impurity scatterings in the proximity
of critical energy scale. Fig.~\ref{Fig-vi-M975-eta-eta-0-away-FP}
collects our primary results for $\eta$. At the first sight, we find
that $M_1$ impurity just brings about slight impacts
on $\eta$ when its strength is very weak as shown in Fig.~\ref{Fig-vi-M975-eta-eta-0-away-FP}(a) and
Fig.~\ref{Fig-vi-M975-eta-eta-0-away-FP}(b). Whereas the values of $\eta$ are pulled
down once the $M_1$ impurity strength is sufficiently enhanced as clearly displayed in
Fig.~\ref{Fig-vi-M975-eta-eta-0-away-FP}(c). Although the parameter $\eta$ does not
receive any corrections at weak $M_3$ impurity, it is worth highlighting
that $M_3$, as opposed to the $M_1$ impurity, is in favor of increasing $\eta$ in the
entire region if the initial impurity strength is strong, such as $v_i=10^{-5}$
in Fig.~\ref{Fig-vi-M975-eta-eta-0-away-FP}(c). Subsequently, we move to study the
changes of $\eta$ due to the single presence of $M_2$ impurity. The blue
curves in Fig~\ref{Fig-vi-M975-eta-eta-0-away-FP} apparently indicate
that the value of $\eta$ is sensitive to the impurity strength,
which is heavily reduced by the $M_2$ impurity and even driven
to zero at the strong impurity strength. Furthermore, one can find
with the help of Fig.~\ref{Fig-vi-M975-eta-eta-0-away-FP} that $M_2$
impurity plays a leading role among other types of impurities once all
three types of impurities are present in the 2D QBCP system. Consequently,
the basic results are analogous to the sole presence of $M_2$ impurity.
Besides, the parameter $\eta$ is also slightly dependent upon its initial value.
The second line of Table~\ref{table-eta} briefly lists the effects of
various types of impurities on the rotational parameter $\eta$.

As aforementioned, the symmetry parameter $\eta$ of 2D QBCP system
is proved to be energy-independent at clean limit.
However, Fig.~\ref{Fig-eta-vi-m5-away-FP} indicates that
impurities would bring considerable influence to
this physical quantity. On one side, the parameter $\eta$ can
be robustly protected against sole $M_0$ impurity. On the other,
it would be either increased or decreased under the presence of other types
of impurities. Under such circumstances, it is of considerable interest
to ask how these impurities affect the rotational
asymmetry of 2D QBCP system and whether the rotational symmetry
can be induced from asymmetry case.

To facilitate our discussions, we provide the black curves in
Fig.~\ref{Fig-vi-M975-eta-eta-0-away-FP} to show the situations that
harbor rotational symmetry. In addition, since the $M_0$ impurity
does not influence $\eta$ at all, we hereby consider the red
lines as the clean-limit cases for convenient
comparisons. With the help of these two kinds of curves, we
find that the space between parameter $\eta$
and black line is susceptible to $M_1$, $M_2$, or $M_3$ impurity
compared to clean limit. As a result, it is of particular interest
to point out that the transition from rotational asymmetry to rotational
symmetry can be triggered by some weak $M_2$ impurity or strong $M_3$ impurity
at certain $\eta_0$, which are denoted by the intersections
dubbed transition point (TP) in Fig.~\ref{Fig-vi-M975-eta-eta-0-away-FP} (a) and (c).
However, we would like to stress that the condition for impurity-induced
rotational symmetry is remarkably strict and in principle the presence
of impurities in the 2D QBCP system is detrimental to the rotational
symmetry in the case of $u_i(0)$ flowing towards Gaussian FP.
In other words, the recovery of rotational symmetry from asymmetry is
only a specific inference (or an intriguing ingredient) of our primary results.

\subsection{QAH FP}

\begin{figure}
\centering
\includegraphics[width=3.2in]{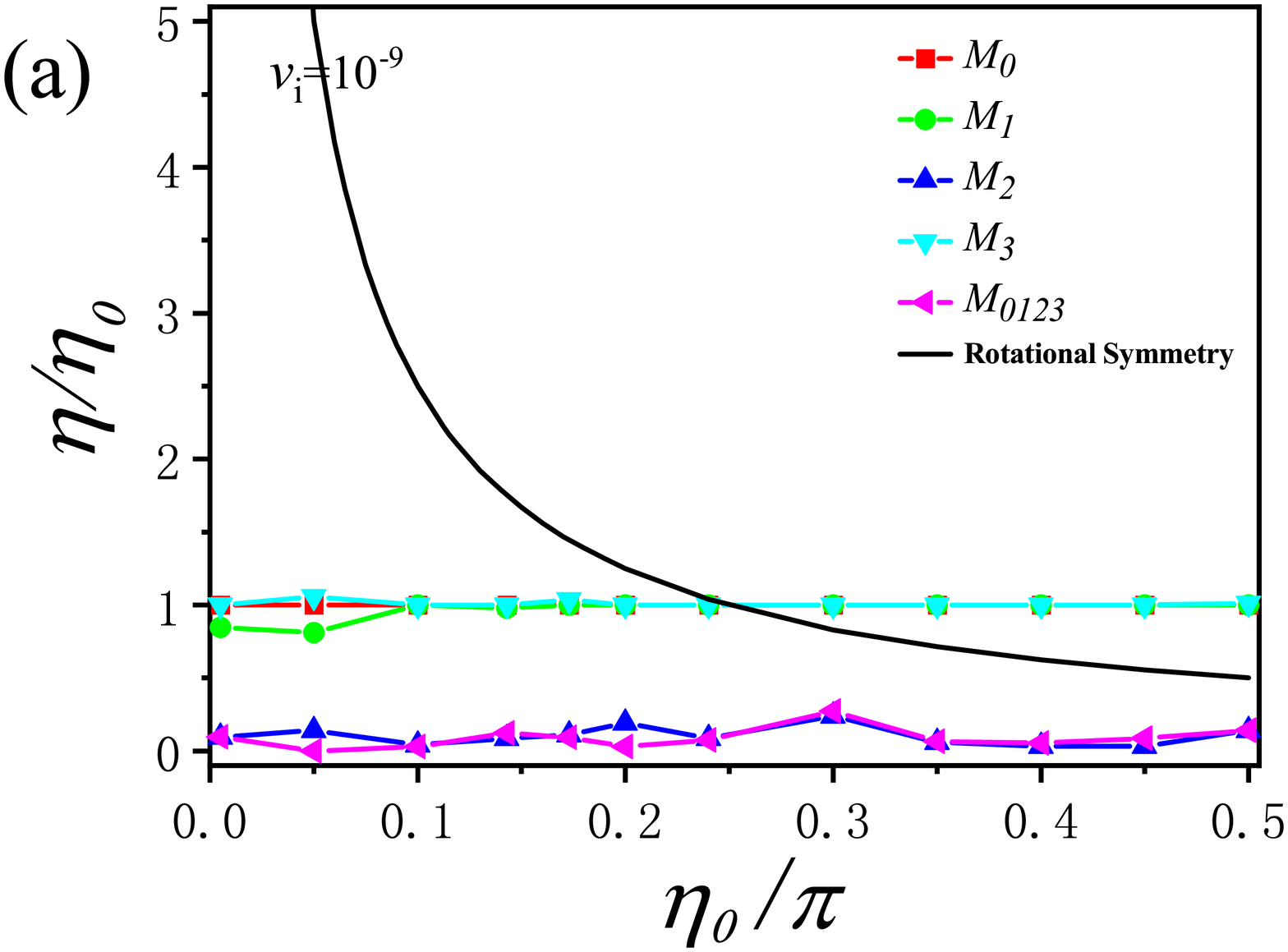}
\vspace{-0.35cm}\\
\includegraphics[width=3.2in]{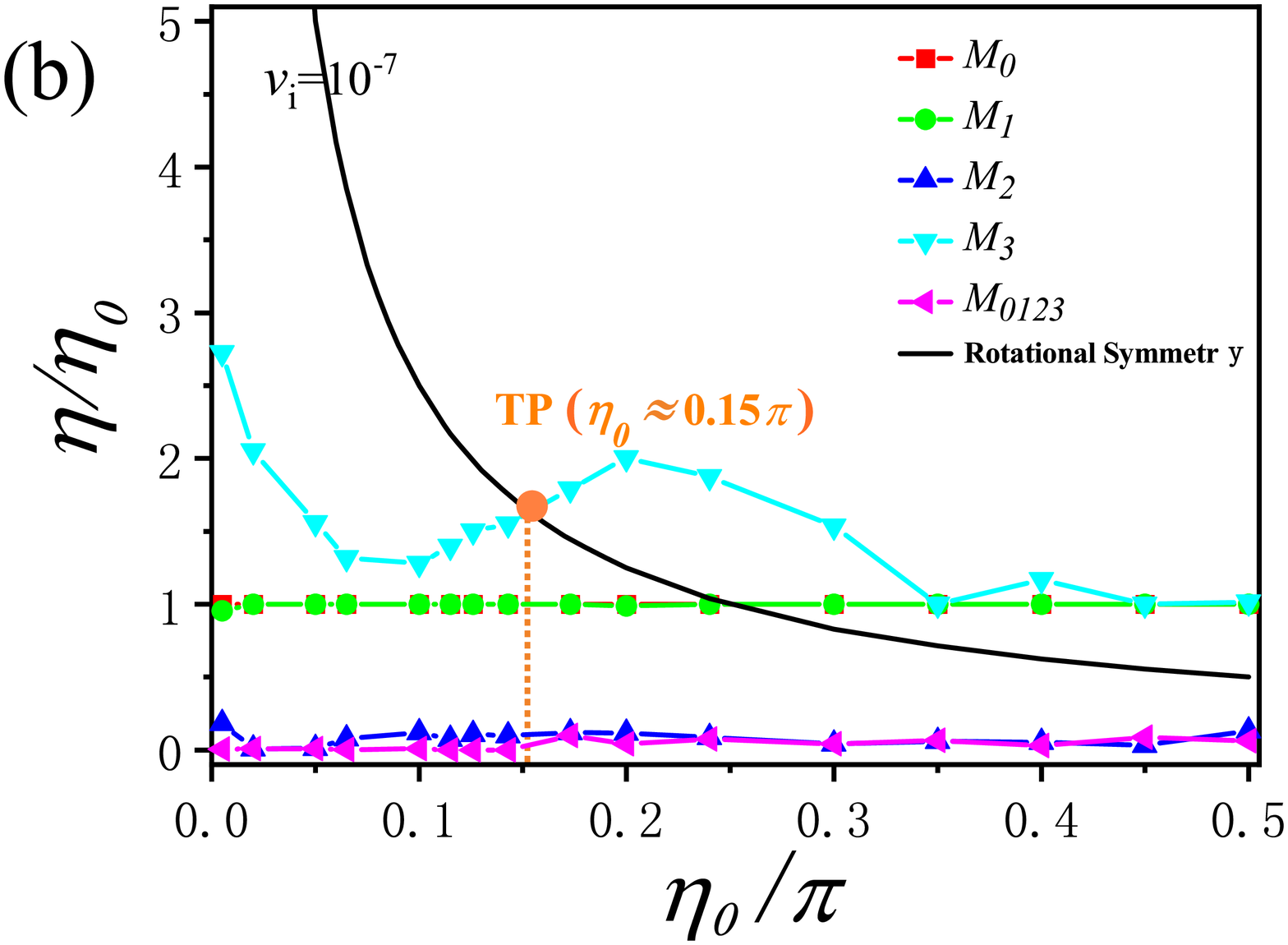}
\vspace{-0.35cm}\\
\includegraphics[width=3.2in]{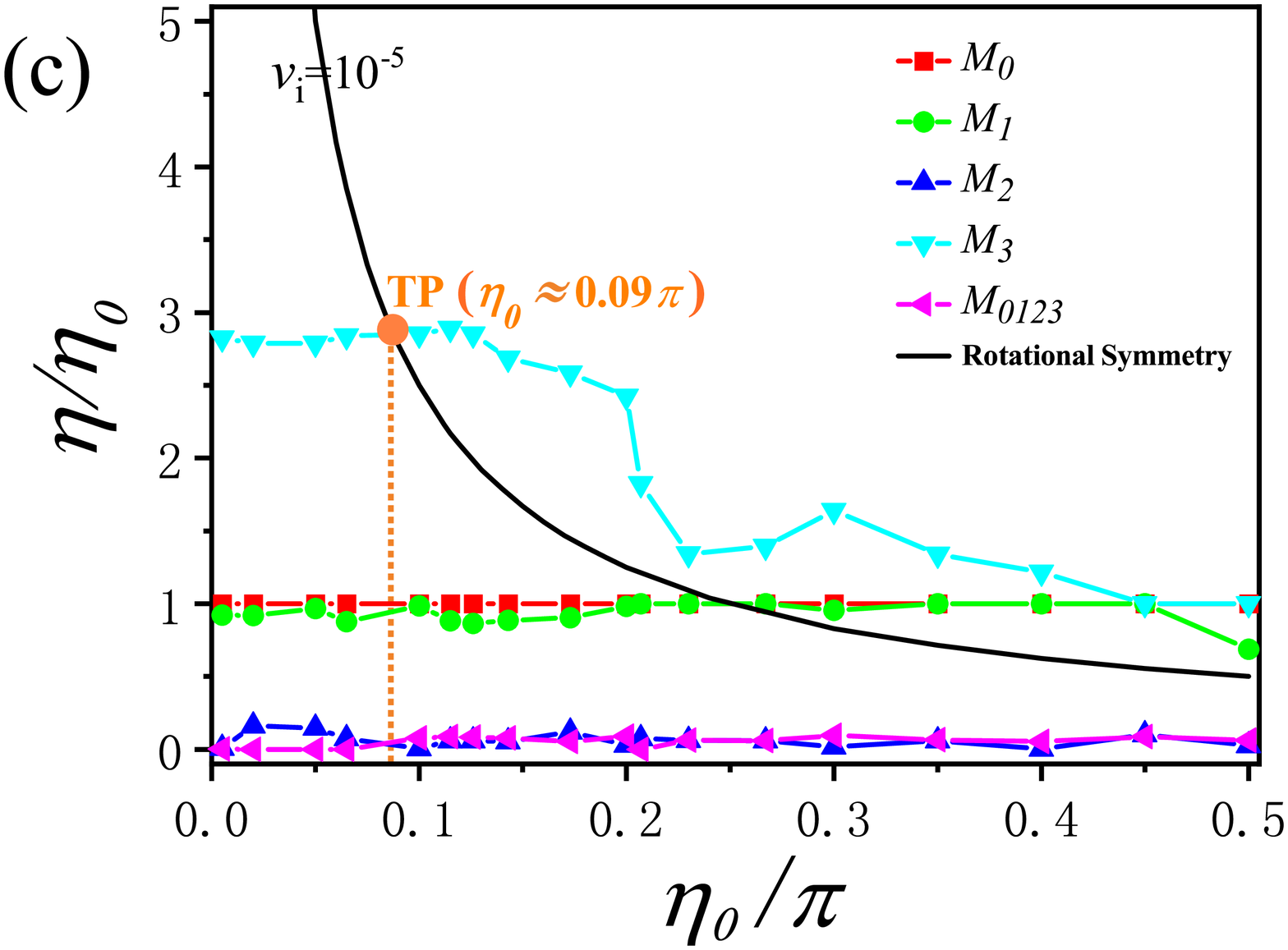}
\vspace{-0.35cm}\\
\caption{(Color online) Behaviors of $\eta(l\rightarrow l_c)/\eta_0$ with
varying initial values of $\eta_0$ against the impurities
for (a) $v_i(0)=10^{-9}$, (b) $v_i(0)=10^{-7}$, and (c) $v_i(0)=10^{-5}$ as
the fermion-fermion couplings flow towards the
NSN FP. The lines for the $M_0$ impurity coincide with their counterparts at
clean limit and ``TP" represents
the very transition point from rotational asymmetry to symmetry.}\label{Fig-vi-M975-eta-eta-0-QSH-FP}
\end{figure}

\begin{figure}
\centering
\vspace{0.7cm}
\includegraphics[width=3.2in]{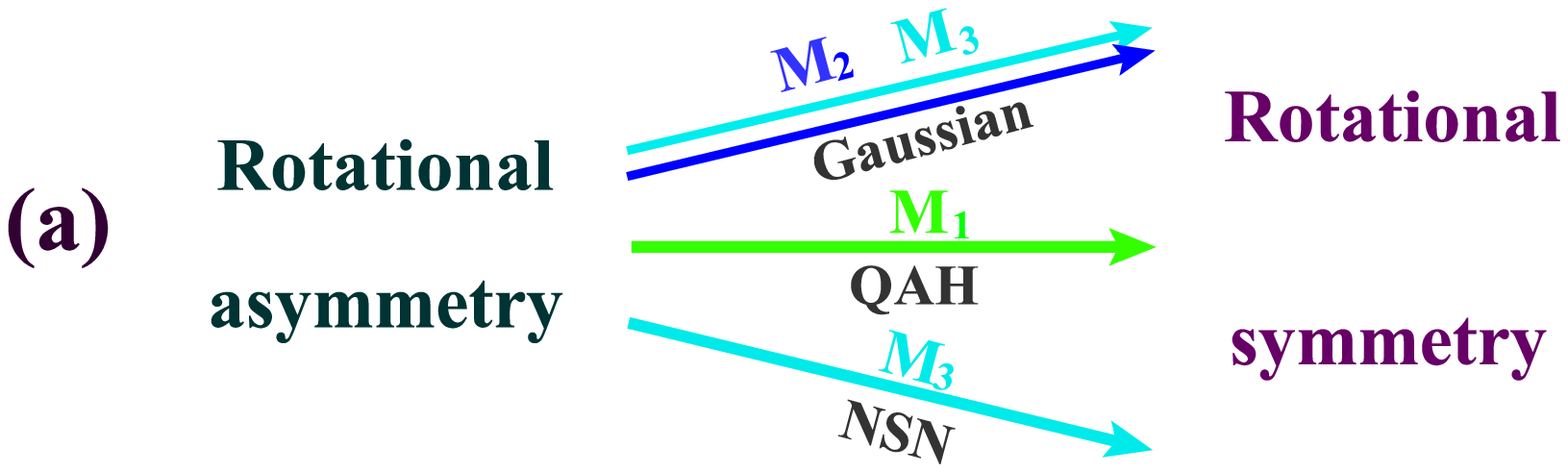}
\vspace{0.5cm}\\
\includegraphics[width=3.2in]{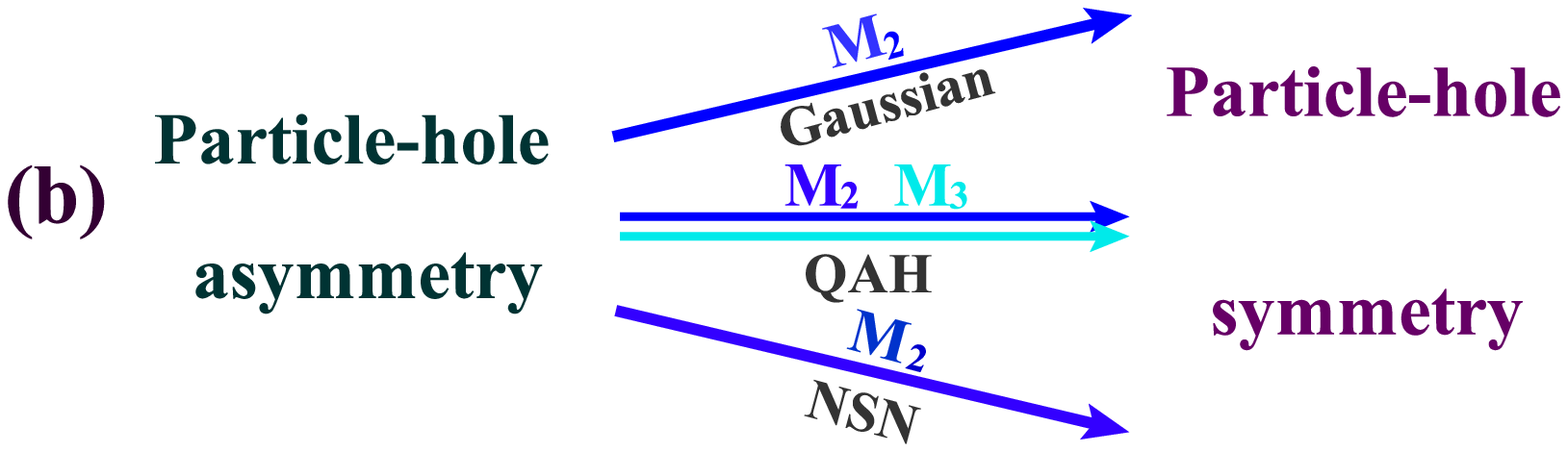}
\vspace{0cm}\\
\caption{(Color online) Potential transitions (a) from rotational
asymmetry to rotational symmetry and (b) from particle-hole asymmetry to
particle-hole symmetry under certain type of impurity as the fermion-fermion
interactions are driven to distinct sorts of FPs.}\label{Fig-symmetry}
\end{figure}

Subsequently, we turn to study how the parameter $\eta$ behaves
against the impurities while the fermion-fermion interactions are
governed by the QAH FP with certain starting values of $u_i(0)$.
Fig.~\ref{Fig-vi-M975-eta-eta-0-nearby-FP} collects our primary results
of $\eta$ caused by impurities at the lowest-energy limit after
performing the numerical evolution. Again, the black curves correspond
to the situations with rotational symmetry and the red lines stand for clean-limit cases.

At first, we find that the rotational parameter $\eta$ is stable under
the weak $M_1$ impurity but would be decreased in the low-energy regime
as long as the impurity strength is sufficiently strong. This is similar
to its Gaussian counterpart. Besides, it is of particular interest to
address that the subtle combination of $M_1$ impurity and fermion-fermion
interaction is endowed with a potential ability to drive an anisotropic system
into an isotropic one with the rotational symmetry at certain very $\eta_0$.
These unique positions denoted by TP at which the phase transitions set in
are illustrated in Fig.~\ref{Fig-vi-M975-eta-eta-0-nearby-FP} (b) and (c).
Next, we inspect the behaviors of $\eta$ under the influence of only $M_2$ impurity.
Fig.~\ref{Fig-vi-M975-eta-eta-0-nearby-FP} singles out that $\eta$ is quickly
diminished and departs from the symmetric curve with tuning up $v_i$. This implies
that it is difficult under $M_2$ impurity to achieve rotational symmetry from anisotropic
case in the 2D QBCP system. Similarly, comparing the blue lines of
Fig.~\ref{Fig-vi-M975-eta-eta-0-away-FP} with Fig.~\ref{Fig-vi-M975-eta-eta-0-nearby-FP},
we find that the values of $\eta$ at QAH FP are reduced and deviated more from rotational symmetry's
than their Gaussian counterparts especially in the presence of weak impurity.
This indicates that $u_i$ flowing towards QAH FP is more detrimental to the
rotational symmetry. As a result, competition between fermion-fermion interaction and $M_2$ impurity
is sensitive to the FP. Moreover, as opposed to the Gaussian situation,
the parameter $\eta$ depicted in Fig.~\ref{Fig-vi-M975-eta-eta-0-nearby-FP} (b) and (c)
would be reduced or even go towards zero with lowering the energy scale in the presence
of the sole $M_3$ impurity. Different trajectories of fermion-fermion interactions
for QAH and Gaussian FPs may be responsible for these distinctions. In other
words, fermion-fermion interactions win against the $M_3$ impurity
while the fermionic couplings are approaching to QAH FP and thus cause the original
contribution of $M_3$ impurity to be neutralized. At last, we realize that
the fates of parameter $\eta$ are similar to the sole presence of
$M_2$ or $M_3$ impurity once there exist all three different types of impurities
in 2D QBCP system. To recapitulate, the presence of impurity is conventionally
harmful to the formation of rotational symmetry except the very point under
the $M_1$ impurity at which the rotational symmetry is preferred.

\subsection{NSN FP}

At last, we go to the case with fermion-fermion interactions evolving towards NSN FP.
In order to unveil the fate of the rotational parameter $\eta$ in the vicinity of NSN FP,
we parallel several tedious but straightforward calculations that are similar to
the procedures for QAH FP and then present our main results
in Fig.~\ref{Fig-vi-M975-eta-eta-0-QSH-FP} from which
one can figure out that the effects of various impurities on $\eta$ are sensitive to
the initial values of $\eta(0)$.

\begin{table}
\caption{Qualitative effects of different types of impurities on the rotational
parameter $\eta$ at $v_i(0)=10^{-5}$ approaching distinct sorts of FPs.
Hereby, ``\textcolor{blue}{--}'' indicates $\eta$ is hardly influenced or remains certain
constant. Instead, ``\textcolor{red}{$\uparrow$}'' and ``\textcolor{green}{$\downarrow$}''
stand for the increase and decrease with lowering the energy scale,
respectively (the double arrows indicate much more increase or decrease).}\label{table-eta}
\vspace{0.39cm}
\centerline{
\begin{tabular}{p{2.2cm}<{\centering}p{1cm}<{\centering}p{1cm}<{\centering}
p{1cm}<{\centering}
p{1cm}<{\centering}p{1.5cm}<{\centering}}
\hline
\hline
\renewcommand{\arraystretch}{1}
\rule{0pt}{15pt}&$M_0$&$M_1$&$M_2$&$M_3$&$M_{0123}$\\
\hline
\rule{0pt}{15pt}Gaussian FP&\textcolor{blue}{--}& \textcolor{green}{
$\downarrow$}&\textcolor{green}{$\downarrow\downarrow$}&
\textcolor{red}{$\uparrow$}&\textcolor{green}{$\downarrow\downarrow$}  \\
\rule{0pt}{15pt}QAH FP& \textcolor{blue}{--}&\textcolor{green}{$\downarrow$}&\textcolor{green}
{$\downarrow\downarrow$}
&\textcolor{green}{$\downarrow\downarrow$}&\textcolor{green}{$
\downarrow\downarrow$} \\
\rule{0pt}{15pt}NSN FP&\textcolor{blue}{--}&\textcolor{blue}{--}&
\textcolor{green}{$\downarrow\downarrow$}&\textcolor{red}{$\uparrow$}
&\textcolor{green}{$\downarrow\downarrow$} \\
\hline
\hline
\end{tabular}
}
\end{table}

At the outset, we find that the contributions of $M_0$ or $M_1$ impurity to
the parameter $\eta$ are negligible even at strong impurity
with lowering the energy scale. This suggests that both of them
prefer to protect the stability of $\eta$ nearby NSN FP.
Then, we turn to examine the effect of the sole presence of $M_2$ impurity.
Studying from Fig.~\ref{Fig-vi-M975-eta-eta-0-QSH-FP}, one can readily realize that
the $M_2$ impurity plays a significant role in pinning down the low-energy
fate of parameter $\eta$ around the NSN FP. Apparently,
$\eta$ is heavily reduced with lowering
the energy scale. As a consequence,
the $M_2$ impurity is considerably detrimental to the rotational symmetry
around NSF FP in 2D QBCP system. Compared to the Gaussian case,
its effects are more manifest and powerful.
This may be ascribed to the intimate coupling between $M_2$
impurity and the fermion-fermion interactions
accessing the NSN FP. Besides the $M_{0,1,2}$
impurities, several intriguing results are also generated
by the sole presence of $M_3$ impurity. In distinction to other sorts of
impurities, the $M_3$ impurity is prone to increase the value of $\eta$ once its initial strength is
suitable, as apparently exhibited by cyan lines of Fig.~\ref{Fig-vi-M975-eta-eta-0-QSH-FP}.
Analogous to the unique phenomena sparked by $M_1$ around QAH FP and $M_3$ nearby
Gaussian FP, Fig.~\ref{Fig-vi-M975-eta-eta-0-QSH-FP}(b) and (c) show that
certain transition from rotational asymmetry to rotational symmetry can
be triggered by the adequately strong $M_3$ impurity in the proximity of NSN FP.
Furthermore, we briefly comment on the concomitant presence of all types of
impurities in 2D QBCP system around NSN FP. Learning from Fig.~\ref{Fig-vi-M975-eta-eta-0-QSH-FP},
one readily figures out that the basic tendencies are consistent with the sole
$M_2$ impurity; namely, $\eta$ is substantially decreased via lowering the energy scale.
This corroborates again that the $M_2$ impurity takes a leading responsibility in
governing the low-energy properties.

To be brief, the effects of impurities on the rotational parameter $\eta$
are very different while the fermion-fermion interactions are controlled by
distinct sorts of FPs. Fig~\ref{Fig-symmetry} (a) and Table~\ref{table-eta}
summarize our primary conclusions for the low-energy behaviors of parameter $\eta$ under
the competitions between fermion-fermion interactions and impurities scatterings.
Before closing this section, we would like to emphasize again that both closely
energy and intimately starting-value dependence of low-energy fates of the asymmetric
parameter $\eta$ constitute our primary conclusions. In comparison, the underlying
transition from rotational to symmetric situation is only an interesting ingredient
of our main results under certain restricted condition.

\section{Fate of particle-hole asymmetry}\label{Sec_lambda}

In the previous section, we address the effects of competitions
between fermion-fermion interactions and impurity scatterings on rotational
asymmetry parameter $\eta$ while the fermion-fermion couplings $u_i$ are
governed by three distinct types of FPs.
To proceed, we within this section endeavor to investigate the behaviors of
particle-hole asymmetry $\lambda$ under these three distinct situations.

\subsection{Warm-up and Gaussian FP}\label{Sec_lambda_GS}

\begin{figure}
\centering
\includegraphics[width=3.2in]{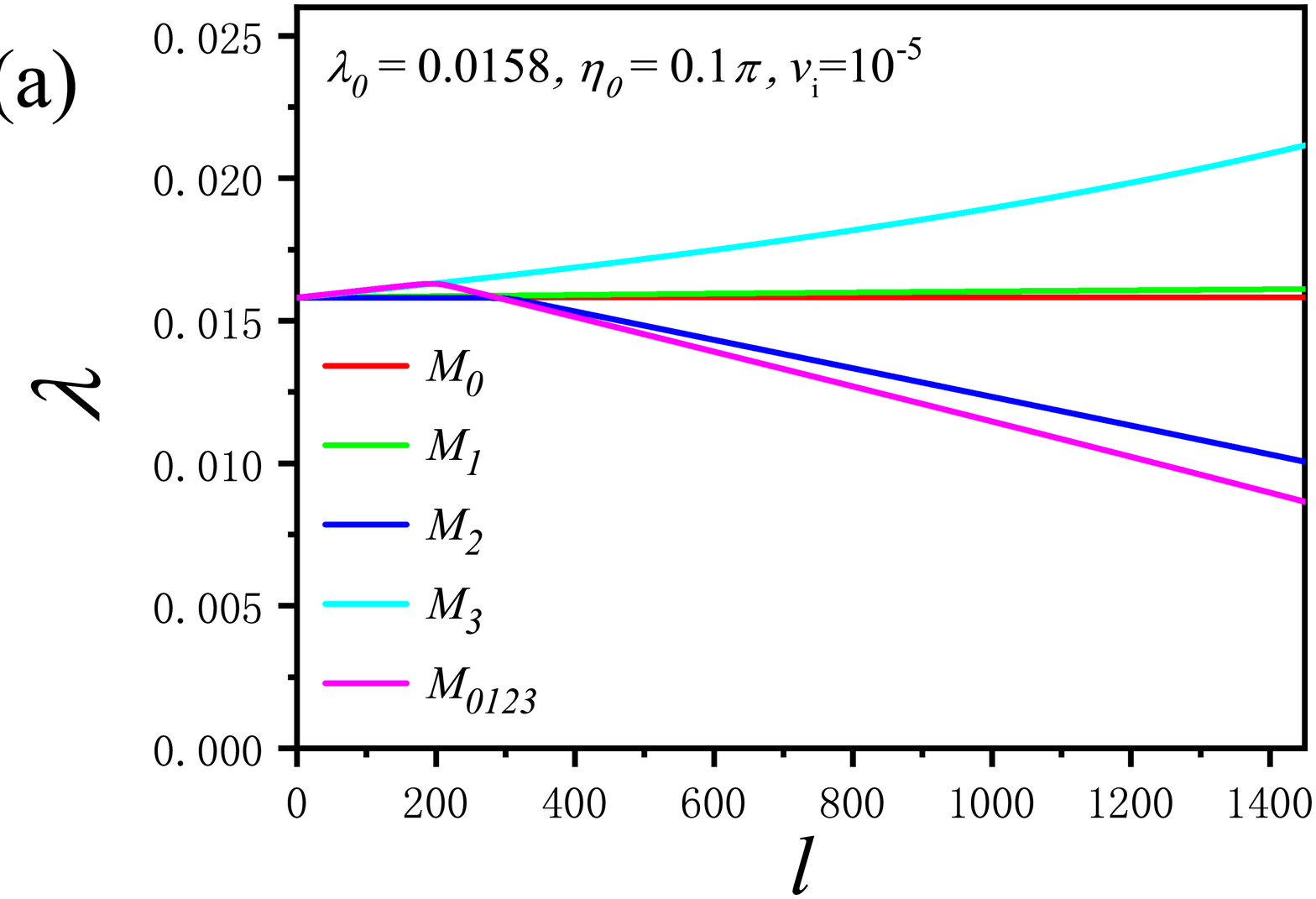}
\vspace{-0.35cm}\\
\includegraphics[width=3.2in]{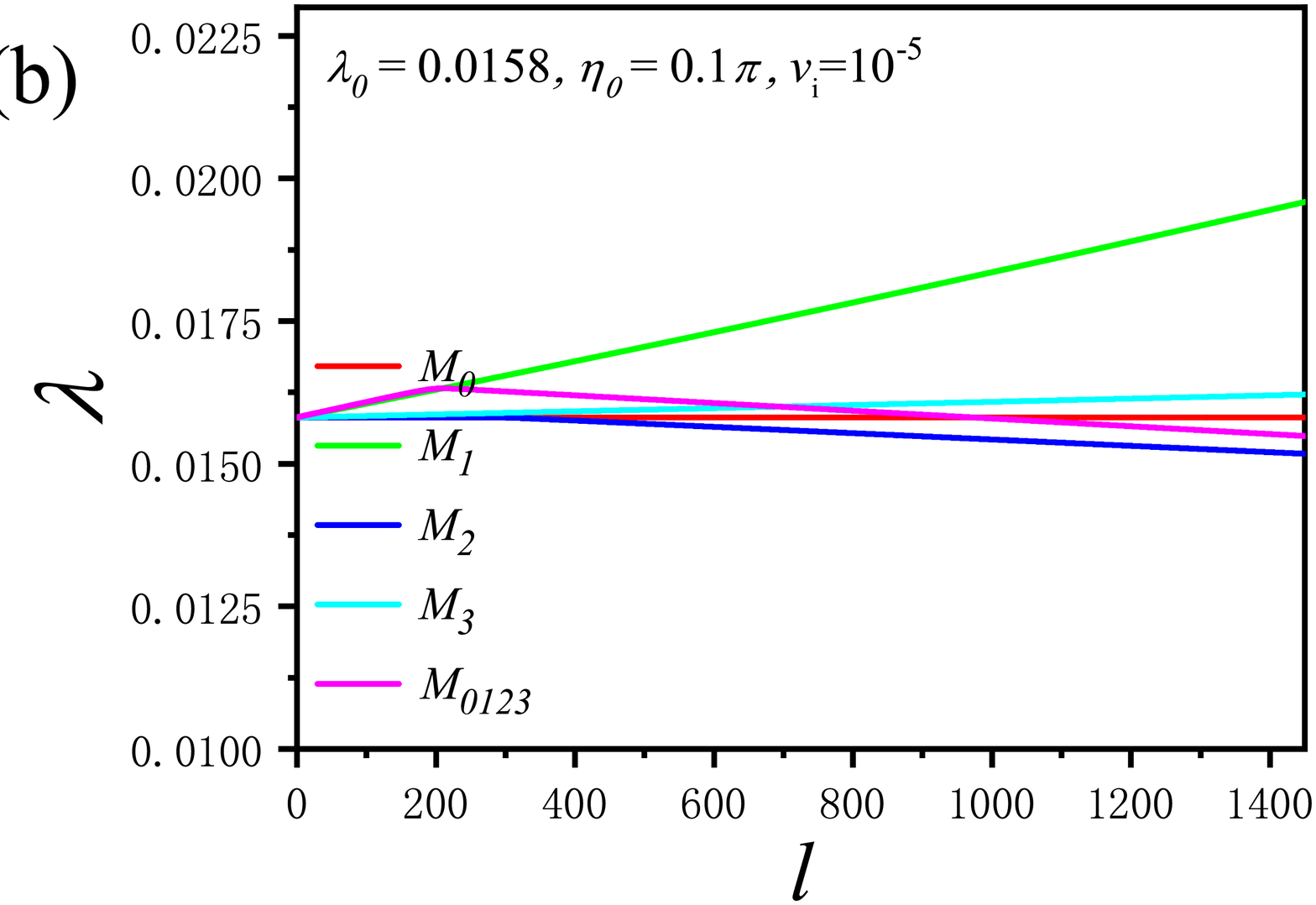}
\vspace{-0.35cm}\\
\caption{(Color online) Evolutions of $\lambda$
against the presence of impurities at $v_i(0)=10^{-5}$
for (a) $\eta_0=0.1\pi$ and (b) $\eta_0=0.4\pi$ with fermion-fermion couplings
flowing towards Gaussian FP.}\label{Fig-lambda-vi-m5-away-FP}
\end{figure}

\begin{figure}
\centering
\includegraphics[width=3.2in]{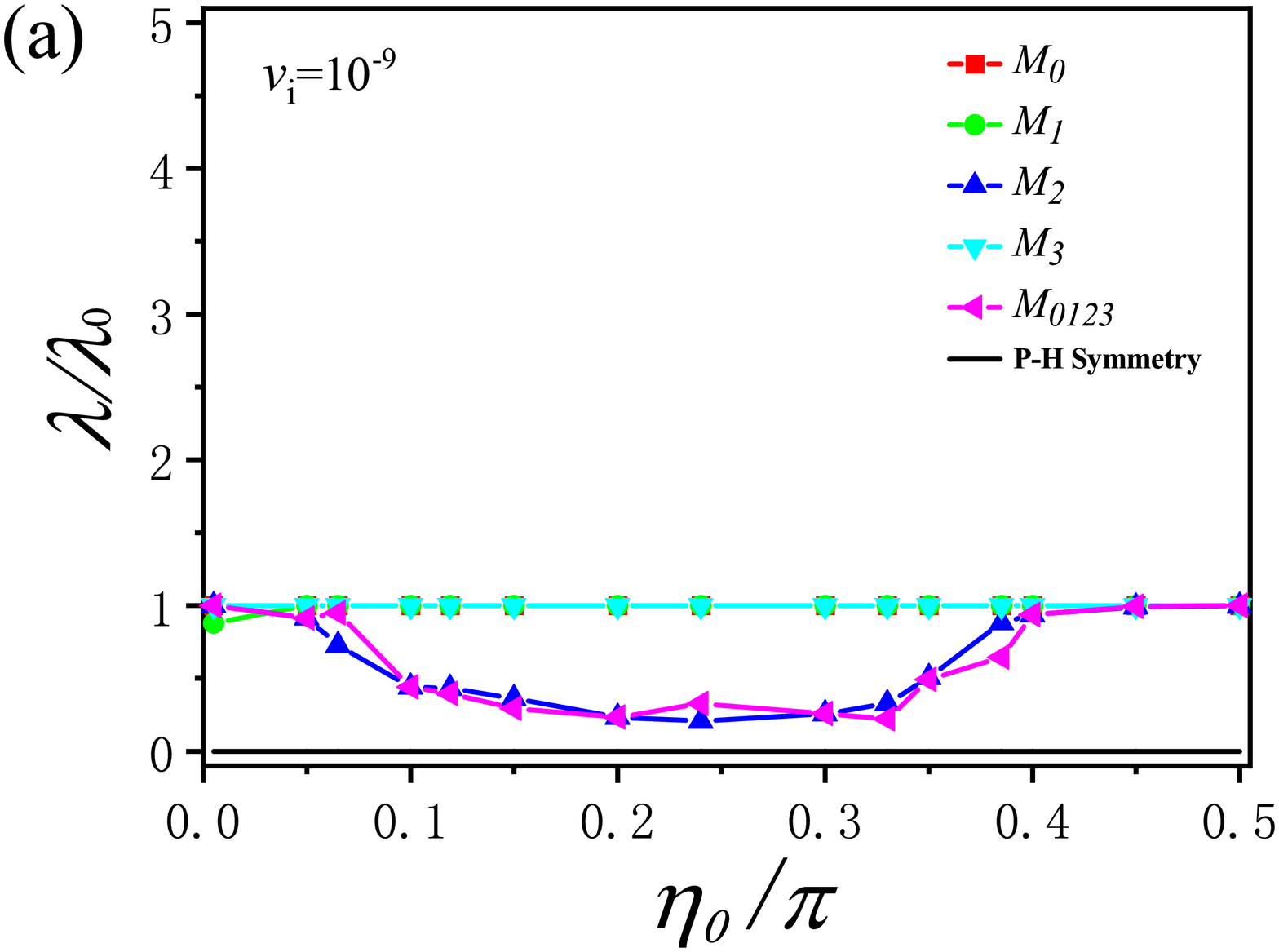}
\vspace{-0.35cm}\\
\includegraphics[width=3.2in]{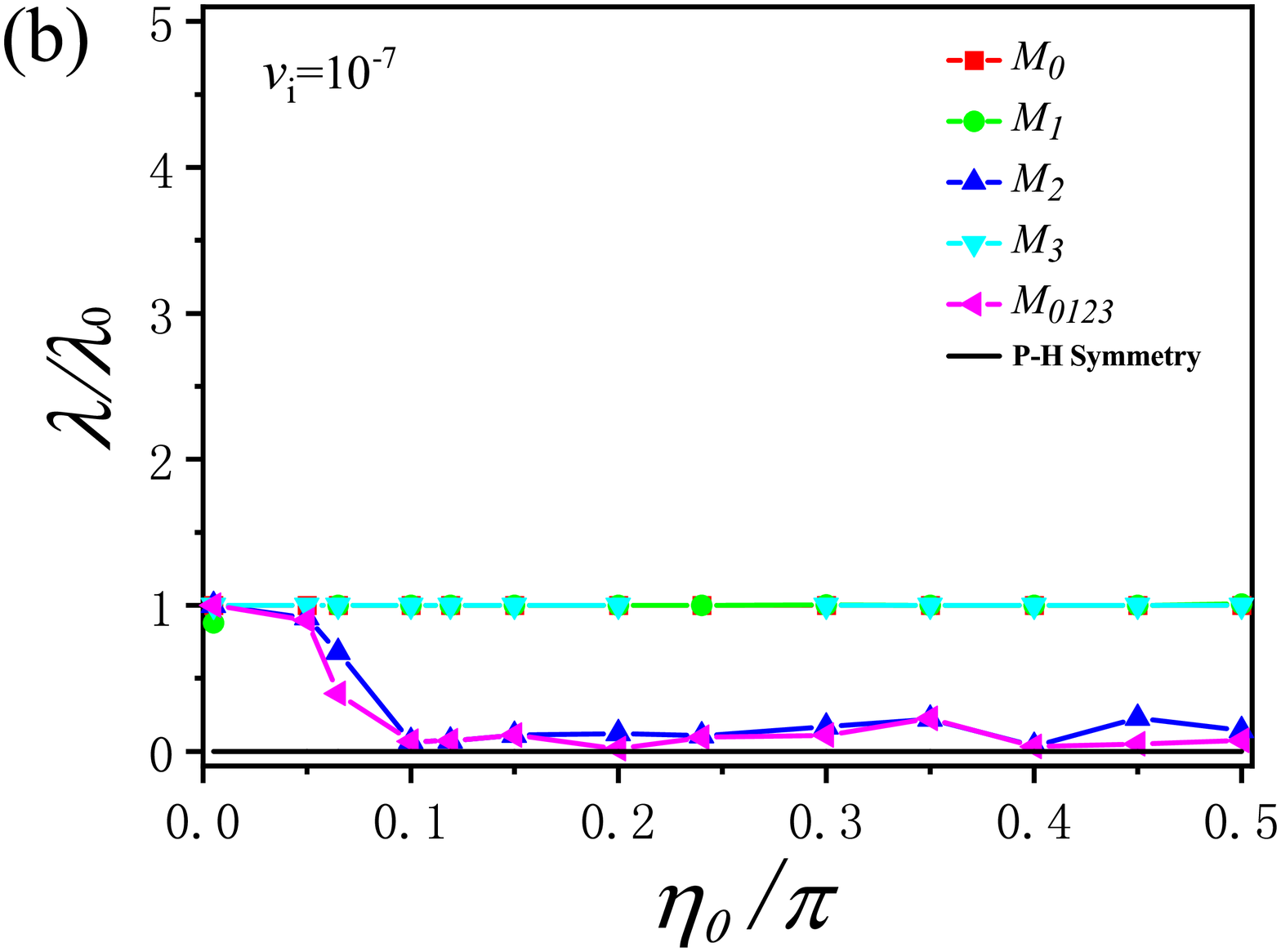}
\vspace{-0.35cm}\\
\includegraphics[width=3.2in]{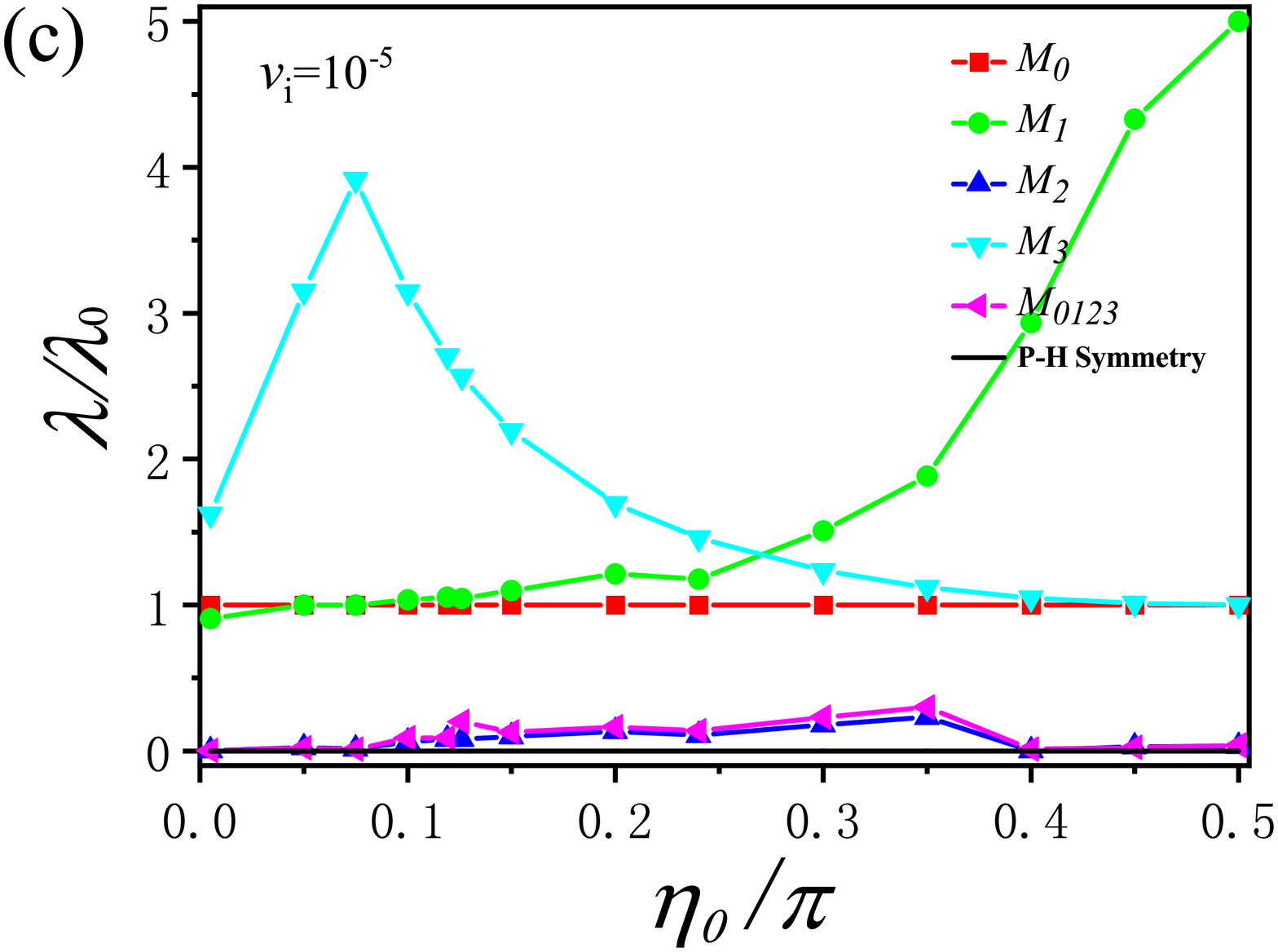}
\vspace{-0.35cm}
\caption{(Color online) Behaviors of $\lambda(l\!\!\longrightarrow\!\!l_c)/\lambda_0$ with
varying initial values of $\eta_0$ against the impurities for (a) $v_i(0)=10^{-9}$,
(b) $v_i(0)=10^{-7}$, and (c) $v_i(0)=10^{-5}$ once the fermion-fermion interactions
flow towards Gaussian FP. The lines for the $M_0$ impurity coincide with their counterparts at
clean limit.}\label{Fig-vi-M975-lambda_i-eta-0-away-FP}
\end{figure}

Before moving further, we would like to make a warm-up again to roughly
check the energy-dependent tendency of the particle-hole parameter $\lambda$
in the presence of impurity. Without loss of generalities, we follow the steps
of Sec.~\ref{Subsection_warm-up} by choosing
specific starting values for interaction parameters and show the influence of
distinct types of impurities on the low-energy behaviors of parameter $\lambda$
in Fig.~\ref{Fig-lambda-vi-m5-away-FP}. Reading off Fig.~\ref{Fig-lambda-vi-m5-away-FP},
we realize that both impurities $M_1$ and $M_3$ prefer to increase
the parameter $\lambda$ as the energy scale is lowered. On the contrary,
it climbs down in the presence of sole $M_2$ or all three
types of impurities.

Again, we would like to highlight that Fig.~\ref{Fig-lambda-vi-m5-away-FP}
only collects the energy-dependent information of parameter $\lambda$ at
$\eta_0=0.1\pi$ and $\eta_0=0.4\pi$. As mentioned in Sec.~\ref{Subsection_eta-Gaussian},
the rotational parameter $\eta$ satisfies $\eta\in\{[-\frac{\pi}{2},0)
\cup(0,\frac{\pi}{2}]\}$ once the QBCP's dispersion is stable.
In order to capture more physical information at the lowest-energy limit,
it is necessary to sort out representative starting values of $\eta$ that are distributed
in the whole restricted range.

Subsequently, we put our focus on the Gaussian FP.
After carrying out the numerical studies with variation of the initial value of $\eta$,
the principal results are summarized in Fig.~\ref{Fig-vi-M975-lambda_i-eta-0-away-FP},
where red and black lines in consistent with Fig.~\ref{Fig-vi-M975-eta-eta-0-away-FP} for the parameter
$\eta$ are exploited to characterize the clean-limit case and 2D QBCP system with
particle-hole symmetry, respectively. Specifically, Fig.~\ref{Fig-vi-M975-lambda_i-eta-0-away-FP}(a)
and (b) uncover that the low-energy values of $\frac{\lambda}{\lambda_0}$ at the weak impurities are
almost unaffected against the presence of a sole $M_1$ or $M_3$ impurity.
However, Fig.~\ref{Fig-vi-M975-lambda_i-eta-0-away-FP}(c) shows that
the parameter $\lambda$ largely climbs up once the impurity strengths
are adequately strong. As a result, the $M_1$ or $M_3$ impurity is harmful to
the particle-hole symmetry. In comparison, the $M_2$ impurity is prone to
heavily reduce the parameter $\lambda$, which is forced to flow towards
zero as the impurity is sufficiently increased. Moreover, when all types of impurities are
present simultaneously, the $M_{1,3}$ and $M_2$ ferociously compete as their contributions
are exactly converse. Eventually, as depicted in Fig.~\ref{Fig-vi-M975-lambda_i-eta-0-away-FP},
$M_2$ wins the competition and hence the basic results
are consistent with the sole presence of the $M_2$ impurity.
In other words, this proposes that $M_2$ impurity, compared to other sorts
of impurities, brings the major contribution to the parameter $\lambda$ in the
low-energy regime. To proceed, we try to investigate whether these
impurities are favorable for particle-hole symmetry around
the Gaussian FP. Given the 2D QBCP system possesses
the particle-hole symmetry exactly at $\lambda=0$,
one, with the help of Fig.~\ref{Fig-vi-M975-lambda_i-eta-0-away-FP}, can infer that
the presence of the $M_2$ impurity alone or three different types of impurities
indeed promote the particle-hole symmetry in 2D the QBCP system.

\subsection{QAH FP}\label{Sec_lambda_QAH}

\begin{figure}
\centering
\includegraphics[width=3.2in]{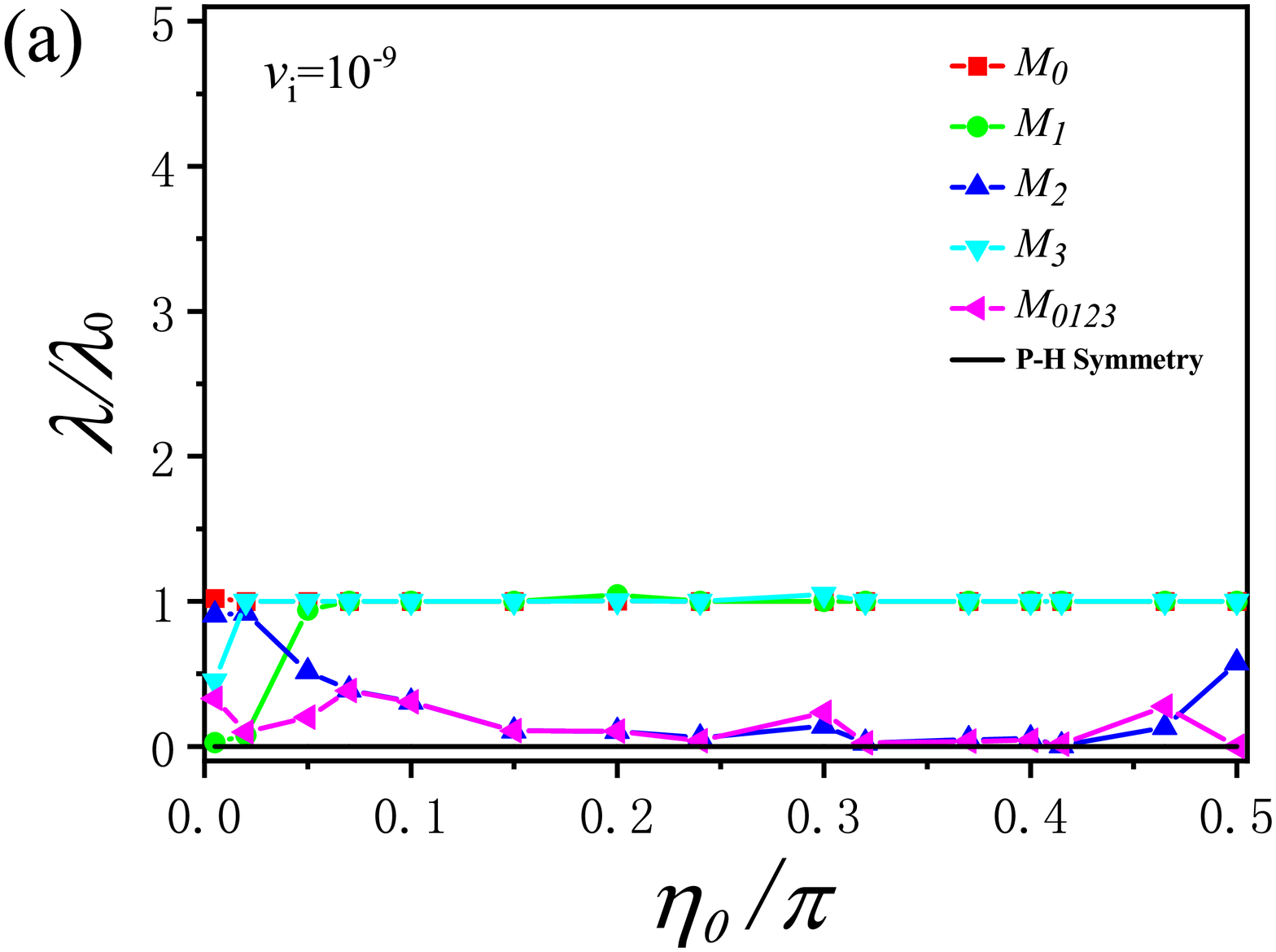}
\vspace{-0.35cm}\\
\includegraphics[width=3.2in]{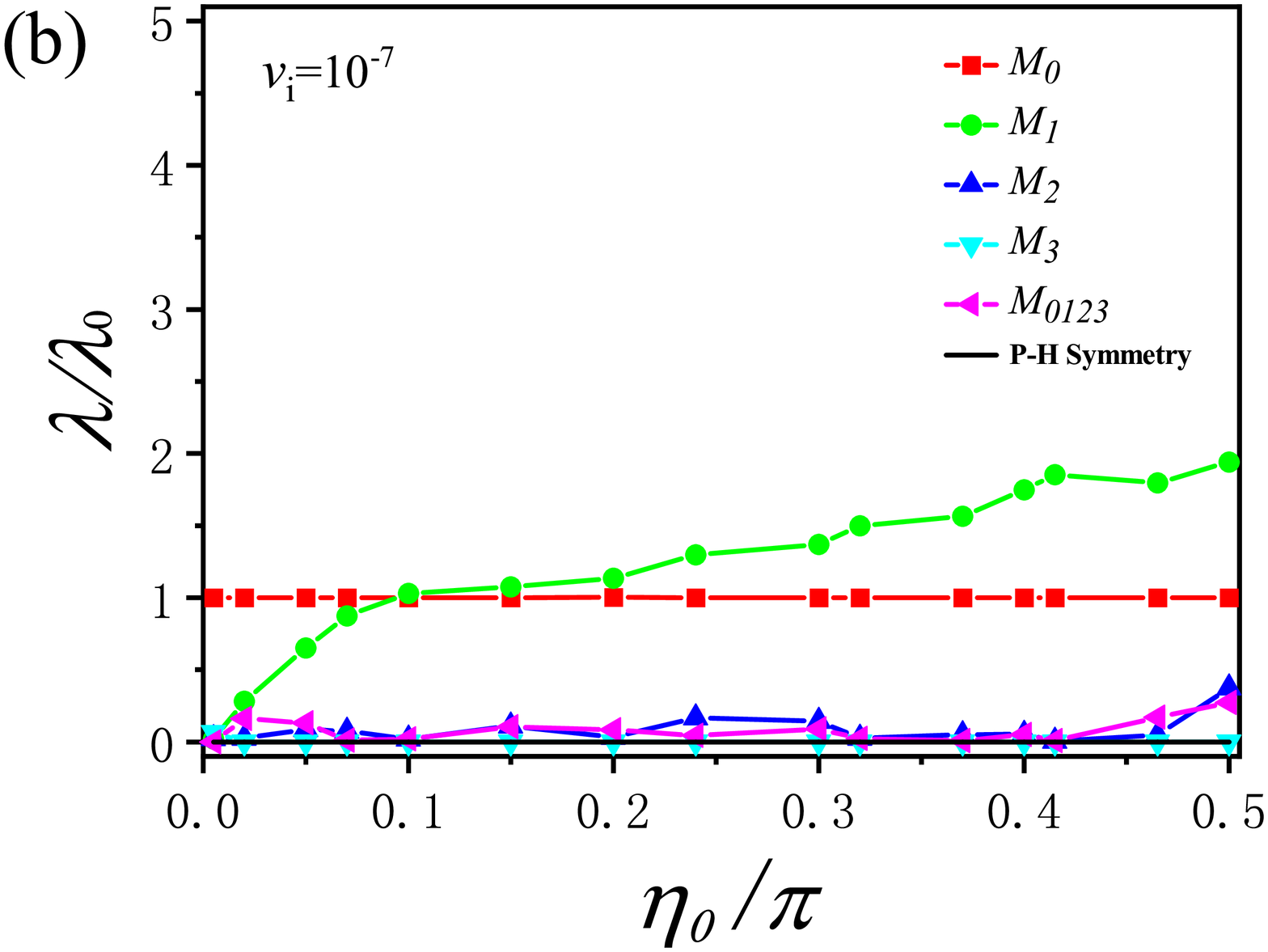}
\vspace{-0.35cm}\\
\includegraphics[width=3.2in]{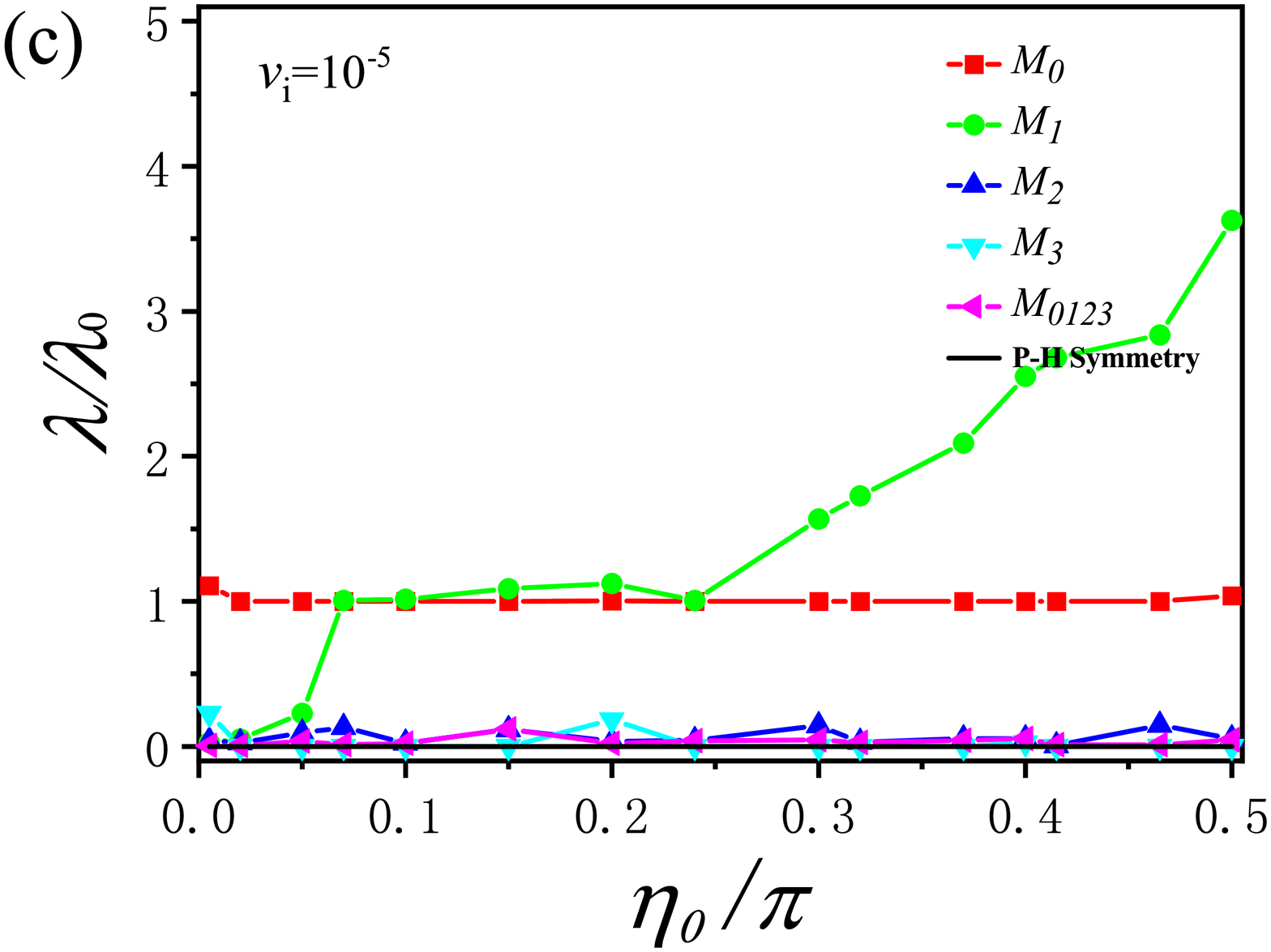}
\vspace{-0.35cm}
\caption{(Color online) Behaviors of $\lambda(l\!\!\longrightarrow\!\!l_c)/\lambda_0$
with varying initial values of $\eta_0$ against the impurities
for (a) $v_i(0)=10^{-9}$, (b) $v_i(0)=10^{-7}$, and (c) $v_i(0)=10^{-5}$
once the fermion-fermion interactions flow towards QAH FP.
The lines for the $M_0$ impurity coincide with their counterparts at
a clean limit.}\label{Fig-vi-M975-lambda_i-eta-0-nearby-FP}
\end{figure}

Subsequently, we move our target to the situation at which the system is
attracted by QAH FP. With the help of coupled RG equations~(\ref{Eq_t_0})-(\ref{Eq_v_3})
and repeating the procedures in Sec.~\ref{Sec_lambda_GS}, we are able to present the
energy-dependent evolutions of particle-hole parameter $\lambda$, which carry the
physical information about the competitions between impurities and fermion-fermion
interactions.

Fig.~\ref{Fig-vi-M975-lambda_i-eta-0-nearby-FP} manifestly characterizes
the physical key points for parameter $\lambda$ against impurities in the proximity
of QAH FP. To be specific, we find from Fig.~\ref{Fig-vi-M975-lambda_i-eta-0-nearby-FP}
that the red curves remain invariant and stable for the whole regime of parameter $\eta$.
In other words, the parameter $\lambda$ is very insensitive to the $M_0$ impurity.
Additionally, $\lambda$ exhibits manifestly $\eta_0$-dependent behaviors under the $M_1$ impurity.
On one hand, the parameter $\lambda$ receives significant revision at $\eta_0<0.07\pi$,
which is insusceptible to impurity strength. On the other,
the effects of weak $M_1$ impurity on the parameter $\lambda$ are
negligible at $\eta_0\geq0.07\pi$ as depicted in
Fig.~\ref{Fig-vi-M975-lambda_i-eta-0-nearby-FP}(a).
However, as the initial strength increases, the $M_1$ impurity
can cause an enhancement of parameter $\lambda$ as shown in
Fig.~\ref{Fig-vi-M975-lambda_i-eta-0-nearby-FP}(b) and (c).
In comparison, more interesting effects on $\lambda$ are triggered by the
sole presence of $M_2$ impurity around the QAH FP.
Fig.~\ref{Fig-vi-M975-lambda_i-eta-0-nearby-FP}
apparently exhibits the parameter $\lambda$ is heavily reduced under the $M_2$ impurity.
In particular, it almost vanishes while the $M_2$ impurity strength is adequate strong as
clearly shown in Fig.~\ref{Fig-vi-M975-lambda_i-eta-0-nearby-FP}(b) and (c).
This indicates that the transformation from particle-hole asymmetry to symmetry
would be expected in that the 2D QBCP system is invariant under the particle-hole
transformation at $\lambda=0$. In other words, $M_2$ impurity is
very much in favor of particle-hole symmetry.
Furthermore, one can readily capture the information from
Fig.~\ref{Fig-vi-M975-lambda_i-eta-0-nearby-FP} that the presence
of all sorts of impurities shares the common basic conclusions with
the sole type of $M_2$ impurity irrespective of impurity strength. This again corroborates
the $M_2$ impurity dominates over other impurities.
At last, we briefly address the effects of $M_3$ impurity
scattering. The influence of weak $M_3$ impurity is consistent
with its $M_1$ counterpart as depicted in Fig.~\ref{Fig-vi-M975-lambda_i-eta-0-nearby-FP}(a).
Rather, in analogy with the $M_2$ impurity illustrated in
Fig.~\ref{Fig-vi-M975-lambda_i-eta-0-nearby-FP}(b) and (c), it
prefers to promote the particle-hole symmetry whilst the
starting value is increased. Compared to the properties around Gaussian FP,
these results unambiguously shed light on the importance of M3 impurity that
behaves like a catalyst to ignite fermion-fermion interactions in boosting
particle-hole symmetry as the system approaches the QAH FP.

\subsection{NSN FP}

\begin{figure}
\centering
\includegraphics[width=3.2in]{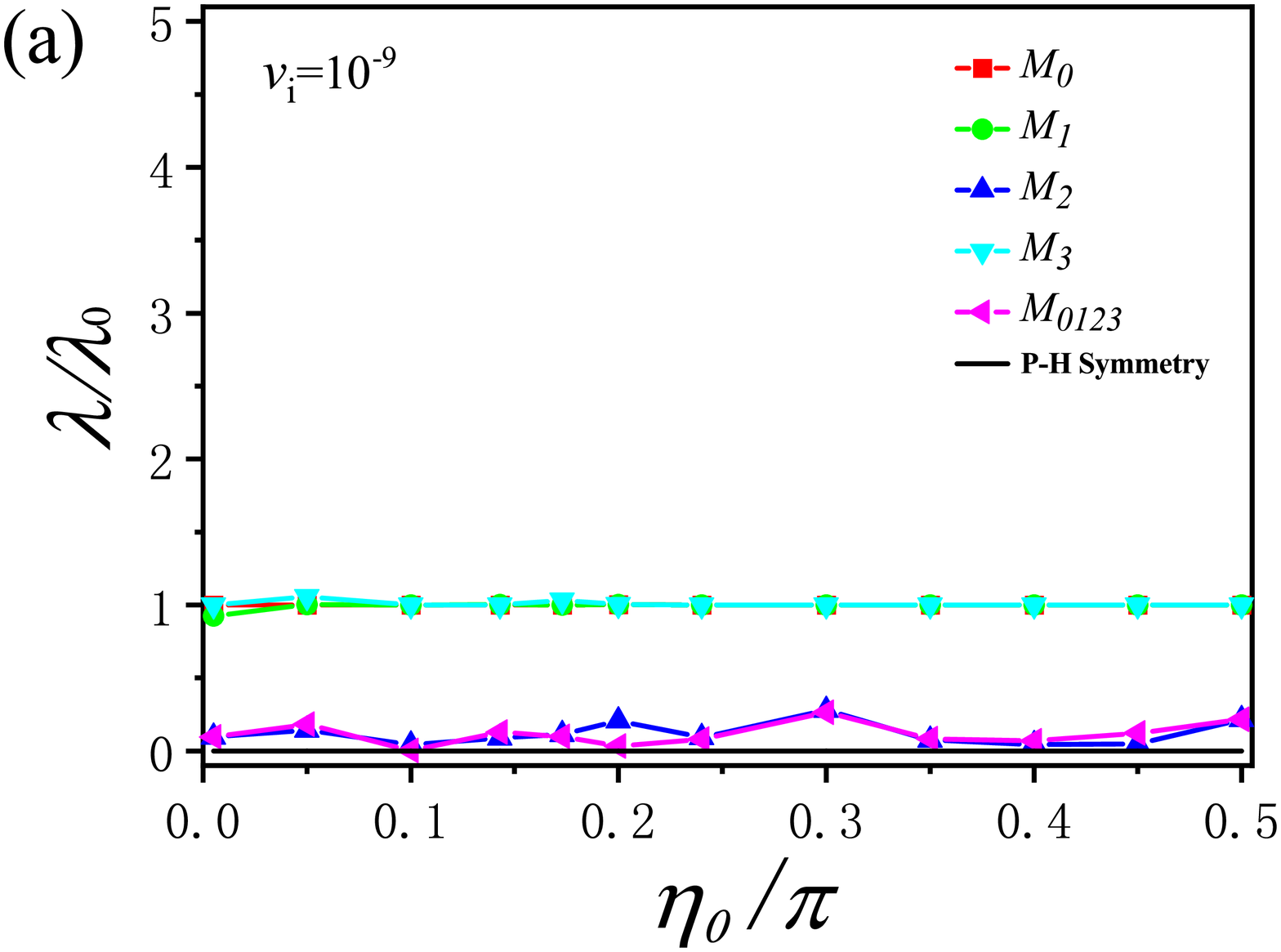}
\vspace{-0.35cm}\\
\includegraphics[width=3.2in]{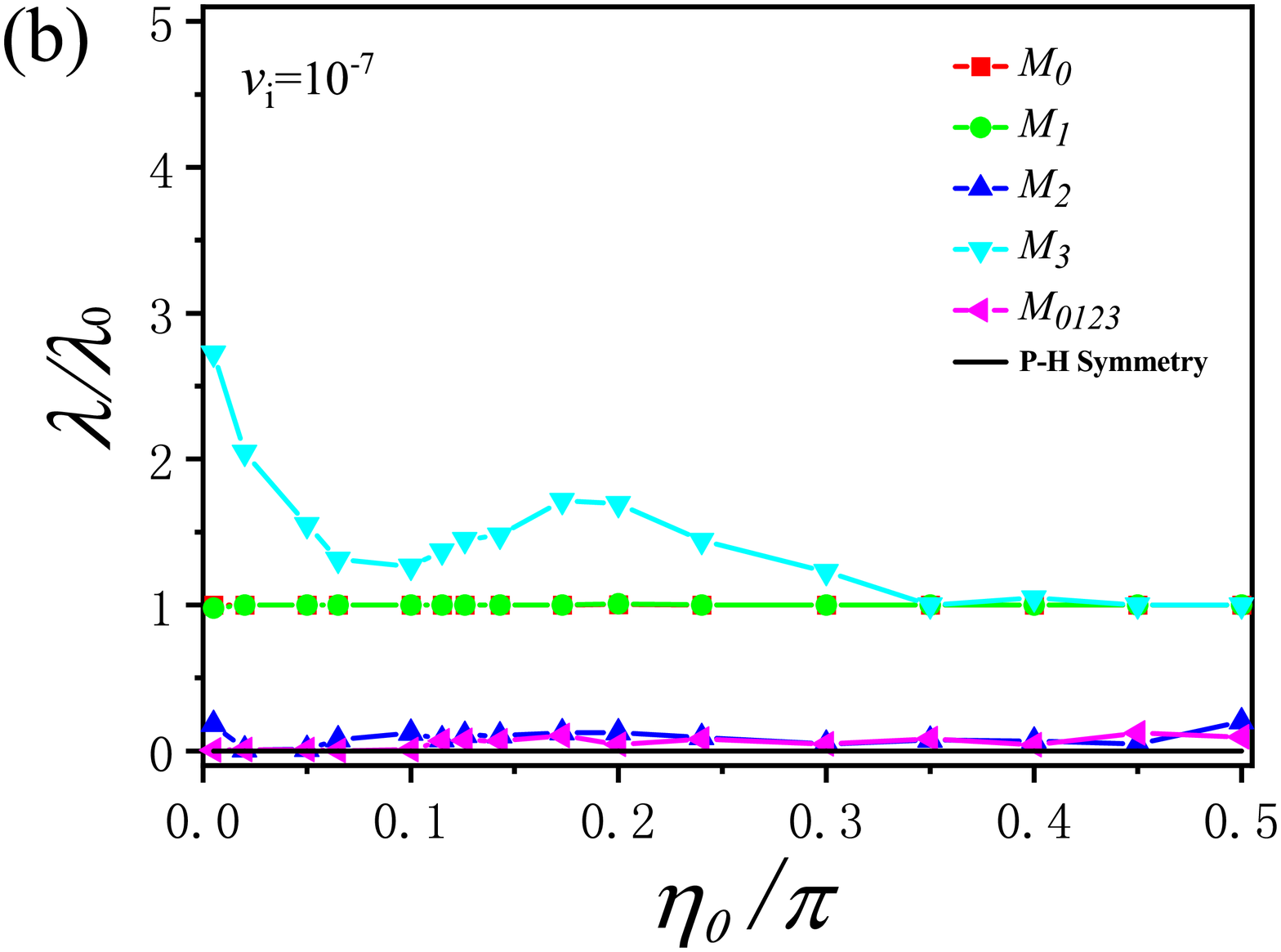}
\vspace{-0.35cm}\\
\includegraphics[width=3.2in]{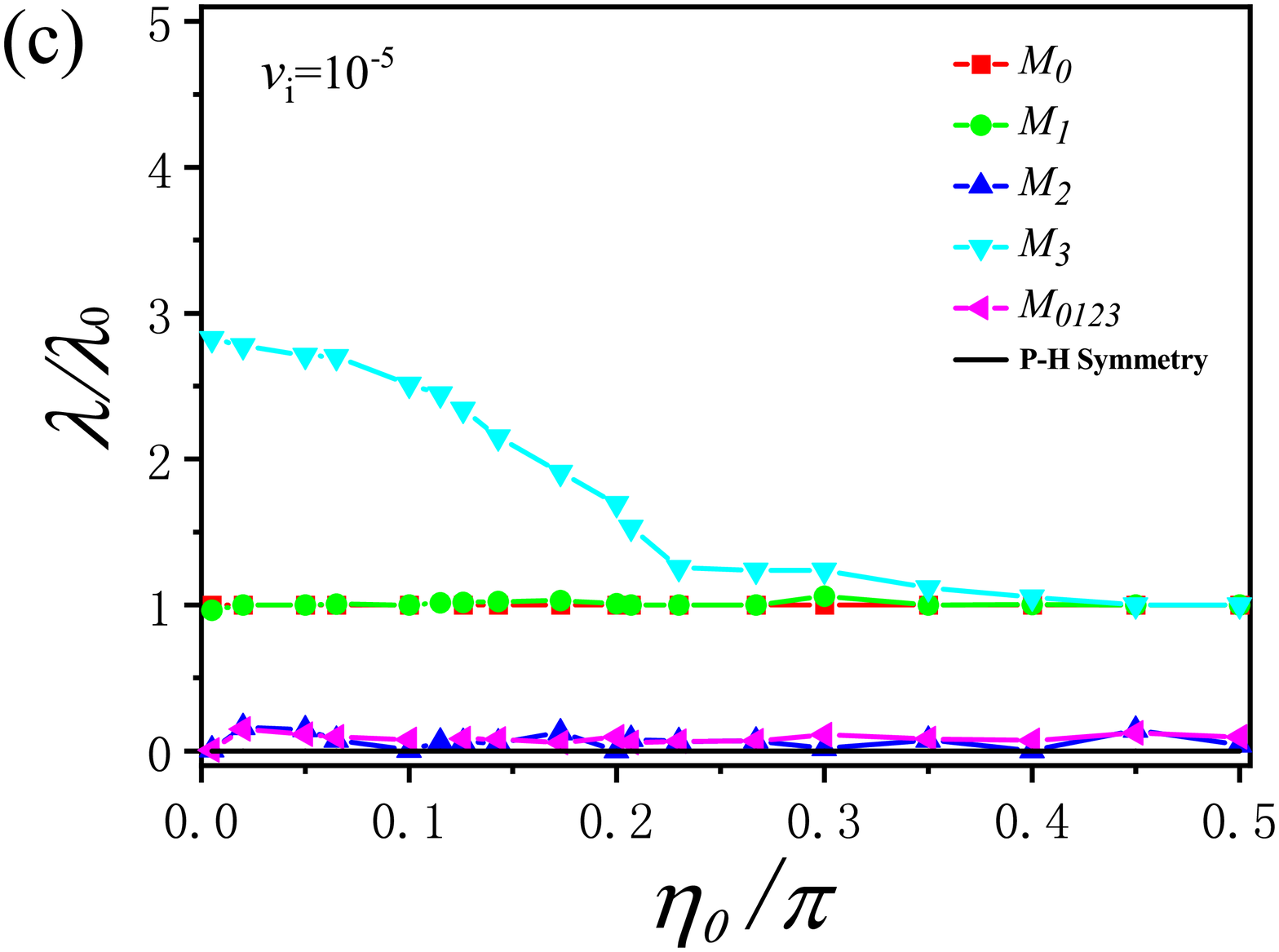}
\vspace{-0.35cm}\\
\caption{(Color online) Behaviors of $\lambda(l\!\!\longrightarrow \!\!l_c)/\lambda_0$ with varying
initial values of $\eta(0)$ against the impurities
for (a) $v_i(0)=10^{-9}$, (b) $v_i(0)=10^{-7}$, and (c) $v_i(0)=10^{-5}$
once the fermion-fermion interactions flow towards the
NSN FP. The lines for the $M_0$ impurity coincide with their counterparts at
clean limit.}\label{Fig-vi-M975-lambda-eta-0-QSH-FP}
\end{figure}

Finally, we turn to extract the low-energy fates of particle-hole
parameter $\lambda$ from the coupled RG equations under
distinct types of impurities in the proximity of NSN FP.

After carrying out tedious but straightforward calculations
with variation of the initial values of $\eta$, the main
results for the low-energy properties of $\frac{\lambda}{\lambda_0}$
are collected in Fig.~\ref{Fig-vi-M975-lambda-eta-0-QSH-FP}.
At the first sight, we get that the values of $\lambda$ at the
critical energy denoted by $l_c$ are very stable if there only exists the
sole $M_0$ or $M_1$ impurity in the 2D QBCP system. This indicates that
the $M_1$ impurity accessing to the NSN FP plays a very distinct role
in determining the fate of parameter $\lambda$ compared to other FPs
studied in Sec.~\ref{Sec_lambda_GS} and Sec.~\ref{Sec_lambda_QAH}.
Distinct energy-dependent trajectories of fermion-fermion couplings
around three types of FPs may be responsible
for this difference. With respect to the $M_2$ impurity,
parameter $\lambda$ displays a sharp drop as shown in blue curves of
Fig.~\ref{Fig-vi-M975-lambda-eta-0-QSH-FP}.
As mentioned for Gaussian and QAH FPs,
this suggests that the transformation from particle-hole asymmetry to
particle-hole symmetry can be manifestly realized in 2D QBCP system
with the help of $M_2$ impurity. On the contrary, as to the $M_3$ impurity, it
inclines to lift the parameter $\lambda$ and therefore hinders the
particle-hole symmetry of the system once the impurity strength is
sufficient strong. Furthermore, Fig.~\ref{Fig-vi-M975-lambda-eta-0-QSH-FP} distinctly
shows that the impact of the presence of all types of impurities on $\lambda$
is in agreement with the sole presence of $M_2$ impurity. This again proves the
leading role of $M_2$ impurity among all sorts of impurities.

To recapitulate, we within this section carefully investigate the influence
of various sorts of impurities on the particle-hole asymmetry parameter
$\lambda$ while the fermion-fermion couplings are governed by three distinct
types of FPs. Several interesting results are obtained. For instance,
certain type of impurity can induce very distinct effects on the
parameter $\lambda$ in the vicinity of different types of FPs.
In addition, the 2D QBCP system may undergo the transition from particle-hole
asymmetry to symmetry in the restricted conditions. Concretely, $M_2$ impurity
prefers to trigger the particle-hole symmetry at all three different types FPs,
while $M_3$ can evoke the symmetry of system only at QAH FP. Fig~\ref{Fig-symmetry}
(b) and Table~\ref{table-lambda} summarize our primary conclusions for the low-energy
behaviors of parameter $\lambda$ under the competitions between fermion-fermion
interactions and impurity scatterings.

\renewcommand\arraystretch{2}
\begin{table}
\caption{Qualitative effects of different types of impurities on the particle-hole
parameter $\lambda$ at $v_i(0)=10^{-5}$ approaching distinct sorts of FPs.
Hereby, ``\textcolor{blue}{--}'' indicates $\lambda$ is hardly influenced or remains certain
constant. Instead, ``\textcolor{red}{$\uparrow$}'' and ``\textcolor{green}{$\downarrow$}''
stand for the increase and decrease with lowering the energy scale,
respectively (the double arrows indicate much more increase or decrease).}\label{table-lambda}
\vspace{0.39cm}
\centerline{
\begin{tabular}{cccccc}
\hline
\hline
\renewcommand{\arraystretch}{1}
&\hspace{-0.2cm}$M_0$\hspace{0.2cm}&$M_1$&\hspace{0.2cm}$M_2$\hspace{0.2cm}&$\hspace{0.2cm}
M_3$\hspace{0.2cm}&$\hspace{0.2cm}M_{0123}$\\
\hline
\hspace{-0.4cm}Gaussian FP&\textcolor{blue}{\hspace{-0.4cm}--}& \textcolor{red}{$\uparrow$}
&\textcolor{green}{$\downarrow\downarrow$}&
\hspace{0.1cm}\textcolor{red}{$\uparrow$}&\textcolor{green}{$\downarrow\downarrow$}  \\
\hspace{-0.4cm}\hspace{0.7cm}\multirow{2}{2cm}{QAH FP}  &\multirow{2}{0.2cm}{\hspace{-0.2cm}\textcolor{blue}{--}}
&\hspace{-0.2cm}\textcolor{green}{$\downarrow$}\hspace{0.2cm}$\frac{\eta_0}{\pi}\in(0,0.07)$&
\hspace{-0.1cm}\multirow{2}{0.2cm}{\textcolor{green}{$\downarrow\downarrow$}}
&\multirow{2}{0.2cm}{\textcolor{green}{$\downarrow\downarrow$}} &\multirow{2}{0.2cm}{\textcolor{green}{$\downarrow\downarrow$}} \\
& &\textcolor{red}{$\uparrow$}\hspace{0.2cm}$\frac{\eta_0}{\pi}\in(0.07,0.5)$& & &\\
\hspace{-0.4cm}NSN FP  &\textcolor{blue}{\hspace{-0.4cm}--}&\textcolor{blue}{--}&\textcolor{green}{$\downarrow\downarrow$}
&\hspace{0.1cm}\textcolor{red}{$\uparrow$} &\textcolor{green}{$\downarrow\downarrow$} \\
\hline
\hline
\end{tabular}
}
\end{table}

\vspace{0.8cm}
\section{Summary}\label{Sec_summary}

In summary, we study the effects of the competitions between
fermion-fermion interactions and impurity scatterings on the low-energy
stabilities of 2D asymmetric QBCP materials by virtue of the RG approach~\cite{Shankar1994RMP,Wilson1975RMP,Polchinski9210046}.
To be specific, we find that the band structure and dispersion
of 2D QBCP systems are considerably robust irrespective of weak or
strong impurities while fermion-fermion interactions flow toward the
Gaussian FP. Rather, they can either be stable or
collapsed if fermionc couplings are attracted by nontrivial QAH and NSN FPs.
Table~\ref{table-stability}
briefly shows our main results for distinct kinds of FPs under distinct sorts of impurities.
Additionally, the low-energy fates of rotational and particle-hole
asymmetries that are characterized by $\eta$ and
$\lambda$ are attentively investigated. With variations of initial
values of fermion-fermion interactions and impurities, these two parameters exhibit
intimately energy-dependent behaviors and can
either be remarkably increased or heavily reduced. It is worth
highlighting that the $M_2$ impurity plays the most important role
among all impurities. Table~\ref{table-eta} and Table~\ref{table-lambda}
collect qualitative results for different sorts
of impurities. In addition to these basic conclusions, we find that,
under certain restricted condition in 2D QBCP systems,
the transition from rotational asymmetry or particle-hole
asymmetry to rotational symmetry or particle-hole symmetry
can be induced by the interplay between fermion-fermion
interactions and impurities.

Furthermore, we hereby present brief comments on symmetric
studies~\cite{Fradkin2009PRL,Murray2014PRB} and our asymmetric case.
With respect to the former that is an ideal limit,
the QBCP band structure is stable and both of two symmetries
are robust as well as the energy dependence of physical implications are absent due to
the trivial properties of structural parameters.
In comparison, we consider a general asymmetric
QBCP system and develop a suitable approach to judge whether the QBCP band
structure is stable and track how the asymmetries behave in the
whole energy region from a starting point to
certain FP. Our results theoretically corroborate the QBCP band structure may be
destroyed and asymmetric parameters can be heavily affected in
the vicinity of FPs. Additionally, we guess that the energy-dependent behaviors
of physical observables upon accessing a specific FP would be principally captured
in the asymmetric situation by taking advantage of the coupled RG equations consisting
of both structural and interaction parameters. These signal that monitoring the stability of
band structure is imperative while one studies the latter case.
To recapitulate, we expect our studies are instructive to provide helpful
clues for enhancing the stability and exploring other singular
behaviors of 2D QBCP systems.


\section*{ACKNOWLEDGEMENTS}

This work was supported by the National Natural Science Foundation of China under Grants
11504360 and 11704278 as well as Natural Science Foundation of Tianjin City under
Grant 19JCQNJC03000. We thank Dr. J.- R. Wang for helpful comments on our paper.

\appendix

\section{one-loop corrections}\label{Appendix_one-loop-corrections}

The one-loop corrections to self-energy, fermion-fermion couplings, and fermion-impurity
vertexes are depicted in Fig.~\ref{Fig_fermion_propagator_correction}, Fig.~\ref{Fig_fermion-fermion_correction} and Fig.~\ref{Fig_fermion-impurity_correction}, respectively.
After long but straightforward calculations, we obtain
\begin{widetext}
\begin{eqnarray}
\delta \Sigma(\omega,\mathbf{k})
\!\!&=&\!\!-\mathcal{N}_1(v^2_0\Delta_0+v^2_1\Delta_1+v^2_2\Delta_2+
v^2_3\Delta_3)(i\omega)
+\mathcal{N}_3(v^2_0\Delta_0+v^2_1\Delta_1+
v^2_2\Delta_2+v^2_3\Delta_3)(t_0\mathbf{k}^2)\nonumber\\
&&\!\!-\!\mathcal{N}_3(v^2_0\Delta_0+v^2_1\Delta_1-
v^2_2\Delta_2-v^2_3\Delta_3)(2t_1k_xk_y
\sigma_1)\!\!-\!\!\mathcal{N}_3(v^2_0\Delta_0-v^2_1\Delta_1-
v^2_2\Delta_2+v^2_3\Delta_3)[t_3(k^2_x\!-\!k^2_y)\sigma_3],
\end{eqnarray}
for one-loop corrections to self-energy derived from Fig.~\ref{Fig_fermion_propagator_correction},
\begin{eqnarray}
\delta \mathcal{S}_{u_0}
\!\!&=&\!\!u_0\int^{+\infty}_{-\infty}\frac{d\omega_1d\omega_2d\omega_3}{(2\pi)^3}
\int^{b}\frac{d^2\mathbf{k}_1d^2\mathbf{k}_2d^2\mathbf{k}_3}{(2\pi)^6}
\psi^\dagger(\omega_1,\mathbf{k}_1)\sigma_0\psi(\omega_2,\mathbf{k}_2)
\psi^\dagger(\omega_3,\mathbf{k}_3)
\sigma_0\psi(\omega_1+\omega_2-\omega_3,\mathbf{k}_1+\mathbf{k}_2-\mathbf{k}_3)\nonumber\\
&&\times\Bigl\{\frac{l}{8\pi u_0\sqrt{2(t^2_1+t^2_3)}}\left[
-\mathcal{C}_1(u^2_0+u^2_1+u^2_2+u^2_3)
-\mathcal{C}_2(u_0u_1+u_2u_3)-\mathcal{C}_3(u_0u_3+u_1u_2)
\right.\nonumber\\
&&\left.-\mathcal{C}_4(u_0u_1-u_2u_3)-\mathcal{C}_5(u_0u_3-u_1u_2)\right]
+2\pi l\sqrt{2(t^2_1+t^2_3)}
\left[v^2_0\Delta_0(\mathcal{D}_0+2\mathcal{N}_1)\right.\nonumber\\
&&\left.+2v^2_2\Delta_2\mathcal{N}_1+v^2_1\Delta_1(\mathcal{D}_1+2\mathcal{N}_1)
+v^2_3\Delta_3(\mathcal{D}_2
+2\mathcal{N}_1)\right]\Bigl\},\\
\delta \mathcal{S}_{u_1}
\!\!&=&\!\!\left(u_1\int^{+\infty}_{-\infty}\frac{d\omega_1d\omega_2d\omega_3}{(2\pi)^3}\int^{b}
\frac{d^2\mathbf{k}_1d^2\mathbf{k}_2d^2\mathbf{k}_3}{(2\pi)^6}
\psi^\dagger(\omega_1,\mathbf{k}_1)\sigma_1\psi(\omega_2,\mathbf{k}_2)
\psi^\dagger(\omega_3,\mathbf{k}_3)
\sigma_1\psi(\omega_1+\omega_2-\omega_3,\mathbf{k}_1+\mathbf{k}_2-\mathbf{k}_3)
\right)\nonumber\\
&&\times\Bigl\{\frac{l}{4\pi u_1\sqrt{2(t^2_1+t^2_3)}}
\Bigl[(2u_0u_1-2u^2_1-2u_2u_1-3u_3u_1-u_0 u_2)\mathcal{C}_3+(u_0u_1+u_2u_3)(\mathcal{C}_2+\mathcal{C}_3)\nonumber\\
&&-\frac{1}{2}(u^2_0+u^2_1+u^2_2+u^2_3)
(\mathcal{C}_2+\mathcal{C}_4)
-(u_0u_1-u_2u_3)(\mathcal{C}_4+\mathcal{C}_5)-(u_1u_3-u_0u_2)
\mathcal{C}_5\Bigl]
\nonumber\\
&&+2\pi l\sqrt{2(t^2_1\!+\!t^2_3)}\!
\Bigl[v^2_0\Delta_0
(\mathcal{D}_3\!+\!2\mathcal{N}_1)\!\!+\!\!v^2_1\Delta_1
(\mathcal{D}_4\!+\!2\mathcal{N}_1)\!\!+\!\!v^2_2\Delta_2
(\mathcal{D}_5\!+\!2\mathcal{N}_1)\!\!+\!\!
v^2_3\Delta_3
(\mathcal{D}_6\!+\!2\mathcal{N}_1)\!\Bigl]\!\Bigl\},\\
\delta S_{u_2}
\!\!&=&\!\!
\left(u_2\int^{+\infty}_{-\infty}\frac{d\omega_1d\omega_2d\omega_3}{(2\pi)^3}\int^{b}
\frac{d^2\mathbf{k}_1d^2\mathbf{k}_2d^2\mathbf{k}_3}{(2\pi)^6}
\psi^\dagger(\omega_1,\mathbf{k}_1)\sigma_2\psi(\omega_2,\mathbf{k}_2)
\psi^\dagger(\omega_3,\mathbf{k}_3)
\sigma_2\psi(\omega_1+\omega_2-\omega_3,\mathbf{k}_1+\mathbf{k}_2-\mathbf{k}_3)
\right)\nonumber\\
&&\times\Big\{
\frac{l}{4\pi u_2\sqrt{2(t^2_1+\!t^2_3)}}\Big[\!(3u_0u_2-2u_1u_2-2u^2_2-2u_3u_2+u_1u_3)
(\mathcal{C}_2+\mathcal{C}_3)-
(u_0u_2-u_1u_3)(\mathcal{C}_4+\mathcal{C}_5)\nonumber\\
&&-(u_1u_2+u_0u_3)\mathcal{C}_2-(u_2u_3+u_0u_1)\mathcal{C}_3
-(u_1u_2-u_0u_3)\mathcal{C}_4-(u_2u_3-u_0u_1)\mathcal{C}_5
\Big]\nonumber\\
&&+\!2\pi l\sqrt{2(t^2_1\!+\!t^2_3)}\!\Big[v^2_0\Delta_0
(\mathcal{D}_7\!+\!2\mathcal{N}_1)\!\!+\!\!v^2_1\Delta_1
(\mathcal{D}_8\!+\!2\mathcal{N}_1)\!\!-\!\!
v^2_2\Delta_2(\mathcal{D}_9\!+\!2\mathcal{N}_1)
\!\!+\!\!v^2_3\Delta_3
(\mathcal{D}_{10}\!+\!2\mathcal{N}_1)\Big]\Big\},\\
\delta S_{u_3}
\!\!&=&\!\!\left(u_3\int^{+\infty}_{-\infty}\frac{d\omega_1d\omega_2d\omega_3}{(2\pi)^3}\int^{b}
\frac{d^2\mathbf{k}_1d^2\mathbf{k}_2d^2\mathbf{k}_3}{(2\pi)^6}
\psi^\dagger(\omega_1,\mathbf{k}_1)\sigma_3\psi(\omega_2,\mathbf{k}_2)
\psi^\dagger(\omega_3,\mathbf{k}_3)
\sigma_3\psi(\omega_1+\omega_2-\omega_3,\mathbf{k}_1+\mathbf{k}_2-\mathbf{k}_3)
\right)\nonumber\\
&&\times\Bigl\{\frac{l}{4\pi u_3\sqrt{2(t^2_1+t^2_3)}}\Bigl[
(2u_0u_3-3u_1u_3-2u_2u_3-2u^2_3-u_0u_2)
\mathcal{C}_2+(u_1u_2+u_0u_3)
(\mathcal{C}_2+\mathcal{C}_3)\nonumber\\
&&-(u_0u_3-u_1u_2)
(\mathcal{C}_4+\mathcal{C}_5)-(u_1u_3-u_0u_2)\mathcal{C}_4-
\frac{1}{2}(u^2_0+u^2_1+u^2_2+u^2_3)(\mathcal{C}_5+\mathcal{C}_3)\Bigl]
\nonumber\\
&&+2\pi l\sqrt{2(t^2_1+t^2_3)}
\!\Bigl[v^2_0\Delta_0(\mathcal{D}_{11}\!+\!2\mathcal{N}_1)\!\!+\!\!v^2_1\Delta_1
(\mathcal{D}_{12}\!+\!2\mathcal{N}_1)-
v^2_2\Delta_2(\mathcal{D}_{13}\!+\!2\mathcal{N}_1)
\!\!+\!\!v^2_3\Delta_3
(\mathcal{D}_{14}\!+\!2\mathcal{N}_1)\!\Bigl]\!\Bigl\}.
\end{eqnarray}
for one-loop corrections to fermion-fermion interactions
based on Fig.~\ref{Fig_fermion-fermion_correction}, and
\begin{eqnarray}
\delta S_{v_0}
\!\!&=&\!\!\left[v_0\int^\infty_{-\infty}\frac{d\omega}{2\pi}\int^b
\frac{d^2\mathbf{k}d^2\mathbf{k}^\prime}{(2\pi)^4}\Psi^\dag(\mathbf{k},\omega)M_0\Psi(\mathbf{k}^\prime,
\omega)D(\mathbf{k}-\mathbf{k}^\prime)\right]\nonumber\\
&&\times\left\{
2\mathcal{N}_1l\left[(v^2_0\Delta_0+v^2_1\Delta_1+v^2_2\Delta_2+v^2_3\Delta_3)
-(u_0+u_1+u_2+u_3)2\pi\sqrt{2(t^2_1+t^2_3)}\right]\right\},\\
\delta S_{v_1}
\!\!&=&\!\!\left[v_1\int^\infty_{-\infty}\frac{d\omega}{2\pi}\int^b
\frac{d^2\mathbf{k}d^2\mathbf{k}^\prime}{(2\pi)^4}\Psi^\dag(\mathbf{k},\omega)M_1\Psi(\mathbf{k}^\prime,
\omega)D(\mathbf{k}-\mathbf{k}^\prime)\right]\nonumber\\
&&\times\left\{2\mathcal{N}_2l\Big[(v^2_0\Delta_0+v^2_1\Delta_1-v^2_2\Delta_2-v^2_3\Delta_3)-
(u_0+u_1-u_2-u_3)2\pi\sqrt{2(t^2_1+t^2_3)}\Big]\right\},\\
\delta S_{v_2}
\!\!&=&\!\!\left[v_2\int^\infty_{-\infty}\frac{d\omega}{2\pi}\int^b
\frac{d^2\mathbf{k}d^2\mathbf{k}^\prime}{(2\pi)^4}\Psi^\dag(\mathbf{k},\omega)M_2\Psi(\mathbf{k}^\prime,
\omega)D(\mathbf{k}-\mathbf{k}^\prime)\right]\nonumber\\
&&\times\left\{2\mathcal{N}_3l\Big[(-v^2_0\Delta_0+v^2_1\Delta_1-
v^2_2\Delta_2+v^2_3\Delta_3)-(-u_0+u_1-u_2+u_3)2\pi\sqrt{2(t^2_1+t^2_3)}
\Big]\right\},\\
\delta S_{v_3}
\!\!&=&\!\!\left[v_3\int^\infty_{-\infty}\frac{d\omega}{2\pi}\int^b
\frac{d^2\mathbf{k}d^2\mathbf{k}^\prime}{(2\pi)^4}\Psi^\dag(\mathbf{k},\omega)M_3\Psi(\mathbf{k}^\prime,
\omega)D(\mathbf{k}-\mathbf{k}^\prime)\right]\nonumber\\
&&\times\left\{2\mathcal{N}_4l\left[(-v^2_0\Delta_0+v^2_1\Delta_1+
v^2_2\Delta_2-v^2_3\Delta_3)-(-u_0+u_1+u_2-u_3)2\pi\sqrt{2(t^2_1+t^2_3)}\right]\right\},
\end{eqnarray}
for one-loop corrections to the fermion-impurity couplings generated by
Fig.~\ref{Fig_fermion-impurity_correction}. Here, all the coefficients appearing in above equations
are designated in Appendix~\ref{Appendix_coefficients}.

\section{Related coefficients}\label{Appendix_coefficients}

\begin{figure*}[htbp]
\centering
\includegraphics[width=0.9in]{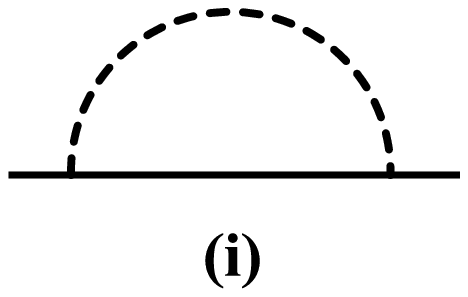}
\hspace{0.5in}
\includegraphics[width=0.9in]{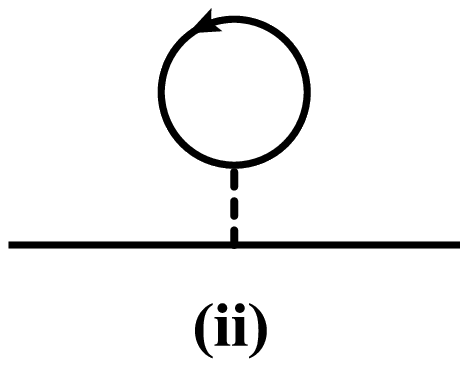}
\hspace{0.5in}
\includegraphics[width=0.9in]{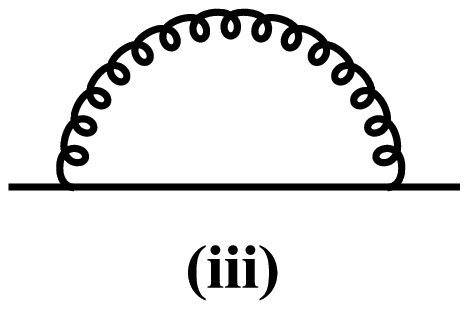}
\hspace{0.5in}
\includegraphics[width=0.9in]{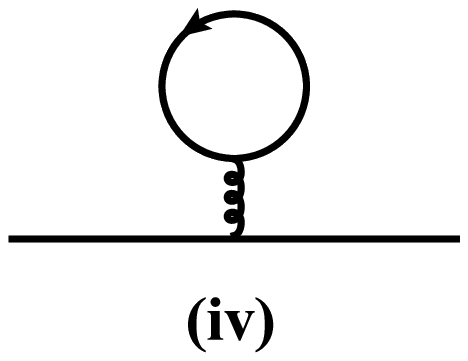}
\caption{One-loop corrections to the fermionic propagator
due to the interplay of fermion-fermion interaction and impurity scattering
(the dash and gluon lines indicate the fermion-fermion interaction and
fermion-impurity interaction, respectively).}\label{Fig_fermion_propagator_correction}
\end{figure*}

\begin{figure*}[htbp]
\centering
\includegraphics[width=0.79in]{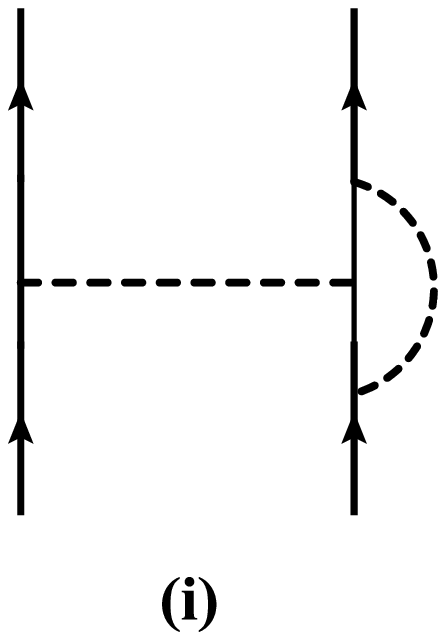}
\hspace{1cm}
\includegraphics[width=0.795in] {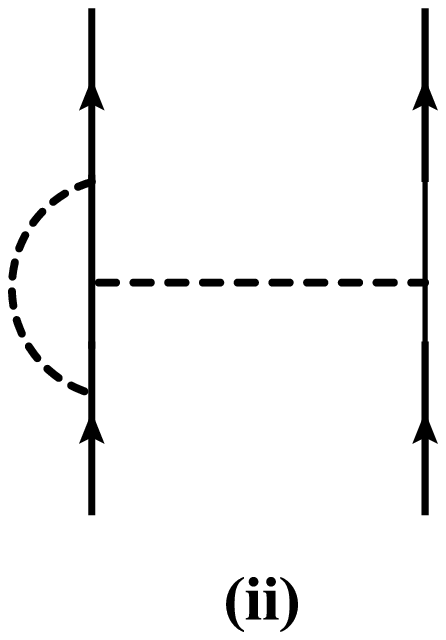}
\hspace{1cm}
\includegraphics[width=0.65in] {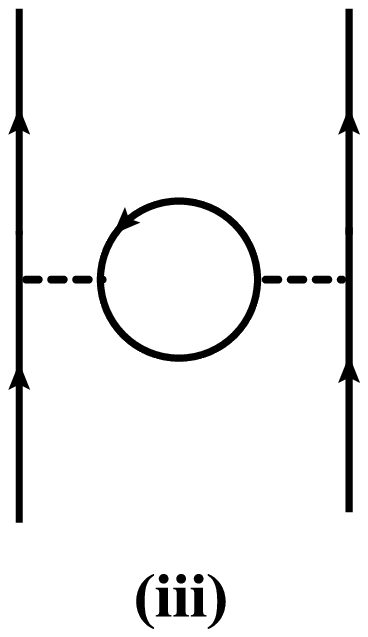}
\hspace{1cm}
\includegraphics[width=0.65in] {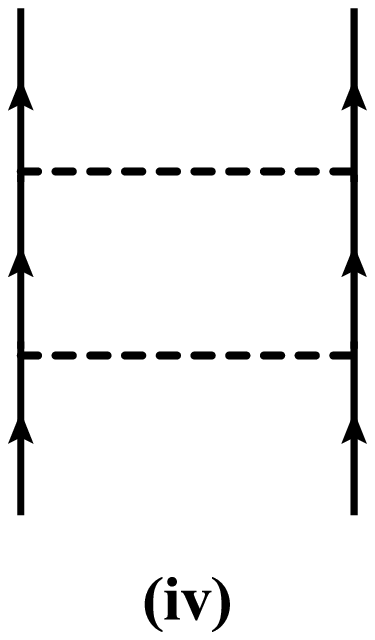}
\hspace{1cm}
\includegraphics[width=0.65in] {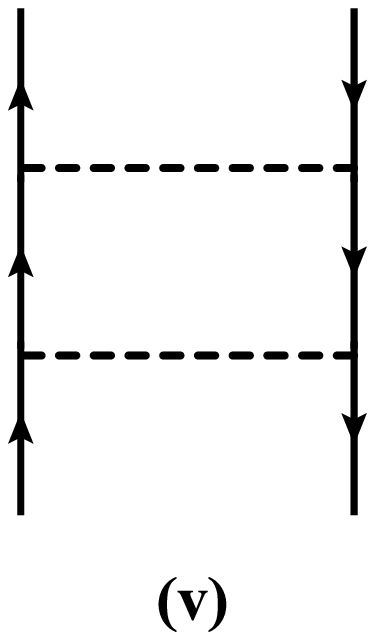}\\
\hspace{0cm}
\includegraphics[width=0.8in] {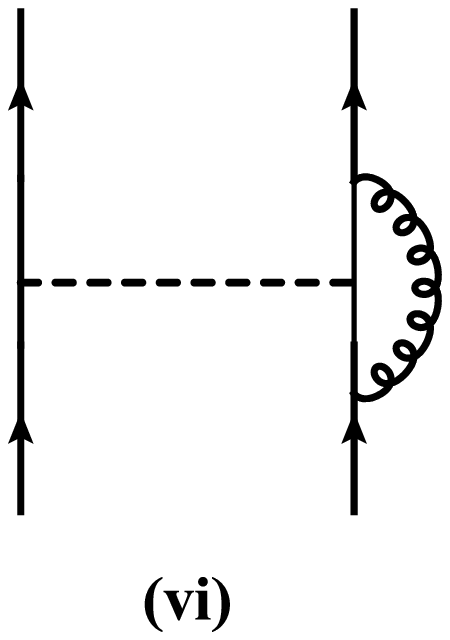}
\hspace{1cm}
\includegraphics[width=0.8in] {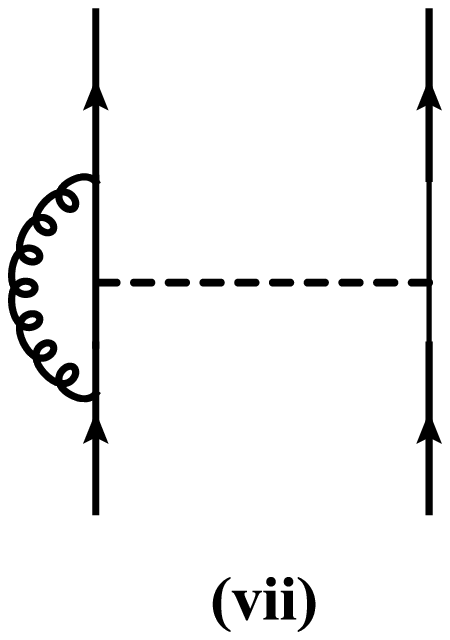}
\hspace{1cm}
\includegraphics[width=0.645in] {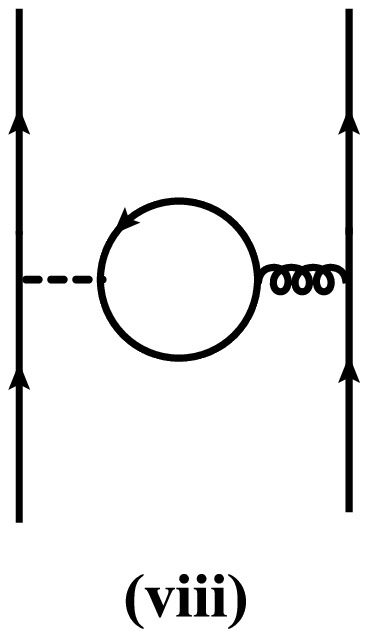}
\hspace{1cm}
\includegraphics[width=0.645in] {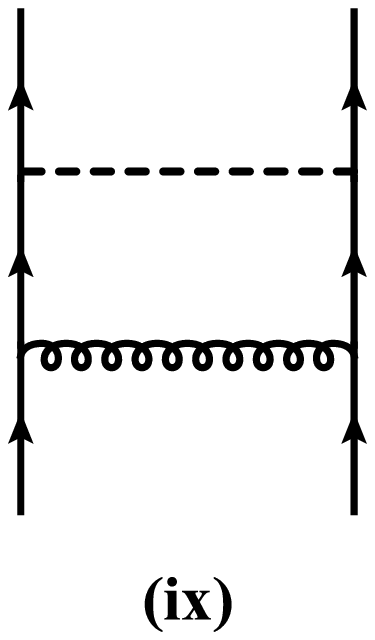}
\hspace{1cm}
\includegraphics[width=0.645in] {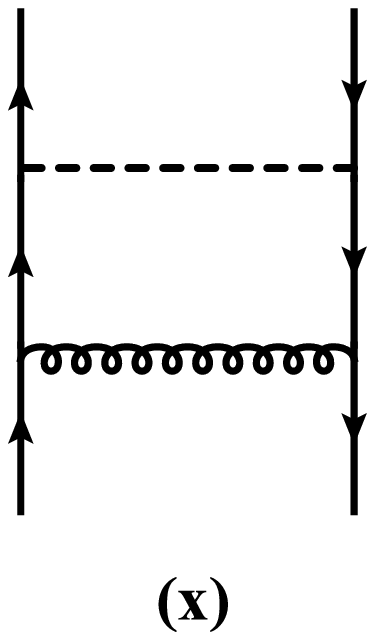}
\vspace{-0.1cm}
\caption{One-loop corrections to the fermion-fermion
interactions (the dash and gluon lines indicate the fermion-fermion interaction and
fermion-impurity interaction, respectively).}\label{Fig_fermion-fermion_correction}
\end{figure*}

\begin{figure*}[htbp]
\centering
\includegraphics[width=0.9in] {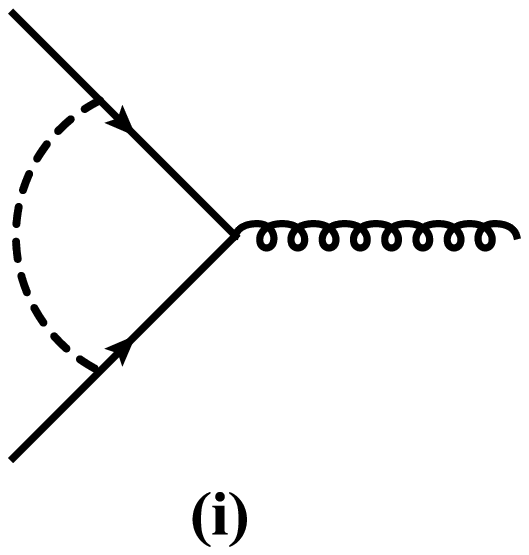}
\hspace{3cm}
\includegraphics[width=0.9in] {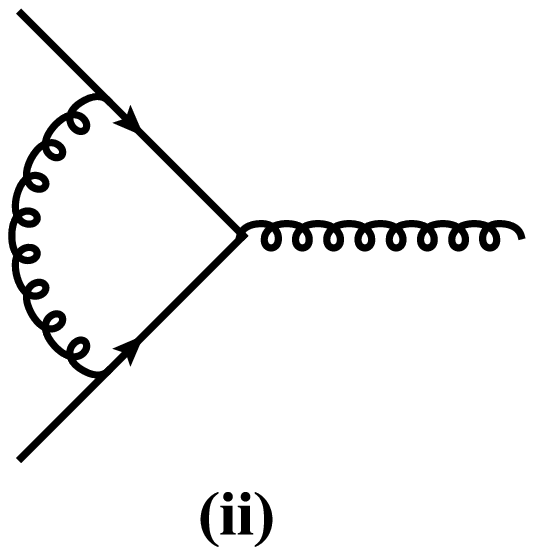}
\caption{One-loop corrections to the fermion-impurity
couplings (the dash and gluon lines indicate the fermion-fermion interaction and
fermion-impurity interaction, respectively). }\label{Fig_fermion-impurity_correction}
\end{figure*}

The related coefficients for the coupled RG equations in the maintext are listed as follows:
\begin{eqnarray}
\mathcal{C}_1&=&\int^{2\pi}_0d\theta
\frac{t^2_0\sqrt{2(t^2_1+t^2_3)}}{4\pi\mathcal{F}(t_0,t_1,t_3,\sin\theta,\cos\theta)
\sqrt{4t^2_1\cos^2\theta\sin^2\theta+t^2_3(\cos^2\theta-
\sin^2\theta)^2}},\label{Eq_C}\\
\mathcal{C}_2&=&\int^{2\pi}_0d\theta
\frac{4t^2_1\cos^2\theta\sin^2\theta\sqrt{2(t^2_1+t^2_3)}}
{{{2\pi[4t^2_1\cos^2\theta\sin^2\theta+t^2_3(\cos^2\theta-\sin^2\theta)^2]}}^\frac{3}{2}},
\mathcal{C}_3=\int^{2\pi}_0\!\!\!\!d\theta
\frac{t^2_3(\cos^2\theta-\sin^2\theta)^2\sqrt{2(t^2_1+t^2_3)}}
{{{2\pi[4t^2_1\cos^2\sin^2\theta+t^2_3(\cos^2\theta-\sin^2\theta)^2]}}^\frac{3}{2}},\\
\mathcal{C}_4&=&\int^{2\pi}_0d\theta
\frac{4t^2_1\cos^2\theta\sin^2\theta\sqrt{2(t^2_1+t^2_3)}}
{2\pi\mathcal{F}(t_0,t_1,t_3,\sin\theta,\cos\theta)
\sqrt{4t^2_1\cos^2\theta\sin^2\theta+t^2_3(\cos^2\theta-\sin^2\theta)^2}},\\
\mathcal{C}_5&=&\int^{2\pi}_0d\theta
\frac{t^2_3(\cos^2\theta-\sin^2\theta)^2\sqrt{2(t^2_1+t^2_3)}}
{2\pi\mathcal{F}(t_0,t_1,t_3,\sin\theta,\cos\theta)
\sqrt{4t^2_1\cos^2\theta\sin^2\theta+t^2_3(\cos^2\theta-\sin^2\theta)^2}}.
\end{eqnarray}
and
\begin{eqnarray}
\mathcal{D}_0&=&\int^{2\pi}_0d\theta
\frac{t^2_0}{2\pi^2\mathcal{F}^2(t_0,t_1,t_3,\sin\theta,\cos\theta)},\hspace{2.6cm}
\mathcal{D}_1=\int^{2\pi}_0d\theta
\frac{4t^2_1\cos^2\theta\sin^2\theta}{2\pi^2\mathcal{F}^2(t_0,t_1,t_3,\sin\theta,\cos\theta)},\\
\mathcal{D}_2&=&\int^{2\pi}_0d\theta
\frac{t^2_3(\cos^2\theta-\sin^2\theta)^2}
{2\pi^2\mathcal{F}^2(t_0,t_1,t_3,\sin\theta,\cos\theta)},\hspace{2.6cm}
\mathcal{D}_3=\int^{2\pi}_0d\theta
\frac{-t^2_3(\cos^2\theta-\sin^2\theta)^2+t^2_0}{
2\pi^2\mathcal{F}^2(t_0,t_1,t_3,\sin\theta,\cos\theta)},\\
\mathcal{D}_4&=&\int^{2\pi}_0d\theta
\frac{4t^2_1\cos^2\theta\sin^2\theta
-t^2_3(\cos^2\theta-\sin^2\theta)^2}{
2\pi^2\mathcal{F}^2(t_0,t_1,t_3,\sin\theta,\cos\theta)},\hspace{1.2cm}
\mathcal{D}_5=\int^{2\pi}_0d\theta
\frac{-4t^2_1\cos^2\theta\sin^2\theta
-t^2_0}{2\pi^2\mathcal{F}^2(t_0,t_1,t_3,\sin\theta,\cos\theta)},\\
\mathcal{D}_6&=&\int^{2\pi}_0d\theta
\frac{-4t^2_1\cos^2\theta\sin^2\theta
+t^2_3(\cos^2\theta-\sin^2\theta)^2-t^2_0}{
2\pi^2\mathcal{F}^2(t_0,t_1,t_3,\sin\theta,\cos\theta)},\hspace{0.2cm}
\mathcal{D}_7=\int^{2\pi}_0d\theta\frac{-1}{
2\pi^2\mathcal{F}(t_0,t_1,t_3,\sin\theta,\cos\theta)},\\
\mathcal{D}_8&=&\int^{2\pi}_0d\theta\frac{4t^2_1\cos^2\theta\sin^2\theta
-t^2_0}{2\pi^2\mathcal{F}^2(t_0,t_1,t_3,\sin\theta,\cos\theta)},\hspace{2.6cm}
\mathcal{D}_9=\int^{2\pi}_0d\theta\frac{t^2_0}{
2\pi^2\mathcal{F}^2(t_0,t_1,t_3,\sin\theta,\cos\theta)},\\
\mathcal{D}_{10}&=&\int^{2\pi}_0d\theta\frac{t^2_3(\cos^2\theta-\sin^2\theta)^2-t^2_0}{
2\pi^2\mathcal{F}^2(t_0,t_1,t_3,\sin\theta,\cos\theta)},\hspace{2.43cm}
\mathcal{D}_{11}=\int^{2\pi}_0
d\theta\frac{-4t^2_1\cos^2\theta\sin^2\theta
+t^2_0}{2\pi^2\mathcal{F}^2(t_0,t_1,t_3,\sin\theta,\cos\theta)},\\
\mathcal{D}_{12}&=&\int^{2\pi}_0
d\theta\frac{4t^2_1\cos^2\theta\sin^2\theta
-t^2_3(\cos^2\theta-\sin^2\theta)^2-t^2_0
}{2\pi^2\mathcal{F}^2(t_0,t_1,t_3,\sin\theta,\cos\theta)},\hspace{0.3cm}
\mathcal{D}_{13}=\int^{2\pi}_0
d\theta\frac{t^2_3(\cos^2\theta-\sin^2\theta)^2+t^2_0}{
2\pi^2\mathcal{F}^2(t_0,t_1,t_3,\sin\theta,\cos\theta)},\\
\mathcal{D}_{14}&=&\int^{2\pi}_0
d\theta\frac{-4t^2_1\cos^2\theta\sin^2\theta
+t^2_3(\cos^2\theta-\sin^2\theta)^2}{
2\pi^2\mathcal{F}^2(t_0,t_1,t_3,\sin\theta,\cos\theta)},\label{Eq_D}
\end{eqnarray}
as well as
\begin{eqnarray}
\mathcal{N}_1\!\!\!&=&\!\!\!\!\int^{2\pi}_0\!\!\!\!\!d\theta\frac{
4t^2_1\cos^2\theta\sin^2\theta+t^2_3(\cos^2\theta-\sin\theta)^2+t^2_0}
{8\pi^2\mathcal{F}^2(t_0,t_1,t_3,\sin\theta,\cos\theta)},\,
\mathcal{N}_2=\!\!\int^{2\pi}_0\!\!\!\!d\theta\frac{
4t^2_1\cos^2\theta\sin^2\theta-t^2_3(\cos^2\theta-\sin^2\theta)^2+t^2_0}
{8\pi^2\mathcal{F}^2(t_0,t_1,t_3,\sin\theta,\cos\theta)},\\
\mathcal{N}_3\!\!\!&=&\!\!\!\!\int^{2\pi}_0\!\!\!\!\!d\theta\frac{
4t^2_1\cos^2\theta\sin^2\theta+t^2_3(\cos^2\theta-\sin^2\theta)^2-t^2_0}
{8\pi^2\mathcal{F}^2(t_0,t_1,t_3,\sin\theta,\cos\theta)},\,
\mathcal{N}_4=\!\!\int^{2\pi}_0\!\!\!\!d\theta\frac{
4t^2_1\cos^2\theta\sin^2\theta-t^2_3(\cos^2\theta-\sin^2\theta)^2-t^2_0}
{8\pi^2\mathcal{F}^2(t_0,t_1,t_3,\sin\theta,\cos\theta)},\\
\mathcal{N}_5\!\!\!&=&\!\!\!\!\int^{2\pi}_0\!\!\!\!\!d\theta\frac{2t^2_0}
{8\pi^2\mathcal{F}^2(t_0,t_1,t_3,\sin\theta,\cos\theta)},
\,\hspace{1.9cm}
\mathcal{N}_6=\!\!\int^{2\pi}_0\!\!\!\!d\theta\frac{2[4t^2_1\cos^2\theta\sin^2\theta
+t^2_3(\cos^2\theta-\sin^2\theta)^2]}
{8\pi^2\mathcal{F}^2(t_0,t_1,t_3,\sin\theta,\cos\theta)},\label{Eq-N-6}
\end{eqnarray}
with $\mathcal{F}(t_0,t_1,t_3,\sin\theta,\cos\theta)$ being designated as
$\mathcal{F}(t_0,t_1,t_3,\sin\theta,\cos\theta)\equiv[4t^2_1\cos^2\theta\sin^2\theta
+t^2_3(\cos^2\theta-\sin^2\theta)^2-t^2_0]$.

\end{widetext}



\end{document}